\documentclass[12pt]{article}
\usepackage{a4wide}
\usepackage{amsmath,amssymb,bbm}
\usepackage{multirow}

\begin{document}
\setcounter{page}{1}
%


%

\def\pct#1{(see Fig. #1.)}

\begin{titlepage}
\hbox{\hskip 12cm KCL-MTH-07-17  \hfil}
\vskip 1.4cm
\begin{center}  {\Large  \bf  $E_{11}$-extended spacetime and gauged supergravities}

\vspace{1.8cm}

{\large \large Fabio Riccioni \ and \ Peter West} \vspace{0.7cm}

{\sl Department of Mathematics\\
\vspace{0.3cm}
 King's College London  \\
\vspace{0.3cm}
Strand \ \  London \ \ WC2R 2LS \\
\vspace{0.3cm} UK}
\end{center}
\vskip 1.5cm

\abstract{We formulate all the five dimensional gauged maximal supergravity theories as non-linear
realisations of the semi-direct product of $E_{11}$  and a set of  generators which transform according to
the first fundamental representation $l$ of $E_{11}$. The latter introduces a generalised space-time which
plays a crucial role for these theories. We derive the $E_{11}$ and $l$ transformations of  all the form
fields and their dynamics. We also formulate the five dimensional gauged supergravity theories using the
closure of the supersymmetry algebra. We show that this closes on the bosonic field content predicted by
$E_{11}$ and we derive the field transformations and the dynamics of this theory. The results are in precise
agreement with those found from the $E_{11}$ formulation. This provides a very detailed check of $E_{11}$ and
also the first substantial evidence for the generalised space-time. The results can be generalised  to all
gauged maximal supergravities, thus providing a unified framework of all these theories as part of $E_{11}$.}

\vfill
\end{titlepage}
\makeatletter \@addtoreset{equation}{section} \makeatother
\renewcommand{\theequation}{\thesection.\arabic{equation}}
\addtolength{\baselineskip}{0.3\baselineskip}

\section{Introduction}

One of the most surprising discoveries in the development of supergravities was the hidden symmetries in the
maximal supergravity theories in lower dimensions. The first to be discovered, in 1978, was the $E_7$
symmetry in four dimensions \cite{1}, while the last to be found in 1983 was the $SL(2,\mathbb{R})$ symmetry
of ten-dimensional IIB supergravity \cite{2}. The highest dimension in which a maximal supergravity multiplet
exists is eleven, and the corresponding theory \cite{3} is unique. This theory compactified on a circle gives
rise to the ten-dimensional IIA supergravity \cite{4}, while the IIB theory \cite{2,5} has no higher
dimensional origin. In any dimension below ten, maximal supergravity theories are unique and can be obtained
by torus dimensional reduction of both the ten dimensional theories and the eleven-dimensional one. The
hidden symmetry increases with the number of compact dimensions, and for instance one gets $E_6$ in five
dimensions \cite{6,7} and $E_8$ in three dimensions \cite{8}, corresponding to compactifying the eleven
dimensional theory on a six-torus and on an eight-torus respectively.

For many years it was universally assumed that these large symmetries were a quirk of dimensional reduction
on a torus and that they were not present in the uncompactified theories. In particular, it was believed that
there was no hidden symmetry in eleven dimensional supergravity. The reason for this is that these hidden
symmetries are associated with the scalars that occur in these theories, and more precisely the hidden
symmetries are non-linearly realised in the scalar sector. The fact that the eleven-dimensional theory has no
scalars was believed not to leave room for any large hidden symmetry. Furthermore, the symmetries found in
the dimensionally reduced theories are internal in that they commute with the spacetime symmetries. It
appeared not to be possible to realise these symmetries in the uncompactified theory, where they would have
to act non-trivially on the gravitational field.

However, in 2001 it was conjectured \cite{9} that eleven dimensional supergravity could be extended so as to
have a non-linearly realised infinite-dimensional Kac-Moody symmetry called $E_{11}$, whose Dynkin diagram is
shown in Fig. 1. We now list the main results supporting this conjecture.
\begin{itemize}
\item
Eleven dimensional supergravity itself can be formulated as a non-linear realisation \cite{10} of an algebra.
This non-linear realisation naturally gives rise to both a 3-form and a 6-form, and the resulting field
equations are first order duality relations, whose divergence reproduces the 3-form second-order field
equations of 11-dimensional supergravity. The eleven-dimensional gravity field describes non-linearly
$SL(11,\mathbb{R})$, which is a subalgebra of this algebra. Indeed, gravity in $D$ dimensions can be
described as a non-linear realisation of the closure of the group $SL(D,\mathbb{R})$ with the conformal group
\cite{10}, as was shown in the four dimensional case in \cite{11}.

\item $E_{11}$ is the smallest Kac-Moody algebra which contains the algebra found in the non-linear realisation
above. This $E_{11}$ algebra is infinite-dimensional, and the $E_{11}$ non-linear realisation contains an
infinite number of fields with increasing number of indices. The first few fields are the graviton, a three
form, a six form and a field which has the right spacetime indices to be interpreted as a dual graviton. This
is the field content of eleven dimensional supergravity, and keeping only the first three of these fields one
finds that the non-linear realisation of $E_{11}$ reduces to the construction discussed in the first point
and so results in the dynamics of  this theory \cite{9}.

\item Theories in $D$ dimensions arise from the $E_{11}$ non-linear realisation by choosing a suitable
$SL(D,\mathbb{R})$ subalgebra, which is  associated with $D$-dimensional gravity. The $A_{D-1}$ Dynkin
diagram of this subalgebra, called the gravity line, must include the node labelled 1 in the Dynkin diagram
of Fig. 1. In ten dimensions there are two possible ways of constructing this subalgebra, and the
corresponding non-linear realisations give rise to two theories that contain the fields of the IIA and IIB
supergravity theories and their electromagnetic duals \cite{9,12}. Below ten dimensions, there is a unique
choice for this subalgebra, and this corresponds to the fact that maximal supergravity theories in dimensions
below ten are unique. Again, the non-linear realisation in each case describes, among an infinite set of
other fields, the fields of the corresponding supergravity and their electromagnetic duals. In each
dimension, the part of the $E_{11}$ Dynkin diagram which is not connected to the gravity line corresponds to
the internal hidden symmetry of the $D$ dimensional theory. This not only reproduces all the hidden
symmetries found long ago in the dimensionally reduced theories, but it also gives an eleven-dimensional
origin to these symmetries.

\item The Weyl transformations of $E_{11}$ are the known U duality symmetries found in the IIA and IIB
supergravity theories and also when these are dimensionally reduced on tori \cite{13}.

\item It is a generic feature of $E_{11}$ and all very extended algebras that their non-linear realisation
contains at low levels the usual fields for the physical degrees of freedom as well as their magnetic duals
\cite{9,12,24}. Amongst the infinitely many fields in the non-linear realisation of $E_{11}$, there is an
infinite preferred set that describes all possible dualisations of the on-shell degrees of freedom of the
eleven-dimensional supergravity theory. This lifts the infinite set of dualities that occur in two dimensions
to eleven dimensions. All the infinitely many remaining fields in eleven dimensions have at least one set of
ten or eleven antisymmetric indices, and therefore they do not correspond to on-shell propagating degrees of
freedom \cite{14}.
\end{itemize}
All the maximal supergravity theories mentioned so far are massless in the sense that no other dimensional
parameter other than the Planck scale is present. In fact, even this parameter can be absorbed into the
fields such that it is absent from the equations of motion. There are however other theories that are also
maximal, {\it i.e.} invariant under 32 supersymmetries, but are massive in the sense that they possess
additional dimensionful parameters. These can be viewed as deformations of the massless maximal theories.
However, unlike the massless maximal supergravity theories they can not in general be obtained by a process
of dimensional reduction and in each dimension they have been determined by analysing the deformations that
the corresponding massless maximal supergravity admits. The first example of such a theory was found in four
dimensions \cite{15}, and it results from gauging an $SO(8)$ subgroup of the global symmetry $E_7$. The
highest dimension for which a massive deformation is allowed is ten, and the corresponding massive theory was
discovered by Romans \cite{16}. This theory possesses a single additional mass parameter and can be thought
of as a deformation of the IIA supergravity theory in which the two-form receives a mass via a Higgs
mechanism.

With the exception of the Romans theory, all the massive maximal supergravities possess a local gauge
symmetry carried by vector fields that is a subgroup of the symmetry group $G$ of the corresponding maximal
supergravity theory, and are therefore called gauged supergravities. In general these theories also have
potentials for the scalars fields which contain the dimensionful parameters as well as a cosmological
constant. Another typical feature of massive maximal supergravities is that their field content is not
usually the same as their massless counterparts. As an example consider the five-dimensional $SO(6)$ gauged
supergravity \cite{17}. While the massless maximal supergravity theory \cite{6,7} contains 27 abelian
vectors, the gauged one describes 15 vectors in the adjoint of $SO(6)$, as well as 12 massive 2-forms
satisfying self-duality conditions. One can regard this as an example of the rearrangement of degrees of
freedom induced by the Higgs mechanism. Given that $E_{11}$ contains in any dimensions all the fields of the
corresponding supergravity together with their magnetic duals, this phenomenon turns out to be automatically
encoded in the $E_{11}$ non-linear realisation.

In recent years there have been a number of systematic searches for gauged maximal supergravity theories and
in particular in nine dimensions and in dimension from seven to three all such theories have been classified
\cite{18,19,20}. A crucial ingredient in the classification is played by the so called {\it embedding
tensor}, that encodes all the possible massive deformations of a given massless theory. This classification
is in perfect agreement with $E_{11}$, and this leads us to the last three points in our list of the main
results supporting the $E_{11}$ conjecture, which are related to the analysis of the $E_{11}$ fields that do
not correspond to the propagating fields of supergravity or to their duals.
\begin{itemize}

\item
The cosmological constant of ten-dimensional Romans IIA theory can be described as the dual of a 10-form
field-strength \cite{21}, and the supersymmetry algebra closes on the corresponding 9-form potential
\cite{22}. This theory was found to be a non-linear realisation \cite{23} which includes a 9-form. This
9-form is automatically encoded in $E_{11}$ \cite{24}, where it arises in the dimensional reduction of the
eleven-dimensional field $A_{a_1 \dots a_{10} , (bc)}$ in the irreducible representation of
$SL(11,\mathbb{R})$ with ten antisymmetric indices $a_1 \dots a_{10}$ and two symmetric indices $b$ and $c$.
Therefore $E_{11}$ not only contains Romans IIA, but it also provides it for the first time with an
eleven-dimensional origin \cite{25}.
\item
The $E_{11}$ non-linear realisation in ten dimensions also predicts the number of spacetime-filling 10-forms
that arise in IIA and IIB supergravities, the result being that there are an $SL(2,\mathbb{R})$ quadruplet
and a doublet of 10-forms in IIB and two 10-forms in IIA \cite{24}. Although these forms are non-propagating
and have no field strength, they are associated to spacetime-filling branes whose presence is crucial for the
consistency of orientifold models. The analysis of 10-forms performed imposing the closure of the
supersymmetry algebra shows perfect agreement with the $E_{11}$ predictions, for both the IIB \cite{26} and
the IIA \cite{27} case. Also, the gauge algebra that supersymmetry implies is exactly reproduced by $E_{11}$
\cite{28}.

\item By studying the eleven-dimensional fields of the $E_{11}$ non-linear realisation, one can
determine all the forms, {\it i.e.} fields with completely antisymmetric indices, that arise from dimensional
reduction to any dimension \cite{29}. In particular, in addition to all the lower rank forms, this analysis
gives all the $D-1$-forms and the $D$-forms in $D$ dimensions. The $D-1$ and $D$-forms predicted by $E_{11}$
can also be derived in each dimension separately \cite{30}. The $D-1$-forms have $D$-form field strengths,
that are related by duality to the mass deformations of gauged maximal supergravities, and the $E_{11}$
analysis shows perfect agreement with the complete classification of gauged supergravities performed in
\cite{19,20}. Therefore $E_{11}$ not only contains all the possible massive deformations of maximal
supergravities in a unified framework, but it also provides an eleven-dimensional origin to all of them.
Indeed, while some gauged supergravities were known to be obtainable using dimensional reduction of ten or
eleven dimensional supergravities, this was not generically the case. As a result the gauged supergravities
were outside the framework of M-theory as it is usually understood. The $D$-forms are associated to
spacetime-filling branes in $D$ dimensions, which again play a crucial role in string theory, and their
classification was not known, apart from the ten-dimensional case.
\end{itemize}
The net upshot of all this is that there is overwhelming evidence for an $E_{11}$ symmetry in the low energy
dynamics of what is often called M theory. The above evidence concerns the adjoint representation of
$E_{11}$, or the part of the non-linear realisation that involves the fields associated with the  $E_{11}$
generators. However, there is also the question of how space-time is encoded in the theory. In the non-linear
realisations mentioned above the  generator of space-time translations $P_a$ was introduced by hand in order
to encode the coordinates of space-time. From the beginning it was understood that this was an ad-hoc step
that did not respect the $E_{11}$ symmetry. It  was subsequently proposed \cite{31} that one could include an
$E_{11}$ multiplet of generators which had as its lowest component the generator of space-time translations.
This was just the fundamental representation of $E_{11}$ associated with the node labelled 1 in the Dynkin
diagram of Fig. 1 and it is denoted in this paper by $l$. The evidence for the relevance of the $l$ multiplet
and this method of introducing space-time is as follows.
\begin{itemize}
\item The infinitely many generators in the $l$ multiplet have an increasing number of eleven-dimensional
space-time indices. The next two generators after the $P_a$'s are objects with two and  five totally
anti-symmetrised indices that can be identified with  the central charges of the eleven-dimensional
supersymmetry algebra, then followed by an infinite number of further elements \cite{31}.

\item  The members of the $l$ multiplet can be identified with the brane charges. This is clearly the
case at the lowest levels, but one can show that to every element of $l$ there  corresponds a field in the
adjoint representation of $E_{11}$ to which the corresponding brane would couple \cite{32}.

\item  By decomposing the $l$ multiplet into
representations of $SL(D,\mathbb{R})\otimes G$, where $SL(D,\mathbb{R})$ is the $E_{11}$ sub-algebra
associated with the $D$-dimensional non-linear realisation of gravity and $G$ the internal symmetry group as
described above, one can find the brane charges predicted in the $D$ dimensional theory. For each type of
brane, {\it i.e.}  point particle, string, etc, one finds charges that are in multiplets of $G$ \cite{33}.
The low level results are summarised in table 1 \cite{33,34}. In fact, the very lowest level brane multiplets
had previously been found \cite{35} by applying the known U-duality rules to a familiar brane charge. The
results from the $l$ multiplet are in complete agreement with those found previously. This check is
comparable to the later one discussed above for the deformation forms of gauged supergravities. As in that
case, $E_{11}$ also provides a previously missing unifying framework for the brane charges, many of which
previously had no higher dimensional origin and could not be identified with charges in the supersymmetry
algebra.

\item The dynamics is taken to be the non-linear realisation based on  $E_{11}\otimes_s l$ which
stands for the semi-direct product between the two algebras. The presence of
the $l$ generators results in an infinite number of coordinates which in eleven dimensions take the form
$$
x^a, \ x_{a_1 a_2}, \ x_{a_1\ldots a_5}, \  x_{a_1\ldots a_7 ,b}, \  x_{a_1\ldots a_8},\ \ldots
$$
As a result, the fields would generically depend on a generalised space-time that is infinite dimensional.
This has the nice interpretation in that one uses a formulation of space-time that includes all possible ways
of measuring it and not just the $x^a$ corresponding to a point particle \cite{31}. The non-linear
realisation mentioned above corresponds to considering the lowest order in the $l$ multiplet, {\it i.e.} only
considering the dependence on the usual coordinate $x^a$ of spacetime. This has similar aspects to the
subsequently proposed generalised geometry, as already pointed out in \cite{37}. The additional coordinates
as seen in $D$ dimensions can be read off from table 1.
\end{itemize}
Thus although there is very good evidence that the $l$ multiplet does correctly account for the brane
charges, there is so far very little evidence for the generalised space-time that should be present in the
non-linear realisation. One of the main results in this paper is to find the dynamics of the five dimensional
gauged supergravities using their formulation as a $E_{11} \otimes_s l$ non-linear realisation. In this
calculation some of the generators of $l$, and their corresponding coordinates, will play a crucial role.

An alternative method of introducing space-time has been considered in the context of $E_{10}$ \cite{36}. In
this approach the fields depend only on time and the spatial derivatives of the fields are conjectured to be
some of  the higher level fields in  $E_{10}$ which are known to have the appropriate $A_9$ structure.

There are two obvious problems in trying to formulate the dynamics of gauged supergravities using non-linear
realisations. The first is that the field-strengths that arise in gauged supergravities contain terms that
have no space-time derivatives while the dynamics which follows from a non-linear realisation  is usually
constructed from the Cartan forms that explicitly contain derivatives. The second problem is that the gauged
supergravity theories involve vector fields that possess non-abelian gauge transformations. Given that
$E_{11}$ is automatically democratic \cite{9,12,24}, in the sense that each form appears with its magnetic
dual, one has to introduce fields that are dual to the non-abelian vectors. For instance these would be
2-forms in five dimensions. These 2-forms transform under the just mentioned Yang-Mills transformations, but
also possess their own gauge symmetry.

In this paper we want to show how the dynamics of gauged supergravities arises in the $E_{11}$ non-linear
realisation. We will see that the two problems above are solved. The first problem is solved by the presence
of the generalised coordinates. One finds terms independent of space-time because some of the derivatives in
the Cartan forms are not those of the familiar space-time, but of the higher coordinates in the generalised
space-time and so they read off the dependence of the fields on these coordinates which as it turns out is
rather constrained. The second problem is solved by considering all the $E_{11}$ form fields and dual form
fields and their corresponding $E_{11}$ transformations in the presence of the generalised coordinates. We
will focus in particular on the five-dimensional case, and show that the non-linear realisation gives the
required gauge-covariant field strengths provided that each form transforms with respect to the gauge
parameter of the form of higher rank. This means that the vector $A_\mu$ has a shift gauge transformation
$\delta A_\mu \sim \Sigma_\mu$ with respect to the gauge parameter $\Sigma_\mu$ of the 2-form, {\it i.e.}
$\delta B_{\mu \nu} = 2\partial_{[\mu} \Sigma_{\nu ]}$, and the 2-form has a shift gauge transformation with
respect to the parameter of the 3-form, and so on. It is this requirement that makes it possible to write a
field strength for the 2-form $B_{\mu\nu}$ that is covariant under the non-abelian gauge transformations
associated with $A_\mu$ and invariant under its own gauge transformations $\Sigma_\mu$, and thus this result
deeply relies on the fact that one has a fully democratic description. Proceeding this way one can write down
gauge-covariant field strengths and gauge-invariant duality relations in all cases. In particular, in the
five-dimensional case the vectors are dual to the 2-forms and the 3-forms are dual to the scalars. This
construction can be generalised to any gauged maximal supergravity, and more generally to any gauged theory
that admits a Kac-Moody description.

In order to provide a check of our $E_{11}$ derivation of gauged supergravities in five dimensions we
consider the supersymmetry transformations on the democratic set of form fields required by $E_{11}$. We find
that the supersymmetry algebra of gauged maximal supergravity in five dimensions does indeed close on the
2-forms and the 3-forms predicted by $E_{11}$, provided that the duality relations between the 2-forms and
the 1-forms, as well as between the 3-forms and the scalars, are satisfied.  We recover precisely the
dynamics implied by the $E_{11}$ non-linear realisation. In fact the features of the gauge algebra associated
to the higher rank fields was discussed in an independent bottom-up approach in \cite{hierarchies}, where the
results of \cite{19,20} were extended to higher rank forms. Our result therefore shows for the first time
that supersymmetry is compatible with this extension.

In order to derive these results, we first have to compare the $E_{11}$ and the supergravity results in the
massless case. We therefore consider the five-dimensional case, and we first determine the massless dynamics
as results from $E_{11}$. We then show that the supersymmetry algebra of massless five-dimensional
supergravity admits a democratic formulation, and we close the algebra on all the forms in the theory with
the exception of the spacetime-filling forms. The results we find using supersymmetry exactly reproduce the
ones obtained from $E_{11}$, and we show in detail how the computations are performed in the two cases, so
that the reader can appreciate how simple they are on the $E_{11}$ side. We then consider the gauged case,
and describe the democratic formulation of gauged maximal supergravity in five dimensions using the
supersymmetry algebra. We finally compare these results with those found from the $E_{11} \otimes_s l$
non-linear realisation and find complete agreement. This analysis shows the crucial role that the $l$
multiplet and its associated generalised coordinates have in describing the dynamics of gauged maximal
supergravities.

The paper is organised as follows. In section 2 we describe how the massless dynamics arises in the $E_{11}$
non-linear realisation. Before considering the five-dimensional case, we review the eleven-dimensional one to
make the reader familiar with the algebra. In section 3 we show that the supersymmetry algebra of massless
maximal supergravity in five dimensions closes on the 2-forms and 3-forms dual to the vectors and the scalars
respectively, and on the 4-forms predicted by $E_{11}$. The field strengths of the 4-forms are dual to the
mass deformation parameters, and thus they vanish in the massless theory. Section 4 is devoted to the
analysis of the supersymmetry algebra of gauged maximal five-dimensional supergravity. We show that the
algebra closes on the 2-forms and the 3-forms, and we determine the duality relation between the field
strengths of the 4-forms and the mass deformation parameters. In section 5 we show how the $E_{11} \otimes_s
l$ non-linear realisation gives rise to gauged maximal supergravities, focusing on the five-dimensional case.
Section 6 is devoted to a detailed discussion of the $E_{11} \otimes_s l$ non-linear realisation for the case
of the five-dimensional $SO(6)$ gauged supergravity. Section 7 contains the conclusions. We also include two
appendices. In appendix A we determine the gauge transformations of the 5-forms of maximal five-dimensional
supergravity from $E_{11}$, and in appendix B we show how the gauging results from generalised coordinates in
the non-linear realisation for a toy model that illustrates the main features of section 5.

\section{$E_{11}$ and massless dynamics}
In this section we will show how the $E_{11}$ non-linear realisation generates the gauge transformations and
the field strengths of all the fields with completely antisymmetrised indices in the five-dimensional case.
As a preliminary step to make the reader familiar with the notation, we will first review the original
$E_{11}$ computation in eleven dimensions \cite{9}, where the gauge transformations and field strengths of
the 3-form and its dual 6-form of eleven-dimensional supergravity were derived.

In \cite{9} it was conjectured that an extension of eleven dimensional supergravity can be described by a
non-linear realisation based on the Kac-Moody algebra $E_{11}$ resulting from the Dynkin diagram of Fig. 1.
\begin{figure}[h]
\begin{center}
\begin{picture}(380,60)
\multiput(10,10)(40,0){10}{\circle{10}} \multiput(15,10)(40,0){9}{\line(1,0){30}} \put(290,50){\circle{10}}
\put(290,15){\line(0,1){30}} \put(8,-8){$1$} \put(48,-8){$2$} \put(88,-8){$3$} \put(128,-8){$4$}
\put(168,-8){$5$} \put(208,-8){$6$} \put(248,-8){$7$} \put(288,-8){$8$} \put(328,-8){$9$} \put(365,-8){$10$}
\put(300,47){$11$}
\end{picture}
\caption{\sl The $E_{11}$ Dynkin diagram corresponding to 11-dimensional supergravity.}
\end{center}
\end{figure}
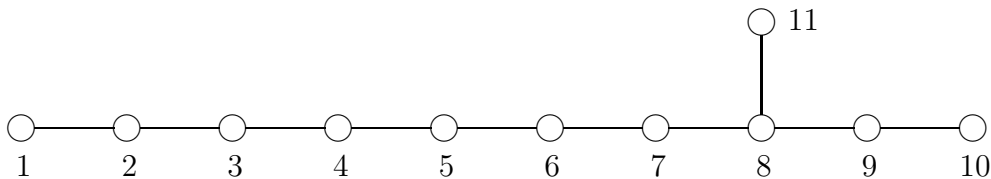
The horizontal line in the Dynkin diagram, associated with the $SL(11,\mathbb{R})$ subalgebra that in the
non-linear realisation is associated to the eleven dimensional gravity sector of the theory, is called the
gravity line.

The generators of $E_{11}$ are essentially the ones of $SL(11,\mathbb{R})$ together with the generators
$R^{abc}$ and $R_{abc}$, in the representations of $SL(11,\mathbb{R})$ with three antisymmetric indices,
associated to the exceptional node, and multiple commutators thereof, subject to the Jacobi identities. More
precisely, $E_{11}$ is defined as a Kac-Moody algebra, which is obtained by multiple commutators of the
Chevalley generators subject to the Serre relations. The multiple commutators of the Chevalley generators of
$SL(11,\mathbb{R})$ lead to all the generators of $SL(11,\mathbb{R})$, while the multiple commutators of
these with the Chevalley generator associated with the exceptional node lead to $R^{abc}$ and $R_{abc}$. All
the other generators are then found from multiple commutators of $R^{abc}$ and $R_{abc}$, subject to the
Serre relations. It is useful to classify the generators of the algebra in terms of the number of times the
generator $R^{abc}$ occurs in the commutators defining them, as was shown in \cite{9}. This was subsequently
called the {\it level}. As an example, the generator with 6 antisymmetric indices occurs in the commutator
  \begin{equation}
  [R^{abc}, R^{def}] = 2 R^{abcdef} \label{2.1}
  \end{equation}
and therefore corresponds to level 2. The generator $R_{abc}$ has level $-1$, and therefore all the generators
with lower indices have negative level. In general, the generators at level $l$ have $3l$ upper indices if $l$
is positive, and $-3l$ lower indices if $l$ is negative. The only $E_{11}$ generators whose $SL(11,\mathbb{R})$
indices are completely antisymmetric are $R^{abc}$ and $R^{a_1 \dots a_6}$, together with their negative level
counterparts.

By definition, the non-linear realisation must be invariant under
  \begin{equation}
  g \rightarrow g_0 \ g \ h \quad , \label{2.2}
  \end{equation}
where $g_0$ is a global $E_{11}$ transformation and $h \in H$ is a local transformation (to be precise, $H$
is the Cartan involution invariant subalgebra, which is the infinite-dimensional generalisation of the
maximal compact subalgebra of finite-dimensional groups). This local transformation can be used to put the
group element in the Borel subgroup, which is the one generated by the Cartan subalgebra and the generators
associated with the positive roots. As a result, there is a one-to-one correspondence between the fields of
the theory and the generators of $E_{11}$ with non-negative level. At level zero, this results in the
description of gravity as a non-linear realisation, and the level zero field is therefore the graviton. The
generator $R^{abc}$ at level 1 corresponds to the 3-form $A_{abc}$ of 11-dimensional supergravity and the
generator $R^{a_1 \ldots a_6}$ at level 2 to its 6-form dual $A_{a_1 \ldots a_6}$. The generator at level 3
$R^{a_1 \dots a_8 , b}$  in the irreducible hooked Young Tableaux representation with 8 antisymmetric indices
corresponds to the dual graviton \cite{9}. At level 3 one might expect also a generator with 9 completely
antisymmetric indices, but this is ruled out due to the Jacobi identities.

In this paper, we want to analyse the gauge transformations and field strengths of the fields with completely
antisymmetric indices in $E_{11}$, and thus in this 11-dimensional case we are interested in the fields up to
and including level 2. We therefore write down only the relevant part of the group element, which is
  \begin{equation}
  g = {\rm exp} (x^\mu P_\mu ) \ g_A  \quad , \label{2.3}
  \end{equation}
where
  \begin{equation}
  g_A = {\rm exp} ({1 \over 6!} A_{a_1 \ldots a_6} R^{a_1 \ldots a_6 } ) \ {\rm exp} ( {1 \over 3!}
  A_{a_1 \ldots a_3}
  R^{a_1 \ldots a_3 } ) \label{2.4}
  \end{equation}
is the part of $g$ that contains the 3-form and the 6-form. This way of writing down the group element
differs from the original one of \cite{10,9} only by terms of higher level, which do not affect the
computation we are reviewing. The momentum generator $P_\mu$ introduces spacetime, and is only the first part
of an infinite dimensional representation of $E_{11}$ which we call the $l$ representation \cite{31}, that
will also be discussed in detail in section 5. A global $E_{11}$ transformation of $g$ acting from the left
leaves the Maurer-Cartan form
  \begin{equation}
  {\cal {V}} = g^{-1}dg \label{2.5}
  \end{equation}
invariant. In particular, we are interested in the $E_{11}$ transformations generated by
  \begin{equation}
  g_0^{(3)} = {\rm exp}({1 \over 3!} a_{a_1 \ldots a_3} R^{a_1 \ldots a_3 })\label{2.6}
  \end{equation}
and
  \begin{equation}
  g_0^{(6)} = {\rm exp}({1 \over 6!} a_{a_1 \ldots a_6} R^{a_1 \ldots a_6 }) \quad , \label{2.7}
  \end{equation}
where $a_{a_1 \dots a_3}$ and $a_{a_1 \dots a_6}$ are infinitesimal and constant. These parameters are global
transformations of the $E_{11}$ fields, and in particular we can read the transformations of the fields
$A_{a_1 \dots a_3}$ and $A_{a_1 \dots a_6}$ in (\ref{2.4}). These transformations will be promoted to gauge
transformations as we will see in the following, and are determined computing the part of $g_0 g_A$ that has
the form $g_{A^\prime}$, where $A^\prime$ are the transformed fields. We use the operator identities
  \begin{eqnarray}
  & & {\rm exp}C \ {\rm exp}B = \dots {\rm exp}(- {1 \over n!}[B,[B \dots [B,[B,C]]\dots ]]) \dots \nonumber \\
  & & {\rm exp}(- {1 \over 2} [B,[B,C]] ) \ {\rm exp}(- [B,C]) \ {\rm exp}B \ {\rm exp}C \label{2.8}
  \end{eqnarray}
and
  \begin{eqnarray}
  & & {\rm exp}C \ {\rm exp}B = \dots {\rm exp}(- {1 \over (n+1)!}[B,[B \dots [B,[B,C]]\dots ]]) \dots \nonumber \\
  & & {\rm exp}(- {1 \over 6} [B,[B,C]] ) \ {\rm exp}(- {1 \over 2}[B,C]) \ {\rm exp}(B+C)  \quad , \label{2.9}
  \end{eqnarray}
where $B$ and $C$ are any operators and  we are only considering first order in $C$, so that we neglect $C^2$
and higher order. Multiplying eq. (\ref{2.8}) by ${\rm exp} (-C)$ one recovers the well-known
Baker-Campbell-Hausdorff formula. Eq. (\ref{2.9}) can be verified order by order expanding the exponentials
and comparing powers of $B$. In our case, the operator $B$ corresponds to $A \cdot R$, and the operator $C$
to $a \cdot R$, and neglecting higher order in $C$ corresponds to the fact that the parameters $a$ are
infinitesimal. When applied to our case, eqs. (\ref{2.8}) and (\ref{2.9}) are particularly useful because
they preserve the form of the group element. Defining $\delta A(x)= A^\prime (x) - A(x)$, we obtain
  \begin{eqnarray}
  & & \delta A_{a_1 \ldots a_3} = a_{a_1 \ldots a_3} \nonumber \\
  & & \delta A_{a_1 \ldots a_6} = a_{a_1 \ldots a_6} + 20 a_{[a_1 \dots a_3}
  A_{a_4 \dots a_6 ]}   \quad .\label{2.10}
  \end{eqnarray}

We now want to determine the field strengths of $A_{a_1 \dots a_3}$ and $A_{a_1 \dots a_6}$ from the
Maurer-Cartan form. To this end, we only need to consider
  \begin{equation}
  g^{-1}_A d g_A  \quad . \label{2.11}
  \end{equation}
We use the operator identities
  \begin{equation}
  e^{-B} d e^B = d B + {1 \over 2} [dB, B] + {1 \over 3!} [[dB,B],B] + {1 \over 4!} [[[dB,B],B],B]+ \ldots
  \label{2.12}
  \end{equation}
and
  \begin{equation}
  e^{-B} D e^B = D + [D, B] + {1 \over 2} [[D,B],B] + {1 \over 3!} [[[D,B],B],B]+ \ldots \label{2.13}
  \end{equation}
valid for any pair of operators $B$ and $D$. Eq. (\ref{2.12}) can be written in the compact form
  \begin{equation}
  e^{-B} d e^B = {1 - e^{-B} \over B} \wedge dB \quad , \label{2.14}
  \end{equation}
where the $\wedge$ product denotes multiple commutators, so that
  \begin{equation}
  B \wedge C = [ B,C] \qquad \quad  B^2 \wedge C = [B,[B,C]] \label{2.15}
  \end{equation}
and so on. Identifying the operators $B$ and $D$ with the relevant $A \cdot R$'s and their derivatives and
commutators, one obtains
  \begin{eqnarray}
  d x^\mu g^{-1}_A \partial_\mu g_A &=& d x^\mu [ {1 \over 3!} \partial_\mu A_{a_1 \ldots a_3} R^{a_1 \ldots a_3
  }\nonumber \\
  &+& {1 \over 6!}(\partial_\mu A_{a_1 \ldots a_6}  + 20 \partial_\mu A_{a_1 \ldots a_3}
  A_{a_4 \ldots a_6}) R^{a_1 \ldots a_6 } + \ldots ] \label{2.16}
  \end{eqnarray}
where the dots correspond to higher level operators. The quantities
  \begin{eqnarray}
  & & G_{\mu a_1 \dots a_3} = \partial_\mu A_{a_1 \ldots a_3} \nonumber \\
  & & G_{\mu a_1 \dots a_6} = \partial_\mu A_{a_1 \ldots a_6}  + 20 \partial_\mu A_{[ a_1 \ldots a_3}
  A_{a_4 \ldots a_6 ]} \label{2.17}
  \end{eqnarray}
are invariant under the transformations of eqs. (\ref{2.10}).

We now want to describe the dynamics out of the $E_{11}$-invariant quantities of eq. (\ref{2.17}). The
requirement is that the system leads to massless equations for the Goldstone fields $A_{a_1 \dots a_3}$ and
$A_{a_1 \dots a_6}$. For this to lead to a consistent dynamics, one needs to promote the global symmetries of
eq. (\ref{2.10}) to local ones, because a massless field of non-vanishing spin requires gauge invariance. It
turns out that if one considers the closure of $E_{11}$ with the eleven-dimensional conformal group, and
considers transformations of the fields that result from multiple commutators of the generators of $E_{11}$
with the conformal ones, the most general transformation that results is a gauge transformation $d \Lambda$,
where $\Lambda$ is an arbitrary function of $x$ \cite{10}. This result is rather remarkable, because it
deeply relies on how the conformal transformations act on the fields. The parameter $a$ can be identified
with the $x$-independent component of $d \Lambda$, and the full transformation can be obtained replacing the
parameter $a$ with $d \Lambda$ in the $E_{11}$ transformations. For fields with totally antisymmetric
indices, the gauge-invariant field strengths are obtained simply antisymmetrising the indices of the $G$'s in
the Maurer-Cartan form.

One therefore obtains from eq. (\ref{2.17}) the field strengths
  \begin{eqnarray}
  & &
  F_{a_1 \dots a_4 } = 4 G_{[ a_1 \dots a_4 ]}= 4 \partial_{[a_1 }
  A_{a_2 a_3 a_4 ]} \nonumber \\
  & & F_{a_1 \dots a_7 } = 7 G_{[ a_1  \dots a_7 ]} = 7
  \partial_{[a_1 } A_{a_2 \dots a_7 ]} + 35 F_{[a_1 \ldots a_4} A_{a_5 a_6 a_7 ]}
  \label{2.18}
  \end{eqnarray}
which are invariant under the gauge transformations
  \begin{eqnarray}
  & &   \delta A_{a_1 \ldots a_3} = 3 \partial_{[a_1 } \Lambda_{a_2 a_3 ]} \nonumber \\
  & &  \delta A_{a_1 \ldots a_6 } = 6  \partial_{[a_1 } \Lambda_{a_2 \ldots a_5 ]}
  + 60 \partial_{[a_1}
  \Lambda_{a_2 a_3 } A_{a_4 a_5 a_6 ]} \quad . \label{2.19}
  \end{eqnarray}
Form the field strengths of eq. (\ref{2.18}), the unique non-trivial first order equation that can be written
is a duality condition of the form
  \begin{equation}
  F_{a_1 \dots a_4} = {1 \over 7!} \epsilon_{a_1 \dots a_4 b_1 \dots b_7} F^{b_1 \dots b_7}
  \quad ,\label{2.20}
  \end{equation}
which leads to second order field equations for both the 3-form and the 6-form \cite{10}.

The supersymmetry algebra of the original 11-dimensional supergravity, which includes a 3-form potential, can
be extended in order to include a 6-form dual to the 3-form. In this democratic formulation, the
supersymmetry algebra on the 6-form closes using the duality relation between the field strengths of the
3-form and the 6-form, and it generates exactly the gauge transformations of eq. (\ref{2.19}), up to field
redefinitions. In the remaining of this section, we want to determine the gauge transformations of the fields
of five dimensional maximal supergravity in the democratic formulation which results from $E_{11}$.

We now consider the $E_{11}$ non-linear realisation giving rise to a five-dimensional spacetime. The
corresponding Dynkin diagram can be drawn as in Fig. 2, where the horizontal line is the gravity line
associated with $SL(5,\mathbb{R})$ and one can see the appearance of the exceptional group $E_{6}$, that is
the internal symmetry group because it is the part of the Dynkin diagram that is not connected to the gravity
line.
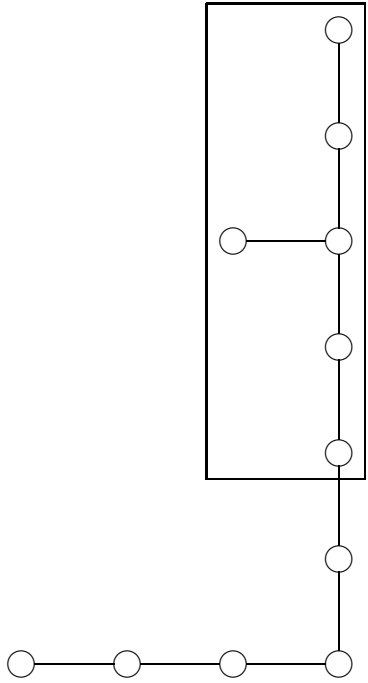
\begin{figure}[h]
\begin{center}
\begin{picture}(140,260)
\multiput(10,10)(40,0){4}{\circle{10}}
\multiput(15,10)(40,0){3}{\line(1,0){30}}\multiput(130,50)(0,40){6}{\circle{10}}
\multiput(130,15)(0,40){6}{\line(0,1){30}} \put(90,170){\circle{10}} \put(95,170){\line(1,0){30}}
\put(80,80){\line(1,0){60}} \put(80,260){\line(1,0){60}} \put(80,80){\line(0,1){180}}
\put(140,80){\line(0,1){180}}
\end{picture}
\caption{\sl The $E_{11}$ Dynkin diagram corresponding to 5-dimensional supergravity. The internal symmetry
group is $E_{6(+6)}$.}
\end{center}
\end{figure}

The generators with completely antisymmetric indices, with the  exception of the space-filling 5-forms, are
given by
  \begin{equation}
  R^\alpha  \qquad R^{a , M} \qquad R^{ab}{}_M \qquad R^{abc , \alpha}  \qquad  R^{abcd}{}_{[MN]}
  \quad , \label{2.21}
  \end{equation}
where  $R^\alpha$, $\alpha = 1 ,\dots ,78$ are the $E_6$ generators, and an upstairs $M$ index, $M= 1 ,\dots
, 27$, corresponds to the ${\bf \overline{27}}$ representation, a downstairs  $M$ index to the ${\bf 27}$ and
a pair of antisymmetric downstairs indices $[MN]$ correspond to the ${\bf  \overline{351}}$ as the tensor
product of ${\bf 27\otimes 27}$ in the anti-symmetric combination only contains the $ {\bf \overline{351}}$
\cite{29}.

We now write the commutators of the $E_{11}$ generators of eq. (\ref{2.21}) \cite{29} as explained above for
the 11-dimensional case. We write the commutation relations for the $E_6$ generators in the form
  \begin{equation}
  [R^\alpha , R^\beta ]= f^{\alpha\beta}{}_{\gamma} R^\gamma \quad , \label{2.22}
  \end{equation}
where $f^{\alpha\beta}{}_{\gamma}$ are the structure constants of $E_6$. The commutator of these generators with
the 1-form generator $R^{a,M}$ is determined by the fact that the Jacobi identity involving $R^\alpha$,
$R^\beta$ and $R^{a, M}$ implies that this generator is in a representation of $E_6$, which is in fact the $
{\bf \overline{27}}$ as noted above, and it is given by
  \begin{equation}
  [R^\alpha , R^{a,M} ]= (D^\alpha )_N{}^M R^{a, N} \quad ,
  \label{2.23}
  \end{equation}
where $(D^\alpha )_N{}^M $ are the generators of $E_6$ in this representation and so obey
  \begin{equation}
  [D^\alpha , D^\beta ]_M{}^N = f^{\alpha\beta}{}_\gamma (D^\gamma )_M{}^N \quad . \label{2.24}
  \end{equation}
The two form generators are in the ${\bf 27}$ representation and so their  commutator with the generators of
$E_6$ is given by
  \begin{equation}
  [R^\alpha , R^{ab}{}_M ]= -(D^\alpha )_M{}^N R^{ab}{}_N \quad .
  \label{2.25}
  \end{equation}
This involves the matrix $(D^\alpha )_M{}^N$ in the way that follows from the fact that if we contract the
indices of a ${\bf \overline {27}}$ with a ${\bf {27}}$ we find an $E_6$ invariant. The $E_6$ commutator of the
$ R^{abc , \alpha}$ is given by
  \begin{equation}
  [R^\alpha , R^{abc ,\beta} ]= f^{\alpha\beta}{}_{\gamma} R^{abc,\gamma} \quad ,
  \label{2.26}
  \end{equation}
as it is in the adjoint representation  while that of the $R^{abcd}{} _{[MN]} $ generator is given by
  \begin{equation}
  [R^\alpha , R^{abcd}{}_{[MN]} ]= -(D^\alpha )_M{}^P R^{abcd}{}_{[PN]} -(D^\alpha )_N{}^P R^{abcd}{}_{[MP]}\quad
  . \label{2.27}
  \end{equation}

The next commutators of the $E_{11}$ algebra to consider are those of the 1-forms which yield a 2-form and are
given by
  \begin{equation}
  [R^{a,M} , R^{b,N} ]= d^{MNP} R^{ab}{}_P \quad ,
  \label{2.28}
  \end{equation}
where $d^{MNP}$ is required by the Jacobi identity involving $R^\alpha$, $R^{a,M}$ and $R^{b,N}$ to be an
invariant tensor transforming in the ${\bf \overline{27}}\otimes {\bf \overline{27}}\otimes {\bf \overline{27}}$
representation and so it is also a symmetric tensor. The  commutator of a 1-form with a 2-form generator is a
3-form generator and the Jacobi identities involving $R^\alpha$, $R^{a,N}$ and $R^{bc}{}_M$ imply that this is
given in terms of the $(D^\alpha )_M{}^N$ matrix as follows:
  \begin{equation}
  [R^{a,N} , R^{bc}{}_M ]= g_{\alpha\beta} (D^\alpha )_M{}^N R^{abc, \beta} \quad
  , \label{2.29}
  \end{equation}
where $g_{\alpha\beta}$ is the Cartan-Killing metric of $E_6$. As mentioned above the 4-form generator is in
the ${\bf \overline {351}}$ representation and as this is the only representation in the  anti-symmetric
tensor product of ${\bf 27\otimes 27}$  it appears on  the right-hand side of the commutators  of two 2-forms
as
  \begin{equation}
  [R^{ab}{}_M , R^{cd}{}_N ]= R^{abcd}{}_{[MN]} \quad . \label{2.30}
  \end{equation}
The commutator of the 1-form with the 3-form also lead to the 4-form, and can be written as
  \begin{equation}
  [R^{a, P}, R^{bcd , \alpha} ]= S^{\alpha P[MN]} R^{abcd}{}_{[MN]} \quad ,
  \label{2.31}
  \end{equation}
where $S^{\alpha P[MN]}$ is an invariant tensor. Using
  \begin{equation}
  g_{\beta\gamma} (D^\alpha )_M{}^N (D^\gamma )_N{}^M = k \delta^\alpha_\beta \label{2.32}
  \end{equation}
one can show that the Jacobi identities constrain the invariant tensor $S^{\alpha P[MN]}$ to satisfy
  \begin{equation}
  S^{\alpha P[MN]} + {1 \over k} g_{\beta\gamma} (D^\alpha D^\beta )_Q{}^P S^{\gamma Q[MN]} =
  - {1 \over k} (D^\alpha )_Q{}^{[M}
  d^{N]PQ}
  \quad . \label{2.33}
  \end{equation}
This equation is solved by
  \begin{equation}
  S^{\alpha P[MN]} = -{1 \over 2} D^\alpha_Q{}^{[M} d^{N]QP } \quad , \label{2.34}
  \end{equation}
and leads to the further identity
  \begin{equation}
  g_{\alpha\beta} D^\alpha_Q{}^{(P} S^{\beta R)[MN]} =  -{1 \over 2} \delta_{Q}^{[M} d^{N]PR}
  \quad . \label{2.35}
  \end{equation}

We now consider the $E_{11}$ group element $g$ in five dimensions. As in 11 dimensions, we neglect the
contribution of all the other generators and we write $g$ in the form
  \begin{equation}
  g = {\rm exp} (x^\mu P_\mu ) \ g_A \ g_\phi  \quad , \label{2.36}
  \end{equation}
where with respect to (\ref{2.3}) we have also included the scalar contribution $g_\phi$, and now
  \begin{eqnarray}
  & &   g_A =
  {\rm exp} (A^{MN}_{a_1 \ldots a_4}
  R^{a_1 \ldots a_4 }_{MN} ) \  {\rm exp} (g_{\alpha\beta} A^{\alpha}_{a_1 \ldots a_3} R^{a_1 \ldots a_3,
  \beta } ) \nonumber \\
  & & {\rm exp}
  (A^{M}_{a_1 a_2} R^{a_1 a_2 }_M ) \ {\rm exp} (A_{a,M} R^{a,M } ) \quad .
  \label{2.37}
  \end{eqnarray}
We determine the $E_{11}$ transformations of each of the fields in (\ref{2.37}) using the same analysis that
was reviewed for the 11-dimensional case at the beginning of this section. Acting with
  \begin{equation}
  g_0^{(4)} = {\rm exp} (a^{MN}_{a_1 \ldots a_4} R^{a_1 \ldots a_4 }_{MN} ) \label{2.38}
  \end{equation}
leads to a transformation of the 4-form field
  \begin{equation}
  \delta A^{MN}_{a_1 \ldots a_4} = a^{MN}_{a_1 \ldots a_4} \quad , \label{2.39}
  \end{equation}
while acting with
  \begin{equation}
  g_0^{(3)} = {\rm exp} (g_{\alpha\beta} a^{\alpha}_{a_1 \ldots a_3} R^{a_1 \ldots a_3 , \beta } )
  \label{2.40}
  \end{equation}
leads to
  \begin{equation}
  \delta A^{\alpha}_{a_1 \ldots a_3} = a^{\alpha}_{a_1 \ldots a_3} \quad . \label{2.41}
  \end{equation}
Indeed one can see from eq. (\ref{2.9}) that each of these group elements can not lead to additional
transformations to any of the fields we are considering, that have at most four indices. The action of
  \begin{equation}
  g_0^{(2)} = {\rm exp} (a^{M}_{a_1 a_2} R^{a_1 a_2}_M ) \label{2.42}
  \end{equation}
instead, produces a transformation of the 2-form as well as the 4-form, as can be deduced from eqs.
(\ref{2.9}) and (\ref{2.8}). Using these equations, together with the commutation relation of eq.
(\ref{2.30}), the form of these transformations is straightforward to determine, and the result is
    \begin{eqnarray}
    & & \delta A^{MN}_{a_1 \ldots a_4} = {1 \over 2} a^{[M}_{[a_1 a_2} A^{N]}_{a_3 a_4 ]} \nonumber \\
    & & \delta A^{ M}_{a_1 a_2 } = a^{ M}_{a_1 a_2 } \quad . \label{2.43}
    \end{eqnarray}
The last transformation we consider is the one generated by
  \begin{equation}
  g_0^{(1)} = {\rm exp} (a_{a,M} R^{a, M}) \quad , \label{2.44}
  \end{equation}
which using eqs. (\ref{2.9}) and (\ref{2.8}) can be seen generating transformations of all the fields. The
result is
  \begin{eqnarray}
  & & \delta A_{a_1 \ldots a_4}^{MN} = a_{[a_1 ,P} A^{\alpha}_{a_2 a_3 a_4 ]} S^{\beta P[MN]} g_{\alpha\beta}- {1
  \over 24} a_{[a_1 ,P} A_{a_2 ,Q} A_{a_3 ,R} A_{a_4 ] ,S} d^{PQT} D^{\alpha}_T{}^R
  S^{\beta S[MN]} g_{\alpha\beta}\nonumber \\
  & & \qquad \quad - {1 \over 4} a_{[a_1 ,P} A_{a_2 ,Q} A^{ [M}_{a_3 a_4 ]} d^{N]PQ}
  \nonumber \\
  & & \delta A^{ \alpha}_{a_1 a_2 a_3 } = a_{[a_1 ,M} A^{N}_{a_2 a_3 ]} D^\alpha_N{}^M + {1 \over 6}
  a_{[a_1 ,M} A_{a_2 , N}  A_{a_3 ], P} d^{MNQ} D^\alpha_Q{}^P \nonumber \\
  & & \delta A^{M}_{a_1 a_2}  = {1 \over 2} a_{[a_1 ,N} A_{a_2 ], P} d^{MNP}\nonumber \\
  & & \delta A_{a, M} = a_{a ,M} \quad . \label{2.45}
  \end{eqnarray}

We now determine from the Maurer-Cartan form the field strengths of all the fields of which we have
determined the $E_{11}$ transformations. As in the 11-dimensional case, we only need to consider $g_A$ in eq.
(\ref{2.37}), and using eqs. (\ref{2.12}) and (\ref{2.13}) one finds
  \begin{equation}
  g_A^{-1} d g_A = dx^\mu [ G_{\mu a ,M} R^{a ,M} + G^{M}_{\mu a_1 a_2} R_M^{a_1 a_2 } +
  G^{\alpha}_{\mu a_1 a_2 a_3} R_\alpha^{a_1 a_2 a_3} + G^{MN}_{\mu a_1 \dots a_4} R_{MN}^{a_1 \dots a_4}
  + \dots ]
 \quad , \label{2.46}
  \end{equation}
where the dots correspond to operators with more than four indices, and the $G$'s are invariant under the
transformations of eqs. (\ref{2.39}), (\ref{2.41}), (\ref{2.43}) and (\ref{2.45}), and are given by
  \begin{eqnarray}
  & & G_{\mu a ,M} = \partial_\mu A_{a ,M} \nonumber \\
  & & G^{M}_{\mu a_1 a_2} = \partial_\mu A^{M}_{a_1 a_2} + {1 \over 2} \partial_\mu A_{[a_1 ,N}
  A_{a_2 ] ,P} d^{MNP} \nonumber \\
  & & G^{\alpha}_{\mu a_1 a_2 a_3} = \partial_\mu A^{\alpha}_{a_1 a_2 a_3} - {1 \over 6} \partial_\mu
  A_{[a_1 ,M} A_{a_2 ,N} A_{a_3 ] ,P} d^{MNQ} D^\alpha_Q{}^P - \partial_\mu A^{M}_{[a_1
  a_2} A_{a_3 ],N} D^\alpha_M{}^N \nonumber \\
  & & G^{MN}_{\mu a_1 \dots a_4} = \partial_\mu A^{MN}_{a_1 \dots a_4} - {1 \over 24} \partial_\mu
  A_{[a_1 , P}  A_{a_2 , Q} A_{a_3 , R} A_{a_4 ] , S} d^{PQT} D^\alpha_T{}^R
  S^{\alpha S [MN]} \nonumber \\
  & & \qquad \qquad - {1 \over 2} \partial_\mu A^{P}_{[a_1 a_2 } A_{a_3 , Q} A_{a_4 ], R}
  D^\alpha_P{}^Q S^{\alpha R[MN]} + {1 \over 2} \partial_\mu A^{[M}_{[a_1 a_2 } A^{N]}_{a_3 a_4 ] }\nonumber \\
  & & \qquad \qquad  +
  \partial_\mu A^{\alpha}_{[a_1 a_2 a_3} A_{a_4 ] , P} S^{\alpha P[MN]} \quad . \label{2.47}
  \end{eqnarray}

As it is clear from eq. (2.36), the Cartan forms actually occur as
  \begin{equation}
  g_\phi^{-1} g_A^{-1} d g_A g_\phi \quad , \label{2.48}
  \end{equation}
which means that they are given by
  \begin{equation}
  g_\phi^{-1} G \cdot R g_\phi \quad , \label{2.49}
  \end{equation}
where the $G$'s are given in eq. (\ref{2.47}). One must also include the Cartan form for the scalars which is
  \begin{equation}
  g_\phi^{-1} \partial_\mu g_\phi \quad . \label{2.50}
  \end{equation}
For example we find for the vector Cartan form
  \begin{equation}
  g_\phi^{-1} \partial_\mu A_{a ,M} R^{a , M} g_\phi = \partial_\mu A_{a ,M} \tilde{V}^M_{ij} R^{a , ij}
  \quad , \label{2.51}
  \end{equation}
where
  \begin{equation}
  g_\phi^{-1} R^{a , M} g_\phi = \tilde{V}^M_{ij} R^{a , ij}\quad , \label{2.52}
  \end{equation}
and the latter is just $R^{a , M}$ decomposed into the $\bf 27$ representation of the local subalgebra
$USp(8)$. Here the $USp(8)$ indices $i,j= 1, \dots ,8$ of $\tilde{V}^M_{ij}$ are antisymmetric and traceless,
giving rise to the ${\bf 27}$ of $USp(8)$. In terms of the parametrisation $g_\phi = e^{\phi_\alpha
R^\alpha}$ the scalars $\tilde{V}^M_{ij}$ are defined by
  \begin{equation}
  \tilde{V}^M_{ij} = {\rm exp}(\phi_\alpha D^\alpha)^M_{ij} \quad , \label{2.53}
  \end{equation}
where we have decomposed the lower index in the {\bf 27} of $E_6$ under $USp(8)$. This decomposition under
$USp(8)$ reflects the fact that the Cartan forms only transform under the local subgroup $USp(8)$ and are
inert under the global group $E_6$. Similar considerations apply to the other Cartan forms. For the case of
the 2-form we have
  \begin{equation}
  g_\phi^{-1} R^{ab}_{M} g_\phi = {V}_M^{ij} R^{ab}_{ij}\quad , \label{2.54}
  \end{equation}
where $V_M^{ij}$ are defined in a similar way to eq. (\ref{2.53}), but now the generators act on the complex
conjugate representation, and therefore the scalars $V_M^{ij}$ are the inverse of $\tilde{V}^M_{ij}$.

In order to obtain massless field equations for the fields, the same arguments that lead to gauge-invariant
equations in eleven dimensions hold here. We thus consider the completion of $E_{11}$ with the conformal
group. This leads to an infinite-dimensional extension of the symmetries, whose net result is to replace the
global parameters $a$ with $d \Lambda$, where $\Lambda$ is a gauge parameter. The corresponding gauge
invariant field-strengths result from the $G$'s of eq. (\ref{2.47}) once all the indices are completely
antisymmetrised.

One therefore obtains from eq. (\ref{2.47}) the field strengths
  \begin{eqnarray}
  & & F_{a  b ,M} = 2 G_{[ ab ] ,M} \nonumber \\
  & & F^{M}_{abc} = 3 G^{M}_{[abc ]} \nonumber \\
  & & F^{\alpha}_{abcd} = 4 G^{\alpha}_{[abcd ]} \nonumber \\
  & & F^{MN}_{abcde } = 5 G^{MN}_{[ abcde ]} \quad ,
  \label{2.55}
  \end{eqnarray}
which are invariant under the gauge transformations obtained from eqs. (\ref{2.39}), (\ref{2.41}),
(\ref{2.43}) and (\ref{2.45}), after having promoted the $E_{11}$ parameters to local ones according to
  \begin{eqnarray}
  & & a_{a ,M} = \partial_a \Lambda_M \nonumber \\
  & & a^{M}_{ab} = 2 \partial_{[a} \Lambda^{ M}_{b ]} \nonumber \\
  & & a^{\alpha}_{abc} = 3 \partial_{[ a} \Lambda^{\alpha}_{bc ]} \nonumber \\
  & & a^{MN}_{abcd}= 4 \partial_{[a} \Lambda^{MN}_{bcd ]} \quad .
  \label{2.56}
  \end{eqnarray}
The actual field strengths are multiplied by factors of $V_M^{ij}$ or $\tilde{V}^M_{ij}$ as explained above
for the Cartan forms.

The unique equations which are invariant under the transformations of the non-linear realisation above and
are Lorentz and $USp(8)$ covariant are
  \begin{equation}
  V_{Mij} F^M_{abc} \sim  \epsilon_{abcde} \tilde{V}^M_{ij} F^{de}_M \qquad
  V_{Mij} \tilde{V}^N_{kl} F^\alpha_{abcd} \sim
  D^\alpha_M{}^N \epsilon_{abcde} (  g_\phi^{-1} \partial^e g_\phi )_{ijkl} \quad .
  \label{2.57}
  \end{equation}
The non-linear realisation also possesses local transformations associated with the Cartan involution
invariant subalgebra. The transformations above, which determine the field strengths, arise from the Borel
subalgebra of $E_{11}$ with the exception of the local $USp(8)$. We believe that also requiring invariance
under the local transformations will fix uniquely the duality relations above.

In the next section we will close the supersymmetry algebra of maximal five-dimensional supergravity in the
democratic formulation, that is including fields and dual fields. This formulation involves the same forms that
were considered in this section, and we will show that, after field redefinitions, the supersymmetry algebra
leads to precisely the same field strengths and the same gauge transformations as predicted by $E_{11}$.

\section{Supersymmetry algebra of the democratic formulation of $D=5$ massless maximal supergravity}
In this section we show that the supersymmetry algebra of maximal supergravity in five dimensions closes on the
fields with totally antisymmetric indices predicted by $E_{11}$. We also show that the resulting gauge algebra
is in precise agreement with the one predicted by $E_{11}$ that was analysed in the previous section.

The 42 scalars belong to the non-linear realisation of $E_6$ with local subgroup $USp(8)$. Taking the
generators to be in the the fundamental {\bf 27} representation of $E_6$, one can write the group element as
$V_{Mij}$, where the indices $i$ and $j$ are antisymmetrised fundamental indices of $USp(8)$ while the lower
index $M$ denotes the ${\bf 27}$ representation of $E_6$, like in the previous section. The justification for
this is that the non-linear realisation is invariant under $V \rightarrow g_0 \ V \ h$, where $g_0 \in E_6$
and $h \in USp(8)$, and as a result the first index of $V$ transforms in the fundamental of $E_6$, while the
second index transforms under the local subgroup $USp(8)$. The decomposition of the {\bf 27} of $E_6$ under
$USp(8)$ gives the {\bf 27} of $USp(8)$, which implies that the scalars satisfy the constraint $\Omega^{ij}
V_{Mij} =0$, where $\Omega^{ij}$ is the invariant metric of $USp(8)$, satisfying
  \begin{equation}
  \Omega^{ki} \Omega_{ij} = - \delta^k_j \quad . \label{3.1}
  \end{equation}

The inverse scalars are denoted by $\tilde{V}^M_{ij}$, and they satisfy the relations
  \begin{equation}
  V_{Mij} \tilde{V}^{Nij} = \delta^N_M \label{3.2}
  \end{equation}
and
  \begin{equation}
  V_{Mij} \tilde{V}^{Mkl} = {1 \over 2} ( \delta^k_i \delta^l_j - \delta^l_i \delta^k_j ) - {1 \over 8}
  \Omega_{ij} \Omega^{kl} \label{3.3}
  \quad .
  \end{equation}
The $USp(8)$ indices are raised and lowered according to
  \begin{equation}
  V^i = V_j \Omega^{ji} \quad , \qquad V_i = \Omega_{ij} V^j \quad . \label{3.4}
  \end{equation}

The other fields in the supergravity multiplet are the vierbein $e_\mu{}^a$, the abelian vectors $A_{\mu M}$
in the ${\bf 27}$ of $E_6$, the gravitino $\psi_{\mu i}$ and the spinor $\chi_{ijk}$ in the ${\bf 8}$ and
${\bf 48}$ of $USp(8)$ respectively. Following \cite{7}, we consider the supersymmetry transformations
  \begin{eqnarray}
  & & \delta e_\mu{}^a = -i \bar{\epsilon}^i \gamma^a \psi_i \nonumber \\
  & & \delta A_{\mu M} = 2i V_{Mij} \bar{\epsilon}^i \psi_\mu^j + {i \over \sqrt{2}} V_{Mij}
  \bar{\epsilon}_k \gamma_\mu \chi^{ijk}
  \nonumber  \\
  & & \delta V_{Mij} = {2i \over \sqrt{2}} V_M^{kl} \bar{\epsilon}_k \chi_{ijl} +
  {2i \over \sqrt{2}}V_M^{kl} \bar{\epsilon}_{[i} \chi_{j]kl} - {i \over \sqrt{2}}
  V_m^{kl} \Omega_{ij} \bar{\epsilon}^m \chi_{klm} + {2i \over \sqrt{2}} V_{M[i}{}^k \bar{\epsilon}^l
  \chi_{j]kl} \nonumber \\
  && \delta \psi_{\mu i } = D_\mu \epsilon_i + Q_{\mu i}{}^j \epsilon_j - {1 \over 6} F_{\nu\rho M}
  \tilde{V}^M_{ij} \gamma^{\nu\rho} \gamma_\mu
  \epsilon^j + {1 \over 3} F_{\mu\nu M} \tilde{V}^M_{ij} \gamma^\nu \epsilon^j \nonumber \\
  & & \delta \chi_{ijk} = \sqrt{2} P_{\mu ijkl} \gamma^\mu \epsilon^l - {3 \over 2 \sqrt{2}} F_{\mu\nu M}
  \tilde{V}^M_{[ij}
  \gamma^{\mu\nu} \epsilon_{k]} - {1 \over 2\sqrt{2}}F_{\mu\nu M} \Omega_{[ij} \tilde{V}^M_{k]l}
  \gamma^{\mu\nu} \epsilon^l \quad , \label{3.5}
  \end{eqnarray}
where
  \begin{equation}
  F_{\mu\nu M} = 2 \partial_{[\mu} A_{\nu ] M}  \quad , \label{3.6}
  \end{equation}
and $P_\mu$ and $Q_\mu$ are defined by
  \begin{equation}
\nabla_\mu V_{Mij} + V_{M kl} P_\mu^{kl}{}_{ij} = \partial_\mu V_{Mij} + 2 Q_{\mu [i}{}^k V_{M \vert k\vert j
]} + V_{M kl} P_\mu^{kl}{}_{ij} = 0 \quad . \label{3.7}
  \end{equation}
We are only considering the bosonic part of the supersymmetry transformation of the fermions. This is because
we are only interested in the terms at lowest order in the fermions. We use conventions similar to those of
\cite{7}, with a mostly minus signature and $\epsilon_{01234}=1$. The antisymmetrised product of gamma
matrices satisfies
  \begin{equation}
  \gamma_{\mu_1 \dots \mu_n} = (-)^{{n(n-1) \over 2}}{1 \over (5-n)!}\epsilon_{\mu_1 \dots \mu_n
  \nu_{n+1} \dots \nu_5}
  \gamma^{\nu_{n+1}  \dots \nu_5} \quad . \label{3.8}
  \end{equation}
The transformations of eq. (\ref{3.5}) where shown in \cite{7} to leave the corresponding action invariant.
We take these transformations as the starting point for our algebraic analysis. In the bosonic sector, the
commutator of two such transformations $[\delta_{\epsilon_1}, \delta_{\epsilon_2}]$ closes on all the local
symmetries of the theory, while in the fermionic sector the same algebra closes only on-shell. We are
interested in studying the supersymmetry algebra on the bosons to lowest order in the fermionic fields. This
leads to the corresponding parameters
  \begin{equation}
  \xi_\mu = - i \bar{\epsilon}_2^i \gamma_\mu \epsilon_{1 i } \label{3.9}
  \end{equation}
for general coordinate transformations, and
  \begin{equation}
  \Lambda_M = 2i V_{Mij} \bar{\epsilon}_2^i \epsilon_1^j - \xi^\mu A_{\mu M} \label{3.10}
  \end{equation}
for gauge transformations.

We now want to generalise this result by introducing dual forms for the bosonic fields above. We want to
close the supersymmetry algebra on these dual fields to lowest order in the fermions, using the fact that
they are related by duality to the bosonic fields already introduced. The duality conditions are first order
equations, which is consistent with the fact that the supersymmetry algebra only closes when these duality
conditions hold. Given the fact that the fermions transform to the field strengths of the bosonic fields
under supersymmetry, and that the algebra closes using the duality relations, the supersymmetry
transformations of the fermions are the ones in eq. (\ref{3.5}) modulo these duality relations. Each form
only transforms with respect to the gauge parameters of lower rank, which means that the closure of the
algebra on each form does not require the knowledge of the transformations of the forms of higher rank. This
resembles the way these gauge transformations result from $E_{11}$, as it is clear from the analysis carried
out in the previous section.

Proceeding this way, we will determine the supersymmetry and gauge transformations of the 2-forms, dual to
the vectors, and the 3-forms, dual to the scalars. Once these transformations are obtained, one can then
determine the 4-forms that supersymmetry allows. These 4-forms are not propagating, and in the ungauged
theory their field-strengths vanish. We thus determine the number of 4-forms requiring that the supersymmetry
algebra closes using the fact that the field-strengths of these forms vanish. In the next section we will
generalise this result to the gauged case, in which the field-strengths of the 4-forms are dual to the mass
deformations of the theory. One could also determine the 5-forms that supersymmetry allows. Although these
fields are not propagating and have no field-strength, they are relevant because they are associated to
spacetime-filling branes, that have a crucial role in orientifold models. We will not determine the 5-forms
from supersymmetry in this section, but appendix A contains the derivation of their gauge transformations
from $E_{11}$.

The method we are using to determine the supersymmetry and gauge transformations of all the forms is
sometimes called democratic formulation of supergravity. In \cite{26} and \cite{27} this method was applied
to IIB and IIA supergravity respectively. It is important to recall here that the $E_{11}$ non-linear
realisation is automatically democratic, and it was the analysis in \cite{26} and \cite{27} that revealed for
the first time that the 10-forms predicted by $E_{11}$ in ten dimensions \cite{24} agree precisely with
supersymmetry. At the end of this section we will compare our results with the results of the previous
section, and we will show that the two perfectly agree.

We start with the 2-forms, and so we close the supersymmetry algebra on the 2-form $B_{\mu\nu}^M$ in the
${\bf \overline{27}}$ of $E_6$. The algebra closes using the fact that the 2-forms are related to the 1-forms
by a duality transformation. It turns out that this uniquely defines the supersymmetry and gauge
transformations of the 2-forms, up to field redefinitions. We use these field redefinitions to choose a
particular form for the gauge transformation of $B_{\mu\nu}^M$ with respect to the parameter $\Lambda_M$,
which we impose to be of the form $\delta B_{\mu\nu}^M \sim \Lambda_N F_{\mu \nu, P}d^{MNP}$. This will be
our general procedure for the rest of this section: we choose the gauge transformation of each field to
contain only the field strengths of the lower rank fields, and there are no derivatives acting on the gauge
parameters, with the exception of the leading one. In this basis the gauge transformations of the fields are
gauge invariant.

We write down the most general supersymmetry transformation, we compute the commutator of two such
transformations and we impose the closure of the algebra. The final result is that the supersymmetry
transformation of $B_{\mu\nu}^M$ is
  \begin{equation}
  \delta B_{\mu\nu}^M = 4i \tilde{V}^M_{ij} \bar{\epsilon}^i \gamma_{[\mu} \psi^j_{\nu]} -
  {i \over \sqrt{2}} \tilde{V}^M_{ij}
  \bar{\epsilon}_k \gamma_{\mu\nu} \chi^{ijk} + 2 d^{MNP} A_{[ \mu N} \delta A_{\nu ] P} \quad
  ,\label{3.11}
  \end{equation}
where the last term contains the supersymmetry variation of the 1-form. The field strength of $B_{\mu\nu}^M$ is
  \begin{equation}
  G_{\mu\nu\rho}^M = 3 \partial_{[\mu } B_{\nu\rho ]}^M + 3 d^{MNP} A_{[ \mu N} F_{\nu\rho ] P} \label{3.12}
  \end{equation}
and it is invariant with respect to the gauge transformations
  \begin{eqnarray}
  & & \delta A_{\mu M} = \partial_\mu \Lambda_M \nonumber \\
  & & \delta B_{\mu\nu}^M = 2 \partial_{[\mu} \Sigma_{\nu ]}^M - d^{MNP} \Lambda_N F_{\mu\nu P} \quad . \label{3.13}
  \end{eqnarray}
The duality between $F$ and $G$ reads
  \begin{equation}
  V_{M ij} G_{\mu\nu\rho}^M = {1 \over 2} \epsilon_{\mu\nu\rho\sigma\tau} \tilde{V}^M_{ij} F^{\sigma \tau}_M
  \quad . \label{3.14}
  \end{equation}
The 1-form gauge parameter generated in the commutator of two supersymmetry transformations is
  \begin{equation}
  \Sigma_\mu^M = -2i \tilde{V}^M_{ij} \bar{\epsilon}_2^i \gamma_\mu \epsilon_1^j + \xi^\nu B_{\mu\nu}^M - 2
  id^{MNP} A_{\mu N}
  V_{Pij} \bar{\epsilon}_2^i \epsilon_1^j \label{3.15}
  \quad .
  \end{equation}
Finally, the invariant symmetric tensor $d^{MNP}$ of $E_6$ satisfies
  \begin{equation}
  \tilde{V}^M_{kl} \tilde{V}^{Nkl} \Omega_{ij} - 4 \tilde{V}^M_{ik}\tilde{V}^{Nk}{}_j +
  4 \tilde{V}^M_{jk}\tilde{V}^{Nk}{}_i + 4 d^{MNP} V_{P ij} =0 \quad , \label{3.16}
  \end{equation}
and similarly for the invariant tensor $d_{MNP}$ with downstairs indices satisfies
  \begin{equation}
  {V}_{Mkl} {V}^{kl}_N \Omega_{ij} - 4 {V}_{Mik}{V}_N^{k}{}_j +
  4 {V}_{Mjk}{V}_N^{k}{}_i + 4 d_{MNP} \tilde{V}^P_{ij} =0 \quad , \label{3.17}
  \end{equation}
which using eq. (\ref{3.3}) implies
  \begin{equation}
  d^{MNP} d_{MNQ} = 5 \delta^P_Q \quad . \label{3.18}
  \end{equation}

We now move to the 3-forms, which are dual to the scalars. We show that the supersymmetry algebra closes on the
3-forms $C_{\mu\nu\rho}^\alpha$, where $\alpha=1,\dots\ 78$ denotes the adjoint representation of $E_6$,
provided that their field strength satisfies a duality condition. The supersymmetry transformation of the 3-form
is
  \begin{eqnarray}
  \delta C_{\mu\nu\rho}^\alpha & = & 12 i D^{\alpha}_M{}^N \tilde{V}^M_{(i \vert k \vert} V_N^k{}_{j)}
  \bar{\epsilon}^i \gamma_{[\mu\nu}
  \psi_{\rho ]}^j - {2i \over \sqrt{2}} D^{\alpha}_M{}^N \tilde{V}^M_{ij} V_{Nkl} \bar{\epsilon}^i
  \gamma_{\mu\nu\rho} \chi^{jkl}
  \nonumber \\
  & - & {2i \over \sqrt{2}} D^{\alpha}_M{}^N \tilde{V}^M_{kl} V_{Nij} \bar{\epsilon}^i \gamma_{\mu\nu\rho} \chi^{jkl}
  +  {2i \over \sqrt{2}} D^{\alpha}_M{}^N \tilde{V}^M_{il} V_{N}^l{}_j \bar{\epsilon}_k \gamma_{\mu\nu\rho}
  \chi^{ijk} \nonumber \\
  & + & 12 D^{\alpha}_M{}^N B_{[ \mu\nu}^M \delta A_{\rho ] N} -6  D^{\alpha}_M{}^N A_{[\mu N}
  \delta B_{\nu\rho ]}^M - 24 S^{\alpha P [MN]} A_{[\mu M}
  A_{\nu N} \delta A_{\rho  ]P} \quad , \label{3.19}
  \end{eqnarray}
where $D^{\alpha}_M{}^N$ are the generators of $E_6$ in the ${\bf \overline{27}}$, satisfying eq.
(\ref{2.24}), and $S^{\alpha P[MN]}$ is the invariant tensor introduced in \cite{29} and satisfying eqs.
(\ref{2.34}) and (\ref{2.35}). The corresponding field strength,
  \begin{equation}
  H^\alpha_{\mu \nu \rho \sigma} = 4 \partial_{[\mu} C_{\nu \rho \sigma]}^\alpha - 24 D^{\alpha}_M{}^N B^M_{[\mu\nu}
  F_{\rho \sigma ]  N} - {8 } D^{\alpha}_M{}^N A_{[\mu N} G^M_{\nu \rho \sigma]} \label{3.20}
  \end{equation}
is invariant under the gauge transformations
  \begin{equation}
  \delta C^\alpha_{\mu\nu\rho} = 3 \partial_{[\mu} \Xi^\alpha_{\nu\rho]} + {2 } D^{\alpha}_M{}^N
  G_{\mu\nu\rho}^M \Lambda_N + 12 D^{\alpha}_M{}^N \Sigma_{[\mu}^M F_{\nu \rho ] N}\label{3.21}
  \end{equation}
together with the ones of eq. (\ref{3.13}). The duality relation between the field strength of eq.
(\ref{3.20}) and the scalars is
  \begin{equation}
  H^\alpha_{\mu\nu\rho\sigma} = D^{\alpha}_M{}^N \tilde{V}^{Mij} V_N^{kl} \epsilon_{\mu\nu\rho\sigma\tau}
  P^\tau_{ijkl} \quad . \label{3.22}
  \end{equation}
We observe that while the 3-forms are in the adjoint of $E_6$, the scalars realise $E_6$ non-linearly, and
therefore their number is ${\rm adj}(E_6 ) - {\rm adj} (USp(8))$. This means that there are ${\rm adj}
(USp(8))$ 3-forms whose field strengths are identically zero using the duality relation of eq. (\ref{3.22}).
This can be seen contracting eq. (\ref{3.22}) with a suitable combination of the scalars $V_{Mij}$. This is a
completely general phenomenon, and was shown for the first time in the IIB case in \cite{38}. The 2-form
gauge parameter that appears in the supersymmetry commutator is
  \begin{eqnarray}
  \Xi_{\mu\nu}^\alpha & =& 4 i D^{\alpha}_M{}^N \tilde{V}^M_{(i \vert k \vert } V_N^k{}_{j)}
  \bar{\epsilon}^i_2 \gamma_{\mu\nu}
  \epsilon_1^j - \xi^\rho C^\alpha_{\mu\nu\rho} \nonumber \\
  & +& {8 i }  D^{\alpha}_M{}^N B_{\mu\nu}^M V_{Nij} \bar{\epsilon}_2^i \epsilon_1^j -
  {8i } D^{\alpha}_M{}^N A_{[\mu N}
  \tilde{V}^M_{ij} \bar{\epsilon}_2^i \gamma_{\nu ]} \epsilon_1^j \quad . \label{3.23}
  \end{eqnarray}

We finally consider the 4-forms. Although these fields are not dual to any of the propagating fields of
five-dimensional supergravity, we proceed in a way analogous to the previous cases, writing down the most
general supersymmetry transformation and requiring the closure of the algebra. The fact that we are
considering a massless theory implies that the 5-form field strengths vanish identically because of the
duality relations, and this requirement is crucial to guarantee the closure of the supersymmetry algebra. In
the next section we will see that this duality relation is modified, and the 5-form field strengths will turn
out to be dual to the mass deformation parameters.

Supersymmetry implies that the 4-forms belong to the {\bf 351} of $E_6$, which corresponds to two
antisymmetrised upstairs fundamental indices. We therefore denote the 4-forms with $D^{MN}_{\mu\nu\rho\sigma}$,
where the antisymmetrisation of the $M$ and $N$ indices is understood. The supersymmetry transformation is
  \begin{eqnarray}
  \delta D_{\mu\nu\rho\sigma}^{MN}&=& 16 i \tilde{V}^M_{(i\vert k \vert} \tilde{V}^{Nk}{}_{j)}
  \bar{\epsilon}^i \gamma_{[\mu\nu\rho}
  \psi_{\sigma ]}^j + {4i \over \sqrt{2}} \tilde{V}^{[M}_{ij} \tilde{V}^{N]}_{kl} \bar{\epsilon}^i
  \gamma_{\mu\nu\rho\sigma}
  \chi^{jkl} \nonumber \\
  & - & 12 g_{\alpha \beta} S^{\alpha P [MN]} C_{[\mu\nu\rho}^\beta \delta A_{\sigma ] P} -
  4 g_{\alpha\beta} S^{\alpha P [MN]}A_{[\mu P} \delta
  C^\beta_{\nu\rho\sigma ]} \nonumber \\
  & +& 36 B_{[\mu\nu}^{[M} \delta B_{\rho \sigma ]}^{N]} - 24 g_{\alpha\beta}
  S^{\alpha P [MN]}D^{\beta}_Q{}^R A_{[\mu P} A_{\nu R} \delta
  B_{\rho\sigma ]}^Q\nonumber \\
  & +& ( 48  g_{\alpha\beta} S^{\alpha P [MN]} D^{\beta}_Q{}^R - 72
  \delta_Q^{[M} d^{N]PR} ) B_{[\mu\nu}^Q A_{\rho P } \delta A_{\sigma ]
  R}\nonumber \\
  & + & 48 g_{\alpha\beta} S^{\alpha P [MN]} D^{\beta}_Q{}^R d^{QST}
  A_{[\mu P} A_{\nu R} A_{\rho S} \delta A_{\sigma ] T} \quad . \label{3.24}
  \end{eqnarray}
The commutator closes after requiring that the field strength
  \begin{eqnarray}
  L^{MN}_{\mu\nu\rho\sigma\tau} & = & 5 \partial_{[\mu} D^{MN}_{\nu\rho\sigma\tau]} -30
  g_{\alpha \beta} S^{\alpha P[MN]} C_{[\mu \nu \rho}^\beta F_{\sigma \tau ] P} \nonumber \\
  & - & 5 g_{\alpha \beta} S^{\alpha P[MN]} A_{[\mu P} H^\beta_{\nu \rho \sigma\tau ]} - 60
  B_{[\mu\nu}^{[M} G_{\rho\sigma\tau]}^{N]} \label{3.25}
  \end{eqnarray}
vanishes identically. Gauge invariance of this field strength imposes for $D^{MN}_{\mu\nu\rho\sigma}$ the gauge
transformations
  \begin{eqnarray}
  \delta D^{MN}_{\mu\nu\rho\sigma} &=& 4 \partial_{[\mu} \Delta^{MN}_{\nu\rho\sigma]} +18
  g_{\alpha\beta}S^{\alpha P[MN]} \Xi^\beta_{[\mu\nu} F_{\rho\sigma] P} \nonumber \\
  &+&
  g_{\alpha\beta}S^{\alpha P[MN]} \Lambda_P H^\beta_{\mu\nu\rho\sigma} + 24 \Sigma_{[\mu}^{[M}
  G_{\nu\rho\sigma]}^{N]} \quad , \label{3.26}
  \end{eqnarray}
and the commutator of two supersymmetry transformations  closes on such gauge transformations. The 3-form gauge
parameter arising from the supersymmetry commutator is
  \begin{eqnarray}
  \Delta_{\mu\nu\rho}^{MN} & =& -4i \tilde{V}^M_{(i \vert k \vert} \tilde{V}^{Nk}{}_{j )} \bar{\epsilon}_2^i
  \gamma_{\mu\nu\rho}
  \epsilon_1^j + \xi^\sigma D_{\mu\nu\rho\sigma}^{MN}
  + 6i g_{\alpha\beta}S^{\alpha P[MN]} C^\beta_{\mu\nu\rho} V_{Pij} \bar{\epsilon}_2^i \epsilon_1^j  \nonumber \\
  &+& 36 i
  g_{\alpha\beta}S^{\alpha P[MN]} A_{[\mu P} D^{\beta}_Q{}^R \tilde{V}^Q_{(i \vert k \vert } V_R^k{}_{j)}
  \bar{\epsilon}^i_2 \gamma_{\nu\rho ]}
  \epsilon_1^j - 36 i B_{[\mu\nu}^{[M} \tilde{V}^{N]}_{ij} \bar{\epsilon}_2^i \gamma_{\rho ]}\epsilon_1^j  \quad
  . \label{3.27}
  \end{eqnarray}

We can now compare these results with the ones of the previous section, which were derived using the $E_{11}$
algebra. The comparison is performed requiring that the field strengths of eq. (\ref{2.55}) are the same as
the ones of eqs. (\ref{3.6}), (\ref{3.12}), (\ref{3.20}) and (\ref{3.25}) up to rescaling. The
field-strengths can always be put in the form that was used in this section up to field redefinitions, so
what we can actually check when we do the comparison are the independent coefficients in each field-strength.
It is straightforward to notice that the 1-forms that result from $E_{11}$ can be chosen to coincide with the
1-forms introduced in this section. For the other fields, this leads to the identifications
  \begin{eqnarray}
  & & A^{M}_{\mu\nu} = {1 \over 4} B^M_{\mu\nu} \nonumber \\
  & & A^{ \alpha}_{\mu\nu\rho} = - {1 \over 72} C^\alpha_{\mu\nu\rho} + {1 \over 6} D^\alpha_M{}^N
  B_{[\mu\nu}^M A_{\rho ] , N} \nonumber \\
  & & A^{MN}_{\mu\nu\rho\sigma} = {1 \over 1152} D^{MN}_{\mu\nu\rho\sigma} + {1 \over 96}
  C^\alpha_{[\mu\nu\rho} A_{\sigma ] , P} g_{\alpha\beta} S^{\beta P[MN]}\nonumber \\
  & & \qquad \qquad  - {1 \over 16} B_{[\mu\nu}^Q A_{\rho ,
  P} A_{\sigma ] , R} g_{\alpha\beta} D^\alpha_Q{}^P S^{\beta R[MN]} \quad . \label{3.28}
  \end{eqnarray}
In particular, once all the possible rescalings of the fields are taken into account, there is one
independent coefficient from $H^\alpha_{\mu\nu\rho\sigma}$ and two independent coefficients from
$L^{MN}_{\mu\nu\rho\sigma\tau}$. The fact that these three coefficients match is therefore non-trivial.

Finally, one can compare the gauge transformations, thus identifying the parameters $a$ of the previous
section with the gauge parameters $\Lambda_M$, $\Sigma_\mu^M$, $\Xi^\alpha_{\mu\nu}$ and
$\Delta^{MN}_{\mu\nu\rho}$ of this section. In eq. (\ref{2.56}) we have identified the parameters $a$ with $d
\Lambda$, where $\Lambda_M$, $\Lambda_\mu^M$, $\Lambda_{\mu\nu}^\alpha$ and $\Lambda_{\mu\nu\rho}^{MN}$ are
the gauge parameters occurring in the $E_{11}$ non-linear realisation. It turns out that the identification
of eq. (\ref{3.28}) is consistent with eq. (\ref{2.56}), and in particular the parameter $\Lambda_M$ in that
equation coincides with the one introduced in this section, while the other parameters are
  \begin{eqnarray}
  & & \Lambda^{M}_{\mu} = {1 \over 4} \Sigma^M_{\mu} - {1 \over 4}\Lambda_N A_{\mu,P} d^{MNP}
  \nonumber \\
  & & \Lambda^{\alpha}_{\mu\nu} =  -{1\over 72} \Xi^\alpha_{\mu\nu} - {1 \over 36}
  D^\alpha_M{}^N B^M_{\mu \nu} \Lambda_N + { 1 \over 9} D^\alpha_M{}^N \Sigma^M_{[\mu} A_{\mu ],N} \nonumber \\
  & & \qquad \qquad -{1 \over 18} \Lambda_M A_{[\mu, N} A_{\nu ] ,P} d^{MNQ} D^\alpha_Q{}^P  \nonumber \\
  & & \Lambda^{MN}_{\mu\nu\rho} = {1 \over 1152} \Delta^{MN}_{\mu\nu\rho} + {1 \over 128}
  \Xi^\alpha_{[\mu\nu} A_{\rho ], P} g_{\alpha \beta} S^{\beta P[MN]} - {1 \over 64} \Sigma^{[M}_{[\mu}
  B^{N]}_{\nu \rho ]}\nonumber \\
  && \qquad \qquad -{1 \over 32} \Sigma^Q_{[\mu} A_{\nu , P} A_{\rho ],R} g_{\alpha \beta} D^\alpha_Q{}^P
  S^{\beta R[MN]} + { 1 \over 1152} \Lambda_P C^\alpha_{\mu\nu\rho} g_{\alpha\beta} S^{\beta P[MN]} \nonumber
  \\
  & & \qquad \qquad + {1 \over 64} B^Q_{[\mu\nu} \Lambda_P A_{\rho ], R} g_{\alpha \beta} D^\alpha_Q{}^P
  S^{\beta R[MN]}+ { 1 \over 192} B^Q_{[\mu\nu} \Lambda_R A_{\rho ], P} g_{\alpha \beta} D^\alpha_Q{}^P
  S^{\beta R[MN]} \nonumber \\
  & & \qquad \qquad + {1 \over 96} \Lambda_S A_{[\mu , T} A_{\nu , R} A_{\rho\texttt{} ] , P} S^{\alpha P [MN]}
  D^\beta_Q{}^R d^{QST} g_{\alpha \beta}  \quad . \label{3.29}
  \end{eqnarray}
All these results show that the predictions of $E_{11}$ are in perfect agreement with the results obtained
imposing the closure of the supersymmetry algebra.

\section{Supersymmetry algebra of the democratic formulation of $D=5$ gauged maximal supergravity}
In this section we extend the results of the previous one in order to account for all the possible massive
deformations of the five dimensional supergravity theory. We will show that the supersymmetry algebra of any
five-dimensional gauged maximal supergravity admits a democratic formulation, in which all the bosonic fields
with antisymmetric indices are introduced together with their magnetic duals. This is the first example of a
democratic formulation of a supergravity theory with a non-abelian gauge symmetry, and this result can be
naturally generalised to any gauged maximal supergravity in any dimension.

We use conventions similar to \cite{20}, where the complete classification of all the gaugings of maximal
five-dimensional supergravity was found. In sections 5 and 6 we will show how the gauging arises in $E_{11}$
independently of the results of this section. We will indeed find that the non-linear realisation reproduces
all the results of this section. In order to make the analogy between the supergravity and the $E_{11}$
results more manifest, we use here the conventions that arise naturally from the $E_{11}$ perspective. It is
for this reason that some of the conventions are slightly different from ref. \cite{20}. The gauge algebra
associated to the higher rank fields was discussed in an independent bottom-up approach in
\cite{hierarchies}, where the results of \cite{19,20} were extended to higher rank forms. Our result
therefore shows that supersymmetry is compatible with this extension.

We first review the results of \cite{20}. In order to describe the gauging of the group $G \subset E_6$, one
introduces the embedding tensor $\Theta^M_\alpha$ so that the generators of $G$ are obtained from the
generators $t^\alpha$ of $E_6$ by
  \begin{equation}
  X^M  = \Theta^M_\alpha t^\alpha \quad . \label{4.1}
  \end{equation}
The $X$'s satisfy the commutation relations
  \begin{equation}
  [X^M, X^N ] = f^{MN}{}_P X^P \quad , \label{4.2}
  \end{equation}
where $f^{MN}{}_P$ are the structure constants of the gauge group. From eqs. (\ref{4.1}) and (\ref{4.2}) it
follows that
  \begin{equation}
  \Theta^M_\alpha \Theta^N_\beta f^{\alpha \beta}{}_\gamma = f^{MN}{}_P \Theta^P_\gamma \quad .\label{4.3}
  \end{equation}
The embedding tensor is invariant under the gauge group $G$, and using eq. (\ref{4.1}) this corresponds to
the condition that the $E_6$ transformation of $\Theta$ vanishes when contracted with $\Theta$. This results
in the equation
  \begin{equation}
  \Theta^M_\alpha ( - f^{\alpha \beta}{}_\gamma \Theta^N_\beta + D^\beta_P{}^N \Theta^P_\gamma ) =0 \quad ,
  \label{4.4}
  \end{equation}
and comparing this equation with (\ref{4.3}) one finds
  \begin{equation}
  X^{MN}_P \Theta^P_\alpha = f^{MN}{}_P \Theta^P_\alpha \label{4.5}
  \quad ,
  \end{equation}
where $X^{MN}_P$ are given by
  \begin{equation}
  X^{MN}_P = \Theta^M_\alpha D^\alpha_P{}^N \label{4.6}
  \quad .
  \end{equation}
Eq. (\ref{4.5}) shows that $X^{MN}_P$ coincides with the structure constant of the gauge group up to terms
that vanish when contracted with the embedding tensor. It can be shown \cite{20} that such terms are
symmetric in $M$ and $N$, and therefore one can write
  \begin{equation}
  X^{[MN]}_P = f^{MN}{}_P \label{4.7}
  \quad,
  \end{equation}
while the symmetric part of $X$ can be written as
  \begin{equation}
  X^{(MN)}_P = - W_{PQ} d^{QMN}\label{4.8}
  \quad ,
  \end{equation}
where $W_{MN}$ is antisymmetric and satisfies the conditions
  \begin{equation}
  W_{MN}\Theta^N_\alpha = 0 \label{4.9}
  \end{equation}
and
  \begin{equation}
  X^{MN}_{[P} W_{Q]N} =0 \label{4.10}
  \quad .
  \end{equation}
Eq. (\ref{4.8}) defines $W_{MN}$. The normalisation in eq. (\ref{4.8}) differs from the one in \cite{20}, and
it is chosen because it arises naturally from the $E_{11}$ analysis, as will become clear in the next
section. The constraints that the embedding tensor satisfies restrict it to belong to the ${\bf
\overline{351}}$ of $E_6$. The same is true for $W_{MN}$, because the ${\bf \overline{351}}$ is indeed the
irreducible representation corresponding to two fundamental antisymmetric lower indices of $E_6$. Eq.
(\ref{4.10}) guarantees that $W_{MN}$ is invariant under the action of the gauge group. The antisymmetric
part of $X^{MN}_P$ is related to $W_{MN}$ by
  \begin{equation}
  X^{[MN]}_P = - 2 d^{MQS} d^{PRT} d_{NQR} W_{ST} \quad . \label{4.11}
  \end{equation}

The scalars, that in the ungauged theory describe the non-linear realisation of $E_6$ with local subgroup
$USp(8)$, are like in the previous section denoted by $V_{Mij}$, antisymmetric and traceless with respect to
the fundamental $USp(8)$ indices $i$ and $j$. In this notation, the gauging of a subgroup of $E_6$
corresponds to a minimal coupling for the scalars $V_{Mij}$, and taking into account eq. (\ref{4.6}) one
writes the condition
  \begin{equation}
  \partial_\mu V_{Mij} + 2 Q_{\mu [i}{}^k V_{M \vert k\vert j
  ]} + g X^{NP}_M A_{\mu , N} V_{Pij}+ V_{M kl} P_\mu^{kl}{}_{ij} = 0 \quad . \label{4.12}
  \end{equation}
This equation is the covariantisation of eq. (\ref{3.7}) with respect to the gauge transformation
  \begin{equation}
  \delta V_{M ij} = - g X^{NP}_M \Lambda_N V_{P ij} \label{4.13}
  \end{equation}
of the scalars.

The variation of the scalars under gauge transformations identifies how all the covariant quantities
transform. In particular, a generic covariant object ${\cal A}^M$ with upstairs indices transforms under the
gauge transformation as
  \begin{equation}
  \delta {\cal A}^M = g X^{NM}_P \Lambda_N {\cal A}^P \quad ,\label{4.14}
  \end{equation}
while an object with downstairs indices transforms according to
  \begin{equation}
  \delta {\cal A}_M = - g X^{NP}_M \Lambda_N {\cal A}_P \quad . \label{4.15}
  \end{equation}

For the gauging of a subgroup $G$ of $E_6$ to occur, a subset of the vectors in the ${\bf 27}$ of $E_6$ have
to collect it the adjoint of $G$, while the rest of the vectors are gauged away by means of a Higgs mechanism
that gives a mass to the 2-forms. More precisely, one requires that the gauge transformation of the vector
becomes the non-abelian one when contracted with the embedding tensor. One thus writes the gauge
transformation of the vectors as
  \begin{equation}
  \delta A_{\mu M} = \partial_\mu \Lambda_M - g X^{[NP]}_M \Lambda_N A_{\mu P} + g W_{MN} \Sigma^N_\mu
  \quad , \label{4.16}
  \end{equation}
where $\Sigma^M_\mu$ are the gauge parameters of the 2-forms introduced in the previous section. Contracting
eq. (\ref{4.16}) with $\Theta^M_\alpha$, the last term vanishes because of eq. (\ref{4.9}), and one is left
with the non-abelian gauge transformation of the vector projected by the embedding tensor.

From eq. (\ref{4.16}) one can write the field strength
  \begin{equation}
  F_{\mu \nu M} = 2\partial_{[\mu} A_{\nu ] M} + g X^{[NP]}_M A_{\mu N} A_{\nu P} - g W_{MN} B_{\mu\nu}^N
  \quad, \label{4.17}
  \end{equation}
that is gauge invariant under $\Sigma^M_\mu$ transformations at order $g$. The normalisation of the last term
in eq. (\ref{4.16}) is chosen in such a way that $F_{\mu \nu M}$ varies under $\Lambda_M$ transformations as
in eq. (\ref{4.15}) at order $g$. Imposing that $F$ transforms covariantly at order $g^2$ partially fixes the
order $g$ gauge transformation of $B_{\mu\nu}^M$ in a way which is consistent with what we will find in the
following. The strategy of ref. \cite{20} was to consider the 2-forms always contracted with $W_{MN}$,
because $W_{MN} B_{\mu\nu}^N$ is the object that appears in the lagrangian. They therefore obtain the part of
the order $g$ transformations of the 2-forms which does not vanish when contracted with $W_{MN}$. As we will
see, our analysis instead will determine the gauge transformations of the 2-forms completely, and we will
also determine the full gauge transformations of the 3-forms dual to the scalars.

In the gauged theory, the supersymmetry transformations of the bosons remains unchanged, while the
transformations of the fermions are modified with respect to eq. (\ref{3.5}) because of two reasons. First of
all, one assumes that the field strengths and the covariant derivatives that occur in the supersymmetry
transformations of the fermions are now covariant with respect to the gauge transformations. This means that
the field-strength $F_{\mu \nu ,M}$ of the vectors is now defined as in eq. (\ref{4.17}), and $Q_{\mu i}{}^j$
and $P_{\mu ijkl}$ are now defined by eq. (\ref{4.12}). Secondly, explicit mass terms appear in the
supersymmetry variation of the fermions. These terms are obtained requiring that the corresponding action is
supersymmetric, and one can show that the results of \cite{20} can be written in a way that makes the scalar
dependence more explicit. The result is
  \begin{equation}
  \delta^\prime \psi_{\mu i} = {1 \over 3} g W_{MN} \tilde{V}^{M}_{ij} \tilde{V}^{Njk}
  \gamma_\mu \epsilon_k \label{4.18}
  \end{equation}
for the gravitino, and
  \begin{equation}
  \delta^\prime \chi_{ijk} = 3 \sqrt{2} g W_{MN} \tilde{V}^M_{[ij} \tilde{V}^N_{k]l}
  \epsilon^l - \sqrt{2}
  g W_{MN} \Omega_{[ij} \tilde{V}^M_{k] }{}^l \tilde{V}^N_{lm} \epsilon^m \label{4.19}
  \end{equation}
for the spinor, where we denote with $\delta^\prime$ the part of the supersymmetry transformations of the
fermions that contain explicit mass terms. Expressing these mass deformations explicitly in terms of $W_{MN}$
and the scalars will turn out to be crucial in the second part of this section, where we will close the
supersymmetry algebra on the 2-forms and the 3-forms dual to the vectors and the scalars respectively, and
where we will derive the duality relation between the 5-form field strengths and the mass parameters.

As in the previous section, we are interested in studying the supersymmetry algebra. Changing the
supersymmetry transformations of the fermions results in additional terms in the commutators of two
supersymmetry transformations $[\delta_{\epsilon_1}, \delta_{\epsilon_2}]$ on the bosons. In particular, the
commutator of two supersymmetry transformations on the scalars produces the gauge transformation of eq.
(\ref{4.13}), while on the vectors it produces the gauge transformation of eq. (\ref{4.16}), where the
parameters $\Lambda_M$ and $\Sigma_\mu^M$ are given by eqs. (\ref{3.10}) and (\ref{3.15}). All the
assumptions in the above construction, and in particular eqs. (\ref{4.14}) and (\ref{4.15}), are very
natural, however the justification for them is that they lead to a supersymmetry algebra which closes and
leads to an invariant action. A more pedagogical but more technically difficult approach  would be to add a
single deformation term, like the first term of order $g$ in eq. (\ref{4.16}), and demand closure of the
supersymmetry algebra by adding terms. One would then recover the same results.

In the above, we have reviewed ref. \cite{20} showing that the supersymmetry algebra of five-dimensional
gauged maximal supergravity closes on the scalars and the vectors. In the rest of this section we will show
how the supersymmetry algebra closes on the 2-forms dual to the vectors, and on the 3-forms dual to the
scalars. This proves that the supersymmetry algebra of gauged maximal supergravities admits a democratic
formulation, in which all the fields are introduced together with their magnetic duals and the algebra closes
using the duality relations. As in the previous section, the analysis is carried out at lowest order in the
fermions, and it generalises the results of the previous section to the case of five-dimensional gauged
supergravity.

We start considering the 2-forms $B_{\mu\nu}^M$. We determine the gauge transformation of $B_{\mu \nu}^M$
requiring that the duality condition of eq. (\ref{3.14}) is gauge invariant. This fixes the gauge
transformation of the field-strength $G_{\mu \nu\rho}^M$ to be
  \begin{equation}
  \delta G_{\mu \nu\rho}^M = g X^{NM}_P \Lambda_N  G_{\mu \nu\rho}^P \quad . \label{4.20}
  \end{equation}
It turns out that this condition determines the gauge transformation of $B_{\mu \nu}^M$ and its field
strength uniquely. In order to facilitate the comparison with the $E_{11}$ results, in this section we always
keep the order of the coupling constant $g$ explicit. This means that we write the gauge transformations and
the field strengths always in terms of the fields and their derivatives, without using the field-strengths of
the lower rank fields, as we did instead in the previous section. We thus write the final result as
  \begin{eqnarray}
  \delta B_{\mu\nu}^M &=& 2 \partial_{[\mu} \Sigma_{\nu]}^M - 2 d^{MNP} \Lambda_N \partial_{[\mu} A_{\nu] P} - {4
  \over 3} g X^{[MN]}_P \Sigma_{[\mu}^P A_{\nu] N} \nonumber \\
  &+& {2 \over 3} g X^{(MN)}_P \Sigma_{[\mu}^P A_{\nu]
  N} -{2 \over 3 } g X^{[MN]}_P \Lambda_N B_{\mu\nu}^P +{4  \over 3 } g X^{(MN)}_P \Lambda_N
  B_{\mu\nu}^P \nonumber \\
  & -& {4 \over 3} g X^{(MQ)}_R d^{RNP} \Lambda_N A_{[\mu , P} A_{\nu ] , Q} + {2 \over 3} g
  X^{[MQ]}_R d^{RNP} \Lambda_N A_{[\mu , P} A_{\nu ] , Q} \nonumber \\
  & +& {1 \over 6} g \Theta^M_\alpha \Xi^\alpha_{\mu\nu} \label{4.21}
  \end{eqnarray}
for the gauge variation of the 2-form and
  \begin{eqnarray}
  G_{\mu\nu\rho}^M  &=& 3 \partial_{[\mu} B_{\nu\rho]}^M + 6 d^{MNP}A_{[\mu N} \partial_{\nu} A_{\rho ] P} +
  2g X^{[MN]}_P B_{[\mu\nu}^P A_{\rho ] N} \nonumber \\
  &-& 4 g X^{(MN)}_P B_{[\mu\nu}^P A_{\rho ] N} + 2g X^{[NP]}_R d^{RQM} A_{[\mu , N} A_{\nu , P } A_{\rho ]
  , Q}
  \nonumber \\
  &-& {1\over 6} g
  \Theta^M_\alpha C^\alpha_{\mu\nu\rho} \label{4.22}
  \end{eqnarray}
for its field strength. It is important to observe that the 2-form varies with respect to the parameter
$\Xi^\alpha_{\mu\nu}$, that is the gauge parameter of the 3-form $C_{\mu\nu\rho}^\alpha$ that we introduced
in the previous section, and that this variation contains the embedding tensor. This has to be compared with
eq. (\ref{4.16}), which shows that the 1-form varies with respect to the gauge parameter of the 2-form by a
term containing $W_{MN}$. The variation of $G_{\mu\nu\rho}^M$ at order $g$ satisfies eq. (\ref{4.20}), and
requiring that this is true also at order $g^2$ partially determines the gauge transformation of the 3-form
in a way that is consistent with what we will find in the following. The gauge transformation of
$B_{\mu\nu}^M$ of eq. (\ref{4.21}) is also consistent with the covariance of $F_{\mu\nu ,M}$ at order $g^2$.

The supersymmetry transformation of $B_{\mu\nu}^M$ is given by eq. (\ref{3.11}), and using the gauged
supersymmetry transformations of the fermions one can show that the supersymmetry algebra closes on
$B_{\mu\nu}^M$, generating the gauge transformation of eq. (\ref{4.21}) with the correct parameters given in
eqs. (\ref{3.10}), (\ref{3.15}) and (\ref{3.23}), and using the duality relation of eq. (\ref{3.14}) where
the field strength of the 2-form is as in eq. (\ref{4.22}). This proves that the gauge transformations we
find are completely consistent with the supersymmetry algebra.

We now consider the 3-forms $C_{\mu\nu\rho}^\alpha$ that are dual to the scalars. The duality relation of eq.
(\ref{3.22}) implies that the gauge transformation of the 4-form field strength $H^\alpha_{\mu\nu\rho\sigma}$
is
  \begin{equation}
  \delta H^\alpha_{\mu\nu\rho\sigma} = g f^{\beta \alpha}{}_\gamma \Theta^M_\beta \Lambda_M
  H^\gamma_{\mu\nu\rho\sigma} \quad . \label{4.23}
  \end{equation}
Starting from the ungauged result of eq. (\ref{3.20}), it turn out that imposing eq. (\ref{4.23}) at order
$g$ completely determines $H^\alpha_{\mu\nu\rho\sigma}$ as well as the gauge transformation of the 3-form
$C_{\mu\nu\rho}^\alpha$. The final result is
  \begin{eqnarray}
  \delta C^\alpha_{\mu\nu\rho} &=& 3 \partial_{[\mu} \Xi^\alpha_{\nu\rho]} + {6 } D^\alpha_M{}^N
  \partial_{[\mu} B_{\nu\rho]}^M \Lambda_N + 24 D^\alpha_M{}^N \Sigma_{[\mu}^M \partial_\nu A_{\rho ] N}
  \nonumber \\
  &+& {12 } D^\alpha_M{}^N d^{MPQ}A_{[\mu P} \partial_{\nu} A_{\rho ] Q} \Lambda_N
  + [3 g D^\alpha_N{}^P W_{PM} - 9 g D^\alpha_M{}^P W_{PN}  ] \Sigma_{[\mu}^M B_{\nu\rho ]}^N \nonumber \\
  &-& [ {3 \over 4 } g f^{\alpha\beta}{}_\gamma \Theta^N_\beta + {1\over 4} g D^\alpha_P{}^N
  \Theta^P_\gamma ] \Lambda_N
  C^\gamma_{\mu\nu\rho}  + [  {9\over 4} g f^{\alpha\beta}{}_\gamma \Theta^N_\beta
  - {1\over 4} g D^\alpha_P{}^N \Theta^P_\gamma ]
  \Xi^\gamma_{[\mu\nu} A_{\rho] N} \nonumber \\
  &+&  [ - {3 \over 2 } g D^\alpha_Q{}^S X^{PN}_S
  + {3 \over 2 } g D^\alpha_Q{}^S X^{NP}_S - {3 \over 2 } g
  D^\alpha_S{}^N X^{PS}_Q - {1 \over 2 } g D^\alpha_S{}^N X^{SP}_Q \nonumber \\
  &-& {9 \over 2 } g D^\alpha_S{}^P X^{NS}_Q - {3 \over 2 } g D^\alpha_S{}^P X^{SN}_Q ] \Lambda_P A_{[\mu N}
  B_{\nu\rho]}^Q \nonumber\\
  & +&  [ g D^\alpha_M{}^N X^{MP}_Q + 3 g D^\alpha_M{}^N X^{PM}_Q - 9 g D^\alpha_Q{}^M X^{PN}_M ]
  A_{[\mu N } A_{\nu P} \Sigma_{\rho]}^Q \nonumber \\
  & +& [ - {4 } g D^\alpha_M{}S X^{(MQ)}_R d^{RNP} + {8 } g D^\alpha_M{}S X^{[MQ]}_R d^{RNP} ]
  \Lambda_N A_{[\mu , P} A_{\nu , Q} A_{\rho ] , S} \nonumber \\
  & +& 12 g D^\alpha_M{}^U W_{UN} S^{P[MN]}_\beta D^\beta_Q{}^R  d^{QST} \Lambda_S A_{[ \mu , T}
  A_{\nu , R} A_{\rho ] , P} + g D^\alpha_M{}^P W_{PN} \Delta^{MN}_{\mu\nu\rho} \label{4.24}
  \end{eqnarray}
for the gauge transformation of the 3-form and
  \begin{eqnarray}
  H^\alpha_{\mu \nu \rho \sigma} &=& 4 \partial_{[\mu} C_{\nu \rho \sigma]}^\alpha - 48 D^\alpha_M{}^N
  B^M_{[\mu\nu}
  \partial_\rho A_{ \sigma ]  N} - {24 } D^\alpha_M{}^N A_{[\mu N} \partial_\nu B^M_{ \rho \sigma]}\nonumber
  \\
  &-& 48 D^\alpha_M{}^N d^{MPQ} A_{[\mu N} A_{\nu P} \partial_\rho A_{\sigma ] Q}
  + 18 g D^\alpha_M{}^P W_{PN} B_{[\mu\nu}^M B_{\rho\sigma]}^N +3g f^{\alpha \beta}{}_\gamma \Theta^N_\beta
  A_{[\mu N } C^\gamma_{\nu\rho\sigma]}\nonumber \\
  & + & g D^\alpha_M{}^P \Theta^M_\beta A_{[\mu P } C^\beta_{\nu\rho\sigma]}
  + [- 18g D^\alpha_Q{}^S X^{[PN]}_S - 18 g D^\alpha_S{}^N X^{PS}_Q \nonumber \\
  & - & 6 g  D^\alpha_S{}^N X^{SP}_Q ] A_{[\mu P}
  A_{\nu N} B_{\rho\sigma ]}^Q
  +12 g X^{[MN]}_R d^{RPS} D^\alpha_S{}^Q A_{[\mu , M} A_{\nu , N} A_{\rho , P} A_{\sigma ] , Q} \nonumber \\
  & -&  g D^\alpha_M{}^P W_{PN}
  D^{MN}_{\mu\nu\rho\sigma} \label{4.25}
  \end{eqnarray}
for its field-strength. Once again, in order to prove gauge covariance it is crucial to impose that the
3-form transforms with respect to the gauge parameter of the 4-form, as the last term in eq. (\ref{4.24})
shows. We also made use of the identity
  \begin{equation}
  f^{\alpha\beta}{}_\gamma \Theta^Q_\beta - D^\alpha_P{}^Q \Theta^P_\gamma = 4 D^\alpha_M{}^P W_{PN} g_{\beta
  \gamma} S^{\beta Q[MN]} \quad , \label{4.26}
  \end{equation}
which shows that the embedding tensor and $W_{MN}$ are related by the invariant tensor $S^{\alpha P[MN]}$,
and thus belong to the same representation of $E_6$. Using eq. (\ref{2.34}) and the invariance of $d^{MNP}$
one can indeed show that this identity leads to the linear constraint of \cite{20}, which is needed to prove
that the embedding tensor belongs to the ${\bf \overline{351}}$ of $E_6$. The variation of
$C_{\mu\nu\rho}^\alpha$ at order $g$ is such that the 3-form field strength $G_{\mu\nu\rho}^M$ of eq.
(\ref{4.22}) is covariant at order $g^2$.

The supersymmetry transformation of $C_{\mu\nu\rho}^\alpha$ is given in eq. (\ref{3.19}), and using the
gauged supersymmetry transformations of the fermions one can compute the commutator of two supersymmetry
transformations on this field at lowest order in the fermions. It turns out that the supersymmetry algebra
closes on $C_{\mu\nu\rho}^\alpha$, generating the gauge transformation of eq. (\ref{4.24}) where the
parameters are as in eqs. (\ref{3.10}), (\ref{3.15}), (\ref{3.23}), and (\ref{3.27}). Like in the massless
case, the supersymmetry algebra closes imposing the duality relation of eq. (\ref{3.22}) where now the field
strength of the 3-form is as in eq. (\ref{4.25}). Therefore the supersymmetry algebra of gauged maximal
supergravity in five dimensions closes on the 2-forms and the 3-forms dual the non-abelian vectors and the
scalars respectively. One could continue this analysis, and show that the supersymmetry algebra closes on the
4-forms, provided that their field-strengths are related by duality to the mass deformation parameters of the
gauged theory. We leave this as an open project. As it is clear from the previous results, in order to
determine the gauge transformation of the 4-forms we would need to know how the 5-forms transform at zeroth
order in $g$. It would be interesting to perform this analysis, and compare the results with the ones of
appendix A, where the gauge transformations of the 5-forms at zeroth order in $g$, that is in the massless
theory, are computed from $E_{11}$.

We do not determine the gauge transformation of the 4-forms $D_{\mu\nu\rho\sigma}^{MN}$ at order $g$, and so
we can not determine the 5-form field strengths $L^{MN}_{\mu\nu\rho\sigma\tau}$ at order $g$ using them.
However, we can still derive the duality relation of these field strengths with the mass deformation
parameters. The supersymmetry transformation of the 4-forms is given in eq. (\ref{3.24}). Using the gauged
supersymmetry transformations of the fermions one can compute the supersymmetry commutator, and from that one
can select the term proportional to the general coordinate transformation parameter given in eq. (\ref{3.9}).
The relevant terms are the ones that arise from performing the variations of eqs. (4.18) and (4.19) in eq.
(\ref{3.9}). This results in the contribution
  \begin{eqnarray}
  & & g i W_{PQ} [ 8 \tilde{V}^{[M}_{ik} \tilde{V}^{N]k}{}_j \tilde{V}^{Pjl} \tilde{V}^Q_{lm} + 8
  \tilde{V}^{[M}_{ij} \tilde{V}^{N]}_{kl} \tilde{V}^{Pjk} \tilde{V}^{Ql}{}_{m}\nonumber \\
  & & + 4 \tilde{V}^{[M}_{ij} \tilde{V}^{N]}_{kl} \tilde{V}^{Pkl} \tilde{V}^{Qj}{}_{m}] (\bar{\epsilon}_2^i
  \gamma_{\mu \nu\rho\sigma} \epsilon_1^m - \bar{\epsilon}_1^i
  \gamma_{\mu \nu\rho\sigma} \epsilon_2^m ) \label{Dcommgauged}
  \end{eqnarray}
to the supersymmetry commutator on $D_{\mu\nu\rho\sigma}^{MN}$. We have to select out of the terms in eq.
(\ref{Dcommgauged}) the part that is proportional to the general coordinate transformation parameter given in
eq. (\ref{3.9}), which means that we have to select the part of the fermionic bilinear that is proportional
to $\Omega^{im}$. This term has to produce the general coordinate transformations of the fields
$D_{\mu\nu\rho\sigma}^{MN}$, and for this to occur the duality relation
  \begin{equation}
  L^{MN}_{\mu\nu\rho\sigma\tau} = g \epsilon_{\mu\nu\rho\sigma\tau} W_{PQ}[\tilde{V}^M_{ij} \tilde{V}^N_{kl}
  \tilde{V}^{Pij} \tilde{V}^{Qkl} -2 \tilde{V}^M_{ik} \tilde{V}^{Nk}{}_{j}
  \tilde{V}^{Pi}{}_{l} \tilde{V}^{Qlj}] \label{4.27}
  \end{equation}
must hold. Here $L^{MN}_{\mu\nu\rho\sigma\tau}$ are the 5-form field-strengths of the gauged theory,
transforming covariantly under gauge transformations and whose zeroth order in $g$ is given in (\ref{3.25}).
The right-hand side of this duality relation is proportional to the scalar potential of \cite{20}. In the
first version of this paper the second term in eq. (\ref{4.27}) was missing. The fact that there was
something odd in that equation was pointed out in \cite{dWNS}. Taking the curl of the duality relation of eq.
(\ref{3.22}) and using eq. (\ref{4.27}) one obtains the second order equation for the scalars, which means
that the scalar potential is encoded in this chain of first order duality relations. The duality relation of
eq. (\ref{4.27}) follows directly from the terms in the supersymmetry transformation of the fermions
containing explicit mass terms, which are given in eqs. (\ref{4.18}) and (\ref{4.19}). These equations indeed
show that $W_{MN}$ should be thought as the mass deformation parameter, and therefore it is natural to expect
that the 5-form field strength is related to $W_{MN}$ by duality, in agreement with our results.

To conclude this section, we want to write the gauge transformations of the gauged theory in terms of the
$E_{11}$ fields and the $E_{11}$ parameters of section 2. We recall that from $E_{11}$ the gauge
transformations of the five-dimensional fields in the massless theory are simply obtained acting on the group
element of eq. (\ref{2.37}) with the elements of eqs. (\ref{2.38}), (\ref{2.40}), (\ref{2.42}) and
(\ref{2.44}) and identifying the $E_{11}$ parameters $a$ with the gauge parameters $\Lambda$ as in eq.
(\ref{2.56}). Performing the redefinitions of the fields and the parameters given in eq. (\ref{3.28}) and
(\ref{3.29}) one derives the gauge transformations for the fields in the massless theory as obtained in the
previous section using supersymmetry. In this section we have shown how these gauge transformations are
modified in the gauged theory using supersymmetry. The fact that all the transformations are first order in
$g$ implies that we can use the zeroth order field and parameter redefinitions as obtained in the previous
section on these gauge transformations, to derive their form in the $E_{11}$ basis. This is consistent with
the fact that $E_{11}$ gives corrections only at order $g$, as will be shown in the next section.

We thus perform the redefinitions of the fields and the parameters given in eq. (\ref{3.28}) and (\ref{3.29})
on the gauge transformations obtained in this section, in order to the determine their form in terms of the
$E_{11}$ fields and parameters. We are only interested in the first order in $g$, because the $E_{11}$
analysis of the massless theory, {\it i.e.} at zeroth order in $g$, has already been performed in section 2.
It turns out that performing these redefinitions, the transformations of eqs. (\ref{4.16}), (\ref{4.21}) and
(\ref{4.24}) drastically simplify, and the final result is
  \begin{eqnarray}
  & & \delta_g A_{\mu , M} = - g \Lambda_P \Theta^P_\alpha D^\alpha_M{}^N
  A_{\mu , N} + 4 g W_{MN} \Lambda_\mu^N \nonumber \\
  & & \delta_g A_{\mu \nu}^M = g \Lambda_P \Theta^P_\alpha D^\alpha_N{}^M  A_{\mu \nu}^N
  + 2 g W_{NQ}\Lambda_{[\mu}^Q  d^{MNP} A_{\nu
  ] , P} - 3 g \Theta^M_\alpha \Lambda^\alpha_{\mu\nu} \nonumber \\
  & & \delta_g A^\alpha_{\mu\nu\rho} = - g \Lambda_P \Theta^P_\beta f^{\alpha\beta}{}_\gamma A^\gamma_{\mu\nu\rho}
  + 4 g W_{MP} \Lambda_{[\mu}^P D^\alpha_N{}^M A_{\nu\rho]}^N \nonumber \\
  && \qquad \quad + {2 \over 3} g W_{MR} \Lambda_{[\mu}^R d^{MNQ}
  D^\alpha_Q{}^P A_{\nu , N} A_{\rho ] , P} - 16 g D^\alpha_M{}^P W_{PN} \Lambda_{\mu\nu\rho}^{MN} \quad ,
  \label{4.28}
  \end{eqnarray}
where $\delta_g$ denotes the part of the gauge transformation which is first order in $g$, and the full
results are recovered adding the zeroth order transformations of eqs. (\ref{2.41}), (\ref{2.43}) and
(\ref{2.45}), where the gauge parameters are given in eq. (\ref{2.56}). Even the reader who is unfamiliar
with $E_{11}$ might get the feeling that there is some hidden structure which is responsible for this drastic
simplification. The rest of this paper will be devoted to showing how the transformations of eq. (\ref{4.28})
result from $E_{11}$. Here we just want to conclude pointing the reader's attention to the similarity between
the gauge transformations of eq. (\ref{4.28}) and the $E_{11}$ transformations of eqs. (\ref{2.41}),
(\ref{2.43}) and (\ref{2.45}).

\section{Generalised spacetime and the $E_{11}$ dynamics of \\ gauged supergravities}

\subsection{Generalised spacetime}

When the $E_{11}$ symmetry was  first conjectured \cite{9} the momentum operator $P_a$ was included in the
group element in order to encode space-time into the non-linear realisation. It was realised that using just
this single generator  does not respect the $E_{11}$ symmetry and thus the momentum operator should be part
of some larger multiplet. The correct procedure \cite{31}, found a few years later, is to introduce a set of
generators that transform as a linear representation of $E_{11}$ which includes the spacetime translations as
its first component. This representation, denoted $l$ here although $l_1$ in the previous literature, is the
fundamental representation of $E_{11}$ associated with the node labelled 1. The next components in the $l$
multiplet in order of increasing level are an anti-symmetric two form $Z^{a_1 a_2}$, an anti-symmetric 5-form
$Z^{a_1\ldots a_5}$, which can be identified with the the central charges in the eleven dimensional
supersymmetry algebra, then   $Z^{a_1\ldots a_7,b}$ associated with the Taub-Nut solution followed by  an
infinite number of components at higher levels.

The dynamics is specified to be a non-linear realisation based on the semi-direct product of the two groups
$E_{11}$ and a group whose elements are those of $l$, and we write this as  $E_{11}\otimes_s l$ \cite{31};
its precise formulation  can be found in \cite{31} and will also be discussed below. The construction of this
non-linear realisation involves a group element that contains the generators of the Borel sub-algebra of
$E_{11}$, once one has taken account of the local sub-algebra, and those of $l$. The coefficients of the
latter include $x^a$, the usual coordinate of space-time but also the coordinates $x_{a_1 a_2}$ and
$x_{a_1\ldots a_5}$ corresponding to  $Z^{a_1a_2}$ and $Z^{a_1\ldots a_5}$ respectively as well as an
infinite number of higher level coordinates all of which can be thought of as constituting a generalised
space-time. The group element can be written in the generic form
  $$
  g= e^{x^aP_a+x_{a_1a_2}Z^{a_1a_2}+x_{a_1\ldots a_5}Z^{a_1\ldots
  a_5}+\dots }e^{A\cdot R} \eqno(5.1.1)
  $$
where $R$ denotes the generators of the Borel sub-algebra of $E_{11}$ and $A$ are the corresponding fields
that depend in general on the generalised space-time. In the past literature the non-linear realisation has
been constructed keeping only the usual coordinate of space-time $x^a$. One of the most pressing problems in
the  understanding of the $E_{11}$ conjecture has been to understand  the precise role that the generalised
space-time plays in the dynamics.  In this section we will show that it plays a central role in the
formulation of the dynamics of the gauged supergravity theories thus providing strong evidence for the role
of the $l$ multiplet in the non-linear realisation and so in M theory.

In fact the $l$ multiplet has a physical interpretation; it is just the multiplet of brane charges \cite{32}.
This is clearly true  at the lowest levels where on finds in order of ascending level the charge of the point
particle, the two brane  charge, the five brane  charge. The dynamics of a $p$  brane is described by an
action which contains a Wess-Zumino term whose  leading term is  a coupling between a rank $p+1$ gauge field
which is one of the non-trivial background fields and a conserved current. This current has a corresponding
charge which is the brane charge and to which the gauge field couples. As such one expects every field in the
$E_{11}$ non-linear realisation to have a corresponding charge in the $l$ multiplet. Indeed this is the case
\cite{32}; the fields in the non-linear realisation are in one to one correspondence with the generators of
the Borel sub-algebra of $E_{11}$ and if one deletes any of the space-time from any one of these  generators
one finds an element in the $l$ representation that has the resulting structure of space-time indices. From
this view point introducing the generalised coordinates corresponds to using coordinates for measuring
space-time using all possible branes and not just those associated with the point particle.

As explained above by choosing different $A_{D-1}$, or $SL(D,\mathbb{R})$ sub-algebras of $E_{11}$ one
identifies different gravity lines and so theories in different dimensions. In this construction one
automatically finds the duality groups long known to be symmetries of the corresponding maximal supergravity
theories, for example $E_7$ in four dimensions \cite{1} and  $SL(2,\mathbb{R})$ for IIB supergravity
\cite{2}. Technically this construction corresponds to decomposing the adjoint representation of $E_{11}$
into those of $A_{D-1}$ direct product with the duality group. The generators of $E_{11}$, and so the fields,
with totally anti-symmetric indices  in the $D$ dimensions were only  rather recently computed. It was found
that they lead to  a totally democratic formulation of the propagating forms together with some forms that
have $D-1$ and $D$ indices. The former correspond to the  gauged supergravities constructed and one finds a
precise match \cite{29,30}  with the pattern of gauged supergravities derived using supersymmetry over very
many years. Thus one finds that $E_{11}$ provides a unified framework for all the maximal supergravity
theories many of which had no higher dimensional origin within the context of traditional supergravity
theories.

When considering the non-linear realisation $E_{11}\otimes_s l$ one must also carry out the  decomposition of
the $l$ multiplet  $A_{D-1}$ direct product with the duality group as well as that for the adjoint
representation of $E_{11}$ in order to determine the theory that results in $D$ dimensions. In fact this
calculation was carried out a few years ago \cite{33,34} and the results for the members of the multiplet
that are forms, that is possess just a set of totally antisymmetrised indices, are summarised in table 1
\cite{33,34}. Comparing with the set of generators of $E_{11}$ table 5 of reference \cite{29} one sees the
above discussed correspondence between charges and fields. The results can be compared with earlier
calculations \cite{35} that assumed U duality symmetries and used the known U duality transformation rules to
compute some of the multiplets of brane charges from a known brane charge. It was observed that the resulting
brane charges were generically more numerous than the central charges in the corresponding supersymmetry and
that many of the charges had a rather exotic structure that did not arise from a reduction of an eleven or
ten dimensional supergravity theory.  The decomposition of the $l$ multiplet gives precise agreement with
these results as can be seen at a glance by comparing the results of table 1 with tables 4.11 and 4.14 of the
third paper in reference \cite{35}. Furthermore, the $l$ multiplet provides a single unifying structure for
all the charges found in lower dimensions many of which have  no higher dimensional origin within the context
of traditional supergravity theories. Thus there is substantial evidence for the relevance of the $l$
multiplet in M theory, however, not so far for the generalised space-time that results from it in the
non-linear realisation. It is the purpose of this section to rectify this short coming.
\begin{table}
\begin{center}
\begin{tabular}{|c|c||c|c|c|c|c|c|c|}
\hline \rule[-1mm]{0mm}{6mm} D & G & $Z$ & $Z^a$ & $Z^{a_1 a_2}$ & $Z^{a_1 \dots a_3}$ & $Z^{a_1 \dots a_4}$
& $Z^{a_1 \dots a_5}$
& $Z^{a_1 \dots a_6}$  \\
\hline \rule[-1mm]{0mm}{6mm} \multirow{3}{*}{8} & \multirow{3}{*}{$SL(3,\mathbb{R}) \times SL(2,\mathbb{R})$}
& \multirow{3}{*}{${\bf ({3}, 2)}$} & \multirow{3}{*}{${\bf (\overline{3},1) }$} & \multirow{3}{*}{${\bf
(1,2)}$} & \multirow{3}{*}{${\bf
({3},1)}$} & \multirow{3}{*}{${\bf (\overline{3},2) }$} & ${\bf (8,1)}$ & ${\bf (\overline{6},2)}$  \\
& & & & & & & ${\bf (1,3)}$ & ${\bf ({3},2)}$   \\
& & & & & & & ${\bf (1,1)}$ & ${\bf ({3},2)}$   \\
 \cline{1-9} \rule[-1mm]{0mm}{6mm} \multirow{3}{*}{7} & \multirow{3}{*}{$SL(5,\mathbb{R})$} & \multirow{3}{*}{${\bf
{10}}$} & \multirow{3}{*}{${\bf \overline{5} }$} & \multirow{3}{*}{${\bf {5}}$} & \multirow{3}{*}{${\bf
\overline{10}}$} &
{${\bf 24 }$} & ${\bf {40}}$   \\
& & & & & & & ${\bf {15}}$   \\
& & & & & & ${\bf 1}$ & ${\bf
{10}}$  \\
 \cline{1-8} \rule[-1mm]{0mm}{6mm}\multirow{2}{*}{6} & \multirow{2}{*}{$SO(5,5)$} & \multirow{2}{*}{${\bf
\overline{16}}$} & \multirow{2}{*}{${\bf 10 }$} & \multirow{2}{*}{${\bf {16} }$} & {${\bf \overline{45}}$} &
{${\bf \overline{144} }$}   \\
& & & & & ${\bf 1}$ & ${\bf
\overline{16}}$ \\
\cline{1-7} \rule[-1mm]{0mm}{6mm} \multirow{2}{*}{5} & \multirow{2}{*}{$E_{6(+6)}$} & \multirow{2}{*}{${\bf
\overline{27}}$} & \multirow{2}{*}{${\bf {27} }$} & {${\bf 78 }$} & {${\bf \overline{351}}$}
  \\
& & & & ${\bf 1}$  &  ${\bf \overline{27}}$ \\
 \cline{1-6} \rule[-1mm]{0mm}{6mm} \multirow{2}{*}{4} & \multirow{2}{*}{$E_{7(+7)}$} & \multirow{2}{*}{${\bf
56}$} & {${\bf 133 }$} & {${\bf 912 }$}  \\
 & & & ${\bf 1}$  & ${\bf 56}$\\
 \cline{1-5} \rule[-1mm]{0mm}{6mm}
 \multirow{3}{*}{3} &  \multirow{3}{*}{$E_{8(+8)}$} &  \multirow{3}{*}{${\bf 248}$} & ${\bf 3875}$
  \\
  & & & ${\bf 248}$\\
 & & & ${\bf 1}$  \\
 \cline{1-4}
\end{tabular}
\end{center}
\caption{\small Table giving the representations of the symmetry group $G$ of the form charges in the $l$
multiplet up to and including rank $D-2$ in $D$ dimensions, in 8 dimensions and below
\cite{33,34}.\label{Table1}}
\end{table}

As  table 1 shows the members of the $l$ multiplet in five dimensions, classified according to $E_6$
multiplets and in order of increasing rank of the totally anti-symmetrised space-time indices, are given by
\cite{33}
  $$
  P_a, \ Z^N,\  Z^a_N, \ Z^{a_1a_2\alpha}, \ Z^{a_1 a_2} , \ Z^{a_1a_2a_3}_{NM} , \ Z^{a_1a_2a_3 N}  ,
  \ldots \eqno(5.1.2)
  $$
As their indices imply these transform according to the ${\bf 1}$, ${\bf \overline{27}}$, ${\bf {27}}$,
adjoint {\it i.e.} ${\bf 78}$, $\bf \overline{351}$ and $\bf \overline{27}$ of $E_6$ respectively. As already
discussed there is a relation between the members of the $l$ multiplet and the fields in the adjoint
representation of $E_{11}$, namely if one deletes a spacetime index from the latter one finds a corresponding
charge in the former.

It is instructive to derive the low level members of the $l$ multiplet in five dimensions from that in eleven
dimensions. In eleven dimensions the $l$ multiplet has the following content \cite{31}
  $$
  P_{\hat a}; Z^{\hat a_1\hat a_2}; Z^{\hat a_1\ldots\hat  a_5}; Z^{\hat a_1\ldots \hat  a_7,b},  Z^{\hat
  a_1\ldots \hat  a_8}; Z^{\hat  a_1\ldots \hat  a_8, \hat b_1\hat b_2\hat b_3}, Z^{\hat  a_1\ldots \hat
  a_9,(\hat b\hat c)}, Z^{\hat  a_1\ldots \hat a_9,\hat b_1\hat b_2},
  $$
  $$
  Z^{\hat  a_1\ldots \hat a_{10},\hat b}, Z^{\hat  a_1\ldots \hat a_{11}}; Z^{\hat  a_1\ldots \hat a_9,\hat
  b_1\ldots \hat b_4,\hat c}, Z^{\hat  a_1\ldots \hat a_8,\hat b_1\ldots \hat b_6}, Z^{\hat  a_1\ldots \hat
  a_9,\hat b_1\ldots \hat b_5},\dots \eqno(5.1.3)
  $$
where $ \hat a=1,\ldots ,11$. To find the content of the five dimensional theory we split the indices range
of $\hat a$ etc  into $\hat a=a, a=1,\dots , 5$ and $\hat a=i+5 ,a=6,\dots , 11$. The latter transform under
$SL(6)$.  If we consider scalars we find at low levels $P_i$, $Z^{ij}$, $Z^{i_1\ldots i_5}$ which are the
$\bf {6}$, $\bf \overline{15}$ and $\bf {6}$ representations of $SL(6,\mathbb{R})$. These collect up into the
$\bf \overline{27}$ of $E_6$, i.e. $Z^N$. For one form elements one finds $Z^{ai}$, $Z^{a i_i\ldots i_4}$,
$Z^{a i_i\ldots i_6,j}$ which belong to the $\bf \overline{6}$, $\bf 15$ and $\bf \overline{6}$
representations of $SL(6,\mathbb{R})$ which collect up into the $\bf \overline{27}$ representation of $E_6$
i.e $Z^a_N$. For the two form  we find
  $$
  Z^{ab} ({\bf 1}), Z^{ab i_1\ldots i_3} ({\bf 20}), Z^{ab i_1\ldots i_5,j} ({\bf35\oplus 1}), Z^{ab i_1\ldots
  i_6} ({\bf 1}), Z^{ab, i_1\ldots i_3} ({\bf 20}), Z^{ab i_1\ldots i_6, j_1\ldots j_6} ({\bf 1})
  \eqno(5.1.4)
  $$
where the number in brackets is the $SL(6,\mathbb{R})$ representation. All these package up into the ${\bf 78
\oplus 1}$ of $E_6$, {\it i.e.} $Z^{a_1a_2\alpha}$ and $Z^{a_1 a_2}$. The latter charge is the Taub-Nut
charge and will play no role in what follows. As such we set it to zero.

This demonstrates how the space-time
generators $P_i$ which occur in the dimensional reduction of conventional supergravity theories are augmented
by the higher members of the $l$ multiplet to form $E_6$ multiplets. In what follows we will see how part of
these multiplets play a crucial role in the construction of the gauged supergravity theories.

We now define in more detail what we mean by the semi-direct product $E_{11}\otimes_s l$ where $l$ is an
algebra whose generators are in one to one relation with the $l$ multiplet.  By definition the commutation
relations  between  the generators $R$ of $E_{11}$ and $Z$ of $l$  are specified by to be of the form
  $$ [R,Z]=U(R) Z
  \eqno(5.1.5)
  $$
where $U(R)$ is the action of the generator $R$ on the generators $Z$ viewed as a representation of $E_{11}$.
Applying this to the $E_6$ sub-algebra we find the commutators
  $$
  [R^\alpha , P_a]=0,\ [R^\alpha , Z^M]= Z^N (D^\alpha)_N{}^M,\ [R^\alpha , Z^a_N]= - (D^\alpha)_N{}^M Z^a_M
  \eqno(5.1.6)$$
  $$
  [R^\alpha , Z^{a_1a_2\alpha}]=f^{\alpha\beta}{}_\gamma Z^{a_1a_2\gamma},\ [R^\alpha , Z^{a_1a_2a_3}_{NM}]=
  -(D^\alpha)_N{}^R Z^{a_1a_2a_3}_{RM} -(D^\alpha)_M{}^R Z^{a_1a_2a_3}_{NR} \ . \eqno(5.1.7)
  $$
One can readily verify that these commutators do satisfy the Jacobi identities found by taking the commutator
with another generator of $E_6$.
\par
The commutators of $E_{11}$ with the space-time translations can only be of the form
$$
[R^{aN}, P_b]=\delta_b^a Z^N,\ [R^{a_1a_2}_{N}, P_b]=2 \delta_{b}^{[a_1}Z^{a_2]}_N,\ \eqno(5.1.8)$$
$$
[R^{a_1a_2a_3\alpha}, P_b]=3 \delta_{b}^{[a_1}Z^{a_2a_3]\alpha},\  [R^{a_1a_2a_3a_4}_{MN}, P_b]=4
\delta_{b}^{[a_1}Z^{a_2a_3a_4]}_{MN} \ . \eqno(5.1.9)$$ The coefficients on the right-hand side are chosen as
above and this fixes the normalisation of the generators that appear on this side of the commutation
relations.
\par
Since all the elements of the $l$ representation can be obtained by taking the commutators of the  $E_{11}$
generators with  $P_a$, the commutators of the remaining generators of $E_{11}$ with those of the $l$
representation can  found by using the Jacobi identities in conjunction with equations (5.1.6), (5.1.7),
(5.1.8) and (5.1.9) as well as $E_{11}$ commutators themselves of equations (2.22), (2.23) and (2.25)-(2.31).
In particular, the Jacobi identity involving $P_a$, $R^{bM}$ and $R^{cN}$  implies the relation
  $$
  [R^{aM}, Z^N]= -d^{MNP} Z^a_P \quad .\eqno(5.1.10)
  $$
Similarly one finds that
$$
[R^{aM}, Z^b_N]= -(D_\alpha)_N{}^M Z^{ab\alpha},\ [R^{a_1a_2}_{M}, Z^N]= -(D_\alpha)_M{}^N Z^{a_1a_2\alpha},\
\eqno(5.1.11)$$
$$
[R^{a_1a_2}_{M}, Z^{a_3}_{N}]=Z^{a_1a_2a_3}_{MN},\ [R^{a_1M}, Z^{a_2a_3\alpha}]=-S^{\alpha M [RS]}
Z^{a_1a_2a_3}_{RS},\ \eqno(5.1.12)$$
$$
[R^{a_1a_2a_3\alpha}, Z^{M}]=-S^{\alpha M [ RS]} Z^{a_1a_2a_3}_{RS} \quad . \eqno(5.1.13)$$

\subsection{The map from  $E_{11}$ into generalised spacetime}
Essential for the construction of the dynamics of the gauged supergravities is the observation that  there
generically exists  a linear map denoted $\Psi$ from $E_{11}$ into the $l$ representation which possesses the
following four  properties (we will give the discussion such that it is valid in any dimension before
implementing it in detail for the five dimensional case):
\begin{description}
\item[A] $ \  $ Let us  denote the image of this map to be $k$, i.e. $\Psi (E_{11})=k$. As $k$ is
part of the representation
$l$ of $E_{11}$ it inherits an action of $E_{11}$ on it. While this will not always act on elements of $k$ so
as to remain in $k$  we demand that the subspace $k$ does carry the adjoint representation of a sub-algebra
$F_{11}$ of $E_{11}$.

\item[B] $ \ $ We demand that the map $\Psi$ be invariant under the action of
$F_{11}$, that is
$$
\Psi (U(T)R)= U(T)\Psi (R) \eqno(5.2.1)$$ where $U(T)$ is the action of the generator $T\in F_{11}$ on the
appropriate space and $R$ is any generator of $E_{11}$.

\item[C] $ \ $  The map $\Psi$ preserves the space-time
nature of the fields, that is the action of the map does not change the number of Lorentz indices the element
carries.

\item [D] $ \ $ The sub-algebra $F_{11}$ is contained in the
Borel sub-algebra of $E_{11}$ together with all of $G$, the internal symmetry algebra which is  $E_{6}$ for
the case of five dimensions.
\end{description}
We now analyse the consequences of these requirements. Let us label the generic elements of $k$ and $F_{11}$
by $V$ and $T$ respectively. Since the adjoint representation of any group is  unique it follows from {\bf A}
that the map $\Psi $ identifies the subspace $k$ of $l$  in a one to one manner  with the  sub-algebra
$F_{11}$ of $E_{11}$ in a way that is preserved by (requirement {\bf B}) the action of $F_{11}$.   To be more
precise given any $T_1\in F_{11}$ it acts on any $T\in F_{11}$ according to the adjoint representation as
$T_1\to [T_1,T]$ while on $k$ the element $T_1$ acts as  $V\to U(T_1)V$.  Given a labeling  of the elements
of $F_{11}$ we may use the correspondence that $\Psi$ provides to similarly label the elements of  $k$.
Indeed, we have a one to one correspondence between $V\in k$ and $T\in F_{11}$ given by $\Psi(T)=V$ such that
$$
U(T_1)V=U(T_1)\Psi(T)=\Psi (U(T_1) T)=\Psi ([T_1,T]) \eqno(5.2.2)$$ for any  $T_1\in F_{11}$. It follows that
the map $\Psi$ induces a map, denoted $\tilde \Psi$,  of $E_{11}$ into itself whose image is the sub-algebra
$F_{11}$ on which it is the identity map.

If we decompose the adjoint representation of $E_{11}$ into representations  of $F_{11}$ then the map $\Psi$
identifies the sub-algebra $F_{11}$ with the subspace $k$ of $l$ as described above and maps to zero all the
other representations in $E_{11}$. Similarly if we decompose the representation $l$ of $E_{11}$ into
representations of $F_{11}$ then only the adjoint representation of $F_{11}$ is in the image of $\Psi$ and
all the other representations are in the complement of $k$.  We may write
  $$
  E_{11}=F_{11}\oplus F_{11}^\perp \ \ {\rm and} \ \ l=k\oplus k^\perp \eqno(5.2.3)
  $$
where $ F_{11}^\perp$ contains all the representations of $F_{11}$ contained in $E_{11}$ other than the
adjoint and similarly for $k^\perp$. Then $\Psi$   maps as $\Psi(F_{11})=k$ and $\Psi(F_{11}^\perp)=0$. We
will label the generic elements of $E_{11}$ and $L$ as $R$ and $l$, those of $F_{11}$ and $k$ as $T$ and $V$,
as was done above, and those of $F_{11}^\perp$ and $k^\perp $ as $S$ and $U$  respectively.  Clearly,
$F_{11}$ acts on $F_{11}^\perp$ to give $F_{11}^\perp$ and on $k^\perp$ to give $k^\perp$.  Also the action
of $ F_{11}^\perp$ on $ F_{11}$ and $k$ must contain all of $ F_{11}^\perp$ and $k^\perp$ respectively as
both the adjoint representation of $E_{11}$ and the representation $l$ are irreducible.

Requirement {\bf C} means that the map $\Psi$ preserves the sub-algebra of $E_{11}$  associated with gravity,
that is $A_{4}$ in the case of five dimensions, and so it maps a generator of $E_{11}$ with a given set of
space-time indices to an element of $k$ with the same set of space-time indices or if it is inside
$F_{11}^\perp$ to zero. As a result, it is useful to subdivide  all the above  spaces  according to the
number of space-time indices their elements possess and indicate this with a suitable superscript, for
example $E_{11}^{(0)}=E_6$, $l^{(0)}=\{ Z^N\}$ for five dimensions.

In this paper we will adopt requirement {\bf D}, but given that the map $\Psi$ can be  non-zero on parts of
the Borel sub-algebra of $E_6$ it natural to expect that $F_{11}$ could include other negative root
generators with non-trivial Lorentz indices.

Clearly, to find such a map one must find elements of $E_{11}$ and $l$ that have the same Lorentz index
structure.  Examining the formulation of $E_{11}$ suitable to eleven dimensions, that is with an $A_{10}$
sub-algebra, of equation (2.1) and that of the $l$ mentioned at the beginning of this section we find that at
low levels there are no such objects and so no map $\Psi$ is possible. However, once one considers lower
dimensions one finds that there are matching elements in $E_{11}$ and $l$ and that  a map with all the above
properties can be constructed. We now concentrate on the case of five dimensions, but it is straightforward
to generalise these considerations to other dimensions.

We will now construct such a map from  the $E_{11}$  generators of equation (2.21) into  the elements of $l$
of equation (5.1.2).  Using requirement {\bf C} and the fact that there is only one object in $l$ with any
space-time indices that are lowered, namely $P_a$, but  no such objects in the Borel sub-algebra of $E_{11}$,
it follows  that $P_a$ must  be in $k^\perp$. Hence, we may write $k^{(-1)\perp}= \{P_a\}$.  For the elements
with no space-time indices  we have a map from $E_{11}^{(0)}=E_6$ to $l^{(0)}=\{Z^N\}$, the $\bf
\overline{27}$ representation of $E_6$, and we define the elements of $F_{11}^{0}$ and $k^{(0)}$ to be given
by
$$
F_{11}^{(0)}=\{T^N: T^N=\Theta^N_\alpha R^\alpha\},\ \  k^{(0)}=\{V^N\} \eqno(5.2.4)$$ respectively, the
elements $T^N$ and  $V^N$  of the two subspaces being in the one to one correspondence
  $$
  \Psi (\Theta^N_\alpha R^\alpha)=V^N \quad . \eqno(5.2.5)
  $$
Here $\Theta^N_\alpha$  is a constant tensor which  enters the theory as the definition of the map $\Psi $ on
the space of elements with no space-time indices, {\it i.e.} $F_{11}^{(0)}$. In fact, the $T^N$ cannot be
linearly independent as this would imply that the $V^N$ were also linearly independent and so would span all
of $l^{(0)}$, it rather describes the way $F_{11}^{(0)}$ is embedded in $E_6$. The complement is  given by
  $$
  F_{11}^{(0)\perp}= \{S \in E_{11}^{(0)}:\ (S,\Theta^N_\alpha R^\alpha)=0\} \eqno(5.2.6)
  $$
If we write $S=c_\alpha R^\alpha$  the orthogonality conditions  it implies that $\Theta^N_\alpha c^\alpha=0$
where $c^\alpha=g^{\alpha\beta}c_\beta$, and $g^{\alpha\beta}$ is the Cartan-Killing metric.

We can now find the restrictions placed on $\Theta^M_\alpha$ by the above requirements. Taking the commutator
of two elements of $F_{11}^{(0)}$, namely $T^M=\Theta^M_\alpha R^\alpha$ and $T^N=\Theta^N_\alpha R^\alpha$
and demanding that $F_{11}^{(0)}$ is a sub-algebra (requirement {\bf A}) with structure constants
$f^{MN}{}_P$, implies that
$$
[T^M, T^N]= [\Theta ^M_\alpha R^\alpha, \Theta ^N_\beta R^\beta] =\Theta ^M_\alpha \Theta ^N_\beta
f^{\alpha\beta}{}_\gamma R^\gamma=  f^{MN}{}_P T^P=f^{MN}{}_P  \Theta ^P_\gamma R^\gamma \eqno(5.2.7)$$ and
so we conclude that
$$
\Theta ^M_\alpha \Theta ^N_\beta f^{\alpha\beta}{}_\gamma= f^{MN}{}_P  \Theta ^N_\gamma \quad .
\eqno(5.2.8)$$
\par
On the other hand demanding that the map $\Psi $ is invariant under $F_{11}^{(0)}$ transformations when
acting on $F_{11}^{(0)}$ (requirement {\bf B})  we find, using equations (5.1.6), (5.2.2)  and (5.2.7),  that
$$
\Psi (U(T^M)T^N)=\Psi ([T^M,T^N])=\Psi (f^{MN}{}_PT^P)
$$
$$
=U(T^M)\Psi(T^N)= U(T^M )V^N=\Theta^M_\alpha V^P(D^\alpha)_P{}^N =X^{MN}_P \Psi(T^P) \eqno(5.2.9)
$$
where we  recall that by  definition $X^{MN}_P =\Theta^M_\alpha(D^\alpha)_P{}^N$.  Hence, we find that
$$
f^{MN}{}_P\Theta^P_\gamma= X^{MN}_P\Theta^P_\gamma \eqno(5.2.10)$$ In deriving this equation we have taken
$V^N$ to transform under $F^{(0)}_{11}$ like $Z^N$. This is because $V^N$ can be obtained from the $Z^N$'s by
a projection in which the leading term is $Z^N$.

Let us now turn to the construction of the map $\Psi$ on elements with one upper space-time index, that is
the map $\Psi$  from $E_{11}^{(1)}=\{R^{aN}\}$ into the space $l^{(1)}=\{Z^a_N\}$. It maps the   $\bf
\overline{27}$ into the $\bf 27$ representation of $E_6$. We define this map by requiring the elements of
$F_{11}^{(1)}$ and $k^{(1)}$ to be given by
$$
F_{11}^{(1)}=\{T_M^a: T_M^a=W_{MN}R^{aN}\},\ \ k^{(1)}=\{V_M^a\} \eqno(5.2.11)$$ where the elements  are in
one to one correspondence
  $$
  \Psi ( W_{MN}R^{aN})=V^a_M \quad .\eqno(5.2.12)
  $$
The constant tensor $W_{NM}$ which defines the map is required to be an anti-symmetric tensor. We find the
complement of $F_{11}^{(1)}$  to be
  $$
  F_{11}^{(1)\perp}=\{S^{aN}\in E_{11}^{(1)}:\ W_{MN}S^{aN}=0\} \eqno(5.2.13)
  $$
where $S^{aN}= L^N{}_P R^{aP}$, with constant $L^N{}_P$'s, are a set of generators that are not linearly
independent. They are in the orthogonal subspace in the sense that $(T,S)=\sum_N T_N^a S^{bN}=0$ and
$L^N{}_P$ can be viewed as projectors.

We can find  the spaces $k^{(0)}$ and $k^{(0)\perp}$ by acting with $F_{11}^{(1)}$ on $P_a$ the lowest
component of the $l$ multiplet. Using equation (5.1.8), we note   that for $T_M^a\in F_{11}^{(1)}$ we find
that
$$
[T_M^a,P_b]=[W_{MN}R^{aN},P_b]=\delta_b^a W_{MN}Z^{N} \eqno(5.2.14)$$ while if $S^{aN}\in F_{11}^{(1)\perp}$
then
$$
[S^{a N}, P_b]= \delta_b^a V^N, \eqno(5.2.15)$$ We note that   $W_{MN}V^{ N}=0$  since   $W_{MN}S^{a N}=0$.
Since the action of $F_{11}$ on $k^\perp$ must lie in $k^\perp$ and the action of $F_{11}^{(1)}$ and
$F_{11}^{(1)\perp}$ on the $P_a$ must lead to all of $l^{(0)}$, we find  that
$$
k^{(0)}=\{V^N\in l^{(0)}:\ \ W_{MN}V^N=0\} \ \ {\rm and }\ \ k^{(0)\perp}=\{U_M:\ U_M=W_{MN}Z^N\} \quad .
\eqno(5.2.16)$$

Examining equation (5.2.5) and using the relation  $W_{MN}V^N=0$ we conclude that $\Psi
(W_{MN}\theta^N_\alpha R^\alpha)=0$ and so
$$
W_{MN}\theta^N_\alpha=0\quad .  \eqno(5.2.17)$$

Taking the commutator of an  element of $F_{11}^{(0)}$, namely $T^N=\Theta^N_\alpha R^\alpha$ and an element
of $F_{11}^{(1)}$, namely $T^a_M=W_{MP} R^{aP}$,   and demanding that they form a closed algebra (requirement
{\bf A}) we find that
$$
[T^N, T^a_M]= \Theta ^N_\alpha W_{MP} [R^\alpha, R^{aP}] =\Theta ^N_\alpha W_{MP}(D^\alpha)_S{}^P R^{aS}
=-X^{NP}_M T^a_P \eqno(5.2.18)$$ provided
$$
X^{NP}_{[S} W_{M]P}=0 \quad .\eqno(5.2.19)$$

Since the algebra $F_{11}$ is generated by $F_{11}^{(0)}$ and $F_{11}^{(1)}$ and we have already specified
the map for these two spaces, $F_{11}$ is completely determined once one takes into account the requirements
{\bf A}, {\bf B}, {\bf C}  and {\bf D}. It only remains to find the consequences for the constant tensors
$\Theta^N_\alpha$ and $W_{MN}$ and the form of the spaces $F_{11}$ and $k$ for the higher rank generators.
Indeed,  at the next level we find, using equation (5.2.17),  that the commutator of two elements of
$F_{11}^{(1)}$ is given by
  $$
  [T^a_N,T^b_M]=W_{NP}W_{MQ}d^{NPQ}R_Q^{ab}=-W_{NP}X^{(NQ)}_M R_Q^{ab} =-{3\over 2} W_{NP} (D^\alpha)_M{}^P
  T^{ab}_\alpha \eqno(5.2.20)
  $$
where
  $$
  T^{ab}_\alpha={1\over 3}\Theta ^S_\alpha R^{ab}_S
  \eqno(5.2.21)$$
and  provided
  $$
  -W_{NP}d^{PQS}=X_N^{(QS)} \quad .\eqno(5.2.22)
  $$
Taking another commutator with an element of $F_{11}^{(1)}$ one finds
  $$
  [T^a_N,T^{bc}_\alpha ]={1\over 3}W_{NP}\Theta_\alpha^S (D^\beta)_S{}^P R^{abc}_\beta ={2\over 3}
  \Theta_\alpha^S T_{NS}^{abc} \eqno(5.2.23)
  $$
where
  $$
  T_{NM}^{a_1a_2a_3}= W_{[N|P}(D^\alpha)_{M ]}{}^P R^{a_1a_2a_3}_{\alpha} \quad . \eqno(5.2.24)
  $$
Hence we conclude that
  $$
  F_{11}^{(2)}=\{T^{ab}_\alpha:T^{ab}_\alpha={1\over 3}\Theta ^S_\alpha R^{ab}_S\} \eqno(5.2.25)
  $$
while
  $$
  F_{11}^{(3)}=\{T_{NM}^{a_1a_2a_3}:T_{NM}^{a_1a_2a_3}= W_{[N|P}(D_\alpha)_{M ]}{}^P R^{a_1a_2a_3\alpha}\}
  \quad .
  \eqno(5.2.26)
  $$

It is straightforward to find the corresponding spaces in $k^\perp$. Taking the commutator of
$T^{ab}_\alpha$ of equation (5.2.21) with $P_c$, and  using equation (5.1.8), we find that
$$
[T^{ab}_\alpha, P_c]={2\over 3}\delta_c^{[a} S_\alpha^{b]}\ \ {\rm where }\ \ S_\alpha^{b}=\Theta_\alpha^N
Z_N^b \eqno(5.2.27)$$ while the commutator of   $T_{NM}^{a_1a_2a_3}$ with $P_c$ gives
$$
[T_{NM}^{a_1a_2a_3},P_c]= 3\delta _c^{[a_1} S_{NM}^{a_2a_3]}\ \ {\rm where }\ \ S_{NM}^{ab}=
W_{[N|P}(D^\alpha)_{M ]}{}^P Z^{a b}_\alpha \eqno(5.2.28)$$ As a result we conclude that
 $$
 k^{(1)\perp}=\{S_\alpha^{b}:S_\alpha^{b}=\Theta_\alpha^N Z_N^b\}\ \ {\rm and }\ \
 k^{(2)\perp}=\{S_{NM}^{ab}:S_{NM}^{ab}= W_{[N|P}(D^\alpha)_{M ]}{}^P Z^{a b}_\alpha \} \eqno(5.2.29)
 $$
It is instructive to carry out the $F_{11}$ transformations on $k^\perp$ and see how it transforms into
itself.

To ensure that the map $\Psi$ satisfies requirement {\bf B} is more involved. For example, requiring the
invariance of the map $\Psi$ under $F_{11}^{(1)}$ transformations acting on $F_{11}^{(0)}$ we also find,
using equations (5.1.10) and (5.2.18), that
  $$
  \Psi(U(T_N^a)T^M)=\Psi([T_N^a,T^M])=\Psi(X^{MP}_N T^a_P)$$ $$ =X^{MP}_N\Psi(T^a_P)=U(T_N^a)\Psi(T^N)=
  U(T_N^a) V^N \quad .\eqno(5.2.30)
  $$
However, to evaluate this last  equation requires us to know how $T_N^a$ acts on $k^{(0)}$. In principle we
know how to evaluate this as we know the action of $E_{11}$ on $l$, but to find a concrete expression
requires us to be able to project from $l^{(0)}$ to $k^{(0)}$ not only in principle, but in practice. It is
very likely that this will lead to the constraint of equation  (5.2.22).

The same pattern occurs at higher levels, and one can compute what the action of $F_{11}^{(0)}$ on $k$ is
using the invariance,  but to derive the required identity one requires a detailed knowledge of the
projector. It would be good to work this out in detail and also investigate precisely what kind of
sub-algebra $F_{11}$ is. In doing this one should recover all the higher identities on $\Theta^N_\alpha$ and
$W_{MN}$ in addition to the ones found above.

\subsection{Field transformations and the dynamics of gauged \\ supergravities}
In this section we will show how the non-linear realisation based on $E_{11}\otimes_s l$ does lead to the
precise dynamics of the gauged supergravities. As we will see an essential role  is played in this
calculation by the higher level coordinates contained in the $l$ representation.

In this  construction of the dynamics an important role is played by a sub-algebra formed from $E_{11}$ and
$l$. The map $\Psi$ described in the above provides an identification of a sub-algebra $F_{11}$ of $E_{11}$
with sub-space of the $l$ representation which we wrote as $\Psi(T)=V$. The sub-algebra of interest is found
by adding together the generators which are identified  by the map, that is we consider the combinations
$$
Y=V+gT,\ {\rm  explicity }\ \ Y^N=V^N+gT^N, \ Y_M^a=V_M^a+gT_M^a,\  T_\alpha^{a b}=V_\alpha^{a
b}+gT_\alpha^{a b}, \ldots \eqno(5.3.1)$$ where $g$ is a constant that will eventually become the coupling
constant associated with the gauged supergravity. The explicit expressions for the $T$'s  are given in
equations (5.2.4), (5.2.11), (5.2.21) and (5.2.24).

In order to  compute the commutators of the $Y$ generators we need those between the $T$ and $V$ generators.
According to equation (5.1.5), and using the invariance condition of $\Psi$ of equation (5.2.2), we find that
$$
[T_1, V_2]= U(T_1) V_2= U(T_1) \Psi(T_2)= \Psi (U(T_1) T_2)$$ $$= \Psi ([T_1,T_2])=\Psi(f_{12}{}^3T_3)=
f_{12}{}^3V_3 \eqno(5.3.2)$$ where $V_i=\Psi(T_i),\ i=1,2,3$ and $[T_1,T_2]=f_{12}{}^3T_3$. Using this
relations it is straightforward to calculate the commutators of two $Y$ generators. The result is
  $$
  [Y_1,Y_2]=[V_1,V_2]+g[T_1,V_2]-g[T_2,V_1]+g^2[T_1,T_2]
  $$
  $$=
  [V_1,V_2]+2gf_{12}{}^3V_3+g^2f_{12}{}^3T_3 =gf_{12}{}^3Y_3 \eqno(5.3.3)
  $$
{\it provided} that one assumes that
  $$
  [V_1,V_2]=-gf_{12}{}^3V_3 \quad . \eqno(5.3.4)
  $$
Thus the generators $Y$ also obey the algebra  $T_{11}$, but with the structure constants rescaled by $g$,
provided we assume the generators $V$ satisfy the same algebra, but with a structure constant rescaled by
$-g$. We will discuss the significance of this commutator of two $V$'s later. At the lowest level we find,
using equations (5.2.14) and (5.2.18), that
  $$
  [Y^N,Y^M]=g f^{NM}{}_PY^P,\ \ [Y^N,Y^a_M]=- g X^{NP}_M Y^a_P,\ldots \eqno(5.3.5)
  $$

Using equations (5.3.2) and (5.3.4) it is easy to also show that
$$
[V_1, Y_2]=0,\ \ [T_1, Y_2]=f_{12}{}^3 Y_3, \eqno(5.3.6)$$
\par
  Starting from  a general group element of  $E_{11}\otimes_s l$ we take
the local sub-algebra to be such that the group element can be brought into the form
$$
g=e^{z\cdot U}e^{y\cdot Y}e^{A(z)\cdot R} \eqno(5.3.7)$$ where we recall that $U\in k^\perp$, $Y=V+gT$, $V\in
k$, $T\in F_{11}$ and $R\in E_{11}$. More explicitly
$$
e^{z\cdot U}=e^{x^aP_a}e^{z^N S_N}e^{z_a^N S^a_N}\ldots \eqno(5.3.8)$$
  $$
  e^{y\cdot Y}= e^{y_N (V^N+gT^N)} e^{y^N_a(V^a_N+gT^a_N)}\ldots \eqno(5.3.9)
  $$
where the coordinates of the generalised space-time are denoted by $z=(x^a,z^N,\dots )$ and
$y=(y_N,y^N_a,\ldots)$. The only $y$ dependence of the group element $g$ is via the generators $Y$, that form
a closed algebra. This is essential for insuring that there is no $y$ dependence in the final equations. In
the expression $e^{A(z)\cdot R}$ the fields $A$ now depend  on the $z$ coordinates. The above form of the
group element differs from the most general one in that it involves only generators from the Borel
sub-algebra of $E_{11}$ and the fields in the last factor only depend on $z$ and not on both $z$ and $y$. As
a result the local transformations must involve those of the Cartan involution invariant sub-algebra as
usual, but in addition $y$ dependent Borel sub-algebra transformations.  We will discuss this later. In fact
we will only retain the $x^a$ coordinate from all the $z$ coordinates, but it is useful to retain the more
general expression for the day when we understand what to do with the higher $z$ coordinates. One can rewrite
the group element $g$ by moving the factors of $e^{g \ y\cdot T}$ in $Y$ through the group element so that
all the generators of $E_{11}$ appear in the order listed in $g$ before the deformation. That is in order of
generators of decreasing  rank. Once this has been done one can interpret the result as taking a fields $A$
to depend generally on $z$,  but in a special  way on $y$.

The Cartan forms are given by
$$
g^{-1}d g= dZ\cdot E\cdot L +dz\cdot G\cdot R+ dy\cdot G\cdot R \eqno(5.3.10)$$ where $L=(U,V)$ are all the
generators of $l$ and $Z=(z,y)$ are  the corresponding coordinates and
$$
dz\cdot G\cdot R=e^{-A(z)\cdot R}d e^{A(z)\cdot R},\ \ dy\cdot G\cdot R=  e^{-A(z)\cdot R}g dy\cdot e\cdot T
e^{A(z)\cdot R} \eqno(5.3.11)$$
$$
dZ\cdot E\cdot L=e^{-A(z)\cdot R}(e^{-y\cdot Y}dz\cdot Se^{y\cdot Y}+ dy\cdot e\cdot V)e^{A(z)\cdot R}\quad .
\eqno(5.3.12)$$ In deriving this equation we have used the fact that $e$ is the group vierbein corresponding
to the algebra $F_{11}$, namely
$$
e^{-y\cdot Y} d e^{y\cdot Y}= dy\cdot e\cdot Y \quad . \eqno(5.3.13)$$

The quantity $dz\cdot G\cdot R$ is just the usual expression for the Cartan forms in the absence of a
deformation. The first such forms were given in equation (2.47) if we replace the $x$ dependence by that of
$z$. The quantity $dy\cdot G\cdot R$ are the $E_{11}$ valued Cartan forms which are in the $y$ direction.
Examining the above expression we find it is of the form
$$
dz\cdot G\cdot R+dy\cdot G\cdot R= (dz,dy) \cdot \left( \begin{array}{cc}
I & 0 \\
0 & e \end{array} \right) \  {\cal G}\cdot R \eqno(5.3.14)$$ where
$$
dy\cdot {\cal G}\cdot R= e^{-A(z)\cdot R}dy\cdot T e^{A(z)\cdot R} \eqno(5.3.15)$$ is independent of $y$. It
is straightforward to compute the first few  Cartan forms using the $E_{11}$ commutation relations given
earlier in the paper. We find that
  $$
  {\cal G}^N{}_{,\alpha}=g\Theta_\alpha^N, $$
  $$ {\cal G}^N{}_{,aM}=gA_{aP}X^{NP}_M,\ \ {\cal G}^N{}_{,a_1a_2
  }^{\ M}=gA_{a_1a_2}^{ P}X^{NM}_p-{g\over 2}A_{[a_1 |Q}A_{a_2] R}X_S^{NR}d^{SQM},\ldots \eqno(5.3.16)
  $$
while
  $$
  {\cal G}_N^b{}_{,\alpha}=0,\ \ {\cal G}_N^b{}_{,aM}=g\delta_a^b W_{NM},\ \ {\cal
  G}_N^b{}_{,a_1a_2}^R=g\delta_{[a_1}^b A_{a_2] P}W_{NM} d^{MPR} \quad . \eqno(5.3.17)$$

The vierbein $E$  of the non-linear realisation is the coefficient of the $l$ generators in the Cartan form
of equation (5.3.10) and it is of the form
$$
dZ\cdot E=dZ\cdot \left( \begin{array}{cc}
I & 0 \\
0 & e \end{array} \right) \  {\cal E} \eqno(5.3.18)$$ where
$$
dZ\cdot {\cal E}\cdot L= e^{-A(z)\cdot R} dZ\cdot L e^{A(z)\cdot R} \eqno(5.3.19)$$ where $dZ=(dz,dy)$. We
observe that  {\cal E} is independent of $y$.  We observe that the inverse quantity is given by
$$
dZ\cdot {\cal E}^{-1}\cdot L= e^{A(z)\cdot R} dZ\cdot Le^{-A(z)\cdot R} \eqno(5.3.20)$$
\par
The first few inverse vierbein components  are readily calculated using equation (5.3.20) and the commutators
of equations (5.1.8-13) and are  given by
$$
{\cal E}^{-1}{}_a{}^\mu=\delta_a^\mu,\ \  {\cal E}^{-1}{}_{a}{}_N=A_{aN},\ \ {\cal E}^{-1}{}_a{}_b^N=2
A_{ab}^N-{1\over 2}A_{[a|M}A_{b]S}d^{SMN},\ldots \eqno(5.3.21)$$ while
$$
{\cal E}^{-1}{}^N{}_M=\delta_M^N, \ \ {\cal E}^{-1}{}^N{}=...\ldots , {\cal
E}^{-1}{}^a_N{}_b^M=\delta_b^a\delta_N^M \quad . \eqno(5.3.22)$$

The above expressions for the Cartan forms and vierbeins omit the fact that they appear multiplied by
$g_{\phi}^{-1}$ on the left and $g_{\phi}$ on the right. The effect of this is just to find the above
quantities but multiplied by factors of $V_{Mij}$ and $\tilde{V}^M_{ij}$ as described below eq. (2.47).

The Cartan form $g^{-1}d g$ is invariant under rigid transformations $g_0$ of $E_{11}\otimes_s l$ which are
of the form $g\to  g_0 g$, but does transform under local transformations, $g\to  g h$ where $h$ is in the
local subgroup as $g^{-1}d g\to h^{-1} g^{-1}d g h+h^{-1}d h$. As such, the only global transformations on
$G$ arise from the global transformations on $dZ$ and, as the full Cartan form is invariant, the
corresponding transformations induced on the first index on $G$.  As such this first index  is a world index
in the sense that it transforms under coordinate transformations induced by the global transformations. To
construct the dynamics one normally makes this index flat using the inverse vierbein $E^{-1}$ of the
non-linear realisation and then the resulting object $\hat G$ is inert under the global transformations and
just transforms under the local sub-algebra. By definition the dynamics is the set of equations which is
invariant under the rigid $g\to  g_0 g$ and the local $g\to  g h$ transformations. Hence,  if we construct
the dynamics  only from $\hat G$ then we need only find equations invariant under the local transformations
as invariance under global transformations is automatic. We note that in our  case
$$
\hat G= E^{-1} G= {\cal E}^{-1}{\cal G} \eqno(5.3.23)$$ where the matrix $E^{-1}$ is understood to act on the
world index of $G$. Clearly, if the dynamics is constructed from the flat $\hat G$'s then it will be
independent of the $y$ coordinates. However, in the case of interest to us here, that is the dynamics of the
gauged supergravities, this is not quite the case as we will explain below. The flat  $\hat G$'s will require
some correction terms, nonetheless the dynamics will be independent of the $y$ coordinates  as one is adding
corrections to terms to the flat $\hat G$'s, which are $y$ independent, as a result of demanding invariance
under $y$ independent transformations. Consequently, although the $y$ coordinates play a key role in
formulating the dynamical equations they are not present in the final result.

Using the expression for the Cartan forms of equations (5.3.16-7) and the inverse vierbein of equations
(5.3.21-2) it is easy to evaluate the $\hat G$; the first one being given by
$$
\hat G_{a\alpha} R^\alpha= {\cal E}_{a}{}^\mu{\cal G}_{\mu\alpha} R^\alpha+ {\cal E}_{a}{}_N {\cal
G}_{N,\alpha} R^\alpha =g_\phi^{-1}(\delta _a^\mu \partial_\mu +g\Theta^N_\alpha R^\alpha A_{aN} )g_\phi
\quad . \eqno(5.3.24)$$ We recognise this expression as the Cartan form associated with the non-linear
realisation $E_6$ with $USp(8)$ local sub-algebra with  a term, proportional to  a deformation parameter $g$,
which describes the coupling of the scalars to the gauge fields $\Theta_\alpha^N A_{aN} R^\alpha$.
Consequently, we find that the gauge group of the non-linearly realised theory has the generators
$\Theta^N_\alpha R^\alpha=T^N$ which we recognise as those of the algebra $F_{11}^{(0)}$. Following the same
arguments it is straightforward to show that
$$
\hat G_{a,bM}=\delta_a^\mu \partial_\mu A_{bM}+g A_{aN}A_{bP}X^{NP}_M +2g A^N_{ab}-{1\over 2}gA_{aT}A_{bS}
d^{STN}W_{NM} \quad . \eqno(5.3.25)$$

We will now calculate the rigid transformations of the deformed theory by starting with  the group element
$g$  of $E_{11}\otimes_s l$ and carrying out the rigid transformations $g\to g_0g$ for $g_0\in
E_{11}\otimes_s l$.  We begin by considering transformations which belong to $k$. These    can be written as
$g_0=e^{b\cdot  V}$ and In carrying out this calculation we will encounter the expression $e^{b\cdot  V}
e^{y\cdot Y}$ which, using equations (5.3.6), we may process as
  $$
  e^{b\cdot  V} e^{y\cdot Y}=e^{b\cdot V+y\cdot Y}= e^{(y+b)\cdot  Y- gb\cdot T}= e^{y^\prime\cdot  V}
  e^{-gb\cdot T} \eqno(5.3.26)
  $$
where
  $$
  e^{y^\prime\cdot  V}= \prod _n e^{-{n\over (n+1)!} g(y\cdot Y)^n\wedge b\cdot Y} e^{(y+b)\cdot  Y}
  \quad . \eqno(5.3.27)
  $$
In carrying out this manoeuvre we have used the equation
  $$
  e^A e^B= \prod _n e^{-{n\over (n+1)!} A^n\wedge B}e^{A+B} \eqno(5.3.28)
  $$
valid for any $A$ and $B$, but only to lowest order in $B$. We recall that $A\wedge B= [A,B]$ and  $A^2\wedge
B= [A,[A,B]]$ etc. Carrying out a $k$ transformation in the non-linear realisation we find that
  $$
  g_0g=e^{b\cdot  V}e^{z\cdot U}e^{y\cdot Y}e^{A(z)\cdot R} =e^{z\cdot U}e^{y^\prime\cdot Y}e^{-gb\cdot
  T}e^{A(z)\cdot R} \quad . \eqno(5.3.29)
  $$
Thus the net effect of a rigid $k$ transformation is to change $y$, and to lead to  the $E_{11}$
transformation $e^{-gb\cdot T}$ on the $E_{11}$ fields. However, as  the dynamics is independent of $y$ we
need only work out the consequences of the latter transformation in eq. (5.3.29). We have assumed that
passing $e^{b\cdot  V}$ through $e^{z\cdot U}$ leads only to changes in $z$ and $y$ which are irrelevant. At
the lowest level we find that taking $g_0=e^{b_N V^N}$ induces the $E_{11}$ transformation $e^{-gb_N T^N}=
e^{- g b_N \Theta^N_\alpha R^\alpha}$,  while taking $g_0=e^{b_a^N V^a_N}$ induces the $E_{11}$
transformation $e^{-gb_a^N T^a_N}= e^{-gb_a^N W_{NM} R^{aM}}$,  taking $e^{b_{a_1a_2}^\alpha
V^{a_1a_2}_\alpha}$ induces the $E_{11}$ transformation $e^{-gb_{a_1a_2}^\alpha T^{a_1a_2}_\alpha} =
e^{-{g\over 3} b_{a_1a_2}^\alpha \Theta^N_\alpha R^{a_1a_2}_N}$ and taking $e^{b_{a_1 a_2 a_3}^{MN} V^{a_1
a_2 a_3}_{MN}}$ induces the $E_{11}$ transformation $e^{-g b_{a_1 a_2 a_3}^{MN} T^{a_1 a_2 a_3}_{MN}}= e^{-g
b_{a_1 a_2 a_3}^{MN} D^\alpha_M{}^P W_{PN} R_\alpha^{a_1 a_2 a_3}  }$.

Using equations (2.8) and (2.9) we find that acting with $g_0=e^{b_N V^N}$ the fields transform as
  $$
  \delta A_{aN}= -g b_S X^{SM}_N A_{aM}, \ \ \ \delta A^N_{a_1a_2}= g b_S X^{SN}_M A^M_{a_1a_2}
  $$
  $$
  \delta A_{a_1a_2a_3}^\alpha = -gb_S\Theta^S_\beta f^{\alpha \beta}{}_\gamma A_{a_1a_2a_3}^\gamma \quad ,
  \eqno(5.3.30)$$
while if we take $g_0=e^{b_a^N V^a_N}$ this results in the transformations of the form of eq. (2.45) with
parameter $a_{aM}=-gb_a^N W_{NM}$. Similarly, acting with $g_0=e^{-gb_{a_1a_2}^\alpha V^{a_1 a_2}_\alpha}$
generates the transformations of the form of eq. (2.43) with parameter $a_{a_1a_2}^N=-{g\over 3}
b_{a_1a_2}^\alpha \Theta^N_\alpha $, and acting with $g_0 = e^{b_{a_1 a_2 a_3}^{MN} V^{a_1 a_2 a_3}_{MN}}$
generates the transformations of the form of eq. (2.41) with parameter $a_{a_1 a_2 a_3}^\alpha = - g b_{a_1
a_2 a_3}^{MN} D^\alpha_M{}^P W_{PN}$.

One can also carry out rigid $k^\perp$ transformations which is  of the form $g_0=e^{c\cdot U}$. Clearly,
taking  $g_0=e^{c^a P_a}$ will only result in the change $x^a\to x^a +c^a$, that is the usual space-time
translations. The higher generators of $k^\perp$ will lead to changes in $z$ and possibly $y$. However, the
coordinates $y$ do not appear in the dynamics and in this paper we will only take the lowest coordinate $x^a$
of the $z$'s. As such, these transformations are irrelevant for the terms computed in this paper.

Now let us carry out a rigid $E_{11}$ transformation of the form $g_0=e^{a\cdot R}$. This gives
  $$
  g_0g=e^{a\cdot R}e^{z\cdot U}e^{y\cdot Y}e^{A(z)\cdot R} =e^{(z\cdot U+ [a\cdot R, z\cdot U])}e^{(y\cdot Y+
  [a\cdot R,y\cdot Y])} e^{a\cdot R}e^{A(z)\cdot R} \quad . \eqno(5.3.31)
  $$
The final factor of $e^{a\cdot R}$ leads to the same rigid transformations on the $E_{11}$ fields that we
found in section 2 for the massless theory.  The commutator $[a\cdot R,y\cdot Y]$ leads to generators of $l$
and $E_{11}$. However, these results in either changes to $z$ and $y$ or additions to the $E_{11}$ fields
$A(z)$ that are $y$ dependent. Such latter terms  do not maintain the form of the group element which must be
brought back to the same form using local $y$ dependent transformations. For the reasons given above we can
in effect forget about these terms. On the other hand the commutator $[a\cdot R, z\cdot U]$ lies in $l$ and
so it contributes to changes in $z$ and $y$.  In the latter case we must rewrite the generators of $k$ in
terms of those of the generators $Y= V+gT$ to find the change in $y$'s and as a result we find additional
$E_{11}$ generators whose effect must be evaluated. Since we are only keeping the coordinates $x^a$ from all
the $z$ coordinates, we only have the factor
  $$
  e^{x^cP_c+ [a\cdot R, x^cP_c]} \quad . \eqno(5.3.32)$$

It is most easy to explain how to process this term  by studying the simplest case from which the general
procedure will become apparent. As such  taking   $g_0= e^{a_{aN}R^{aN}}$, we find that  the factor of
equation (5.3.32) is equal to
$$
e^{x^cP_c}e^{ x^c a_{cN}V^N} \eqno(5.3.33)$$ where we have thrown away the part of $Z^N$ that belongs to
$k^\perp$ and taken $[P_c , V^N ]=0$. The net result is a rigid  $k$ transformation with parameter $x^c
a_{cN}$. Following our discussion above for such transformations we find that acting with $g_0=
e^{a_{aN}R^{aN}}$ leads to a group element of the form
$$
e^{x^cP_c} e^{y^\prime\cdot Y} e^{-gx^c a_{cN}T^N}e^{a_{aN}R^{aN}} e^{A(x)\cdot R} \eqno(5.3.34)$$ which can
be evaluated using the $E_{11}$ commutators as we did for the massless theory. A similar calculation taking
$g_0=e^{a_{a_1a_2}^N R^{a_1a_2}_N}$, $g_0=e^{a_{a_1a_2a_3}^\alpha R^{a_1a_2a_3}_\alpha}$ and $g_0 = e^{a_{a_1
\dots a_4}^{MN} R^{a_1 \dots a_4}_{MN}}$ leads to effective $k$ transformations with $k$ parameters  $2x^c
a_{ca}^N$, $3x^c a_{ca_1a_2}^\alpha$ and $4 x^c a_{c a_1 \dots a_3}^{MN}$.

Examining equation (5.3.33)  we conclude that a rigid $E_{11}$ transformations results in the $x$ independent
transformations of the massless theory as well as $x$ dependent transformations that can be interpreted as
effective $k$ transformations. As such we can account for the latter transformations by replacing the $x$
independent parameters $b$ of the $k$ transformations by $b(x)$ where
  $$
  b_N(x) =b_N+ x^c a_{cN},\ \  b_a^N(x) =b_a^N+ 2x^c a_{ca}^N,$$
  $$ b_{a_1a_2}^\alpha(x) =b_{a_1a_2}^\alpha+ 3x^c
  a_{c a_1a_2}^\alpha , \ \ b_{a_1 a_2 a_3}^{MN}(x) = b_{a_1 a_2 a_3}^{MN}+ 4 x^c a_{c a_1 \dots a_3}^{MN}
  \dots \eqno(5.3.35)$$
Thus the rigid transformations of $E_{11}\otimes_s l$ lead to the same rigid transformations of the massless
theory as well as $k$ transformations that have the $x$ dependent parameters of equation (5.3.35).

The resulting  $E_{11}\otimes_s l$  transformations of the the $E_{11}$ fields are given by
$$
\delta A_{aN}= \partial_a b_N(x)-g b_S(x) X^{SM}_N A_{aM}+gW_{MP} b^P_a(x) \ , \eqno(5.3.36)$$
$$
\delta A^N_{a_1a_2}= {1\over 2} \partial_{[a_1}b^N_{a_2]} (x) +{1\over 2} \partial_{[a_1}b_S (x) A_{a_2 ]T}
d^{STN} +gb_S(x) X^{SN}_M A_{a_1a_2}^M
$$
$$+{1\over 2}W_{SP}b^P_{[a_1}(x) A_{a_2
]T}d^{STN} -{1\over 3} g b_{a_1a_2}^\alpha(x) \Theta _\alpha^N  \ , \eqno(5.3.37)$$
  $$
  \delta A_{a_1a_2a_3}^\alpha= {1 \over 3} \partial_{[a_1} b_{a_2 a_3 ]}^\alpha (x)+  \partial_{[a_1} b_M (x)
  A^{N}_{a_2 a_3 ]} D^\alpha_N{}^M
  $$
  $$+ {1 \over 6}
  \partial_{[a_1} b_M (x) A_{a_2 , N}  A_{a_3 ], P} d^{MNQ} D^\alpha_Q{}^P
  - g b_P (x) \Theta^P_\beta f^{\alpha\beta}{}_\gamma A^\gamma_{a_1 a_2 a_3}$$
  $$
  + g W_{MP} b_{[a_1}^P (x) D^\alpha_N{}^M A_{a_2 a_3 ]}^N
  + {1 \over 6} g W_{MR} b_{[a_1}^R (x)  d^{MNQ}
  D^\alpha_Q{}^P A_{a_2 , N} A_{a_3 ] , P} $$
  $$- g D^\alpha_M{}^P W_{PN} b_{a_1 a_2 a_3}^{MN} (x) \quad ,
  \eqno(5.3.38)
  $$
where we have used the identities
  $$
  \partial_a b_N(x)=a_{a N} \quad \quad {1\over 2} \partial_{[a_1}b_{a_2]} ^N(x)=a_{a_1 a_2}^N
  $$
  $$
  {1\over 3} \partial_{[a_1}b_{a_2a_3]} ^\alpha(x)=a_{a_1 a_2a_3} ^\alpha \quad \quad
  {1\over 4} \partial_{[a_1}b_{a_2a_3 a_4]}^{MN}(x)=a_{a_1 \dots a_4}^{MN} \eqno(5.3.39)
  $$
to rewrite the transformations that are the same as in  the massless theory.

The transformations of equations  (5.3.36) to (5.3.38) uniquely determine the corresponding invariant field
strengths as they are both only first order in derivatives. These are obtained adding the order $g$
corrections to the field strengths of eq. (\ref{2.55}) of the massless theory. The results is
  $$
  F_{a_1 a_2 ,M} = 2 \partial_{[a_1} A_{a_2 ] ,M} + g X^{[NP]}_M A_{[a_1 ,N} A_{a_2], P} - 4 g W_{MN}A_{a_1
  a_2}^N \eqno(5.3.40)
  $$
  $$
  F^{M}_{a_1 a_2 a_3} = 3\partial_{[a_1} A^{M}_{a_2 a_3 ]} + {3 \over 2} \partial_{[a_1} A_{a_2 ,N}
  A_{a_3 ] ,P} d^{MNP} - 6 g X^{(MN)}_P A_{[a_1 a_2}^P A_{a_3 ], N}
  $$
  $$
  +{1 \over 2} g X^{[NP]}_R d^{RQM} A_{[a_1 , N} A_{a_2 , P} A_{[a_3 ] , Q} + 3 g \Theta^M_\alpha A_{a_1 a_2
  a_3 }^\alpha \eqno(5.3.41)
  $$
  $$
  F^{\alpha}_{a_1 \dots  a_4} = 4 \partial_{[a_1} A^{\alpha}_{a_2 \dots a_4 ]} - {2 \over 3} \partial_{[a_1}
  A_{a_2 ,M} A_{a_3 ,N} A_{a_4 ] ,P} d^{MNQ} D^\alpha_Q{}^P - 4\partial_{[a_1} A^{M}_{a_2
  a_3} A_{a_4 ],N} D^\alpha_M{}^N
  $$
  $$
  + 4 g D^\alpha_M{}^P \Theta^M_\beta A_{[a_1 , P} A^\beta_{a_2 \dots a_4 ]} + 16 g  D^\alpha_M{}^P  W_{PN}
  A^{MN}_{a_1 \dots a_4} -4 g  D^\alpha_M{}^P W_{PN} A^M_{[a_1 a_2} A^N_{a_3 a_4 ]}
  $$
  $$
  -4 g  D^\alpha_M{}^P X^{(MR)}_Q A_{[a_1 , P} A_{a_2 , R} A_{a_3 a_4 ]}^Q - {1 \over 6} g X^{[MN]}_R d^{RPS}
  D^\alpha_S{}^Q A_{[a_1 , M} A_{a_2 , N} A_{a_3 , P} A_{a_4 ] , Q}  \quad . \eqno(5.3.42)
  $$

Requiring the closure of $E_{11}$ with the conformal group has the net effect of promoting the parameters
$b(x)$ to be arbitrary functions of $x$. Given that $b(x)$ contain the term $x \cdot a $ in eq. (5.3.35), the
identification $a = d \Lambda$ in eq. (\ref{2.56}) gives the normalisation of $b(x)$ in terms of $\Lambda$ as
  $$
  \Lambda_M = b_M (x) \quad \Lambda_a^M = {1 \over 4} b_a^M (x) \quad \Lambda_{a_1 a_2}^\alpha = {1 \over
  9} b_{a_1 a_2}^\alpha (x) \quad \Lambda_{a_1 a_2 a_3}^{MN} = {1 \over 16} b_{a_1 a_2 a_3}^{MN} (x)\quad .
  \eqno(5.3.43)
  $$
Substituting this into the transformations of eqs. (5.3.36), (5.3.37) and (5.3.38) we find
  $$
  \delta A_{aN}= \partial_a \Lambda_N -g \Lambda_S X^{SM}_N A_{aM}+ 4 gW_{MP} \Lambda^P_a \ , \eqno(5.3.44)$$
  $$
  \delta A^N_{a_1a_2}= 2 \partial_{[a_1}\Lambda^N_{a_2]}  +{1\over 2} \partial_{[a_1} \Lambda_S  A_{a_2 ]T}
  d^{STN} +gb_S(x) X^{SN}_M A_{a_1a_2}^M
  $$
  $$+2W_{SP} \Lambda^P_{[a_1} A_{a_2
  ]T}d^{STN} - 3 g \Lambda_{a_1a_2}^\alpha \Theta _\alpha^N  \ , \eqno(5.3.45)$$
  $$
  \delta A_{a_1a_2a_3}^\alpha= 3 \partial_{[a_1} \Lambda_{a_2 a_3 ]}^\alpha +  \partial_{[a_1} \Lambda_M
  A^{N}_{a_2 a_3 ]} D^\alpha_N{}^M
  $$
  $$+ {1 \over 6}
  \partial_{[a_1} \Lambda_M  A_{a_2 , N}  A_{a_3 ], P} d^{MNQ} D^\alpha_Q{}^P
  - g \Lambda_P  \Theta^P_\beta f^{\alpha\beta}{}_\gamma A^\gamma_{a_1 a_2 a_3}$$
  $$
  + 4 g W_{MP} \Lambda_{[a_1}^P  D^\alpha_N{}^M A_{a_2 a_3 ]}^N
  + {2 \over 3} g W_{MR} \Lambda_{[a_1}^R   d^{MNQ}
  D^\alpha_Q{}^P A_{a_2 , N} A_{a_3 ] , P} $$
  $$- 16 g D^\alpha_M{}^P W_{PN} \Lambda_{a_1 a_2 a_3}^{MN}  \quad .
  \eqno(5.3.46)
  $$

We can now compare the field strengths and gauge transformations obtained here with those found from
supersymmetry in section 4. To do this, we carry out the field redefinitions and the corresponding
redefinitions of parameters of eqs. (3.28) and (3.29), both of which are determined completely from the
massless theory. We find complete agreement, and in particular one can check that the order $g$ terms in eqs.
(5.3.44-46) are identical to the order $g$ transformations found from supersymmetry and reformulated in terms
of the $E_{11}$ fields and parameters in eq. (4.29). The relation of eq. (3.29) between the gauge parameters
obtained from supersymmetry and the $E_{11}$ parameters $\Lambda$ has been carried out in such a way that the
variation of the field $A_n$ is in both cases of the form $\delta A_n = n \partial \Lambda_{n-1}$. This
ensures that the parameters are normalised in the required way. All the remaining coefficients in eqs.
(5.3.44-46) are then determined independently by both calculations, thus giving 12 independent checks.

As noted above, since the transformations from $k$ involve only part of the $l$ multiplet, the corresponding
generators satisfy constraints and as a result the associated parameters have an ambiguity. For example, as
$W_{MN}V^N=0$ the parameter $b_N$ is ambiguous up to $b_N\to b_N+W_{MN}c^N$ for any constants $c^N$.
Examining the transformations of equation (5.3.36) to (5.3.38) we indeed see that such ambiguities do not
affect the transformations of the fields as a result of  identities such as that of equation (5.2.13).

The unique equations which are invariant under the transformations of the non-linear realisation above and
are Lorentz and $USp(8)$ covariant are of the form of eq. (2.57). The result is
  $$
  V_{Mij} F^M_{abc} = {1 \over 8} \epsilon_{abcde} \tilde{V}^M_{ij} F^{de}_M \qquad
  V_{Mij} \tilde{V}^N_{kl} F^\alpha_{abcd} = {1 \over 72}
  D^\alpha_M{}^N \epsilon_{abcde} (g_\phi^{-1} \partial^e g_\phi )_{ijkl} \quad ,\eqno(5.3.47)
  $$
which is the same as eqs. (3.14) and (3.22). The non-linear realisation also possesses local transformations
associated with the Cartan involution invariant subalgebra. The transformations above, which determine the
field strengths, arise from the Borel subalgebra of $E_{11}$ with the exception of the local $USp(8)$. We
believe that also requiring invariance under the local transformations will fix uniquely the duality
relations above, including the coefficients in eq. (5.3.47).

We conjecture that the duality relation between the field strength of the 4-forms and the mass deformation
parameters arise from equating the Cartan form in the $x^a$ direction proportional to $R^{a_1 \dots
a_4}_{MN}$ and the Cartan form $G^b_{M,aN} = g W_{MN} \delta^b_a$ in equation (5.3.17), in the $y^M_b$
direction and proportional to $R^{a , N}$. This leads to the duality relation
  $$
  V_{M ij} V_{N kl} F_{abcde}^{MN} = {1 \over 5760}\epsilon_{abcde} [ \tilde{V}^M_{ij} \tilde{V}^N_{kl}
  - 2 \tilde{V}^M_{[i \vert m} \tilde{V}^{Nm}{}_{[k} \Omega_{j]l]}]   G^a_{M,aN}
  $$
  $$
  = {1 \over 1152} g \epsilon_{abcde} W_{MN} [ \tilde{V}^M_{ij} \tilde{V}^N_{kl} - 2 \tilde{V}^M_{[i \vert m}
  \tilde{V}^{Nm}{}_{[k} \Omega_{j]l]}]
  \quad , \eqno(5.3.48)
  $$
which is equal to the duality relation of eq. (4.28) that was obtained imposing the closure of the
supersymmetry algebra. The overall coefficient and the $USp(8)$ structure of the terms are fixed from eq.
(4.28), but requiring that the Cartan form transforms correctly under the full local subalgebra will fix this
duality relation uniquely and independently of the supersymmetry result. Thus we find that the non-linear
realisation of $E_{11}\otimes_s l$ does indeed correctly account for the dynamics of the gauged
supergravities. It is important to note that the $k$ part of the $l$ multiplet, that is some of the
generalised coordinates, play an essential role. The above calculations artificially truncated the remaining
part of the $l$ multiplet and it would be very interesting to find out what is the effect of these additional
coordinates. Some considerations on this can be found in reference \cite{37}.

In the usual non-linear realisations of $E_{11}\otimes_s l$ an extension to include the closure with the
conformal group has been used. This had the virtue of making local all the global transformation of $E_{11}$,
the rigid parameter being the part of the local gauge transformation that is linear in $x$. However, as
already mentioned, for the above case we found that the rigid parameters of $E_{11}$ combined with the $k$
transformations into a parameter which has a constant and a linear term in $x$. Combining with the conformal
group would add all the higher $x$ dependent terms of a completely local transformation.

Using  a non-linear realisation in which part of the $l$ multiplet of generators plays a non-trivial role
leads to additional rigid transformations corresponding to the part of the $l$ multiplet that is non-trivial,
i.e. $k$. As we have seen these combine with the induced $x$ dependent $E_{11}$ transformations to form a set
of  parameters that has a term which is constant and one that is linear in $x$. This parameter does not occur
in the massless theory where one only has the constant $E_{11}$ parameters which once one closes  with the
conformal group becomes replaced by  local gauge parameters which contain the constant $E_{11}$ parameter as
the term linear in $x$. The transformations of the fields then only contain the derivative of the gauge
parameter. The situation in the deformed theory is different in that the closure with the conformal group
will lead to local (gauge) parameters which have a constant part that contains the parameters of $k$
transformations and a part that is linear in $x$ which contains the $E_{11}$ parameters. However, the
transformations of the fields contain not only the derivative of the gauge parameter,  but also the gauge
parameter itself. Indeed the presence of these latter terms can be viewed as a consequence of the existence
of a non-trivial role for some of the generalised coordinates.

Comparing the field strengths of equations (5.3.44) with the ``flat'' Cartan forms of equation (5.3.25) we
see that they have the required from but the numerical coefficients are not quite the same. In fact the
expression of equation (5.3.25) is not quite invariant under the rigid transformations. This is in contrast
with the Cartan form of equation (5.3.24), which is invariant. The problem is that the form of the group
element, and so the Cartan forms, has been fixed using the local symmetries and having made a rigid
transformation one must make a compensating local transformation. However, such a local compensating
transformation does not leave the ``flat'' Cartan forms invariant and in particular this is the case for the
local transformations which are $y$ dependent. Taking this into account one recovers invariant expressions
that are in agreement with the ones mentioned above and derived by using the explicit transformations of the
fields. In the general procedure to find the dynamics from the non-linear realisation using the ``flat''
Cartan forms the final step is to find expressions which are invariant under the local transformations which
automatically include any local compensating transformations. As such following this procedure will also lead
to the same result as we found using the explicit variations of the fields.

Essential for the derivation of the dynamics of the gauged supergravities was the choice of group element of
equation (5.3.7). This can be obtained from the most general group element by taking a sufficiently large
local sub-algebra. The most general group element  differs from that of equation  (5.3.7) only in that the
fields $A$ are functions of $z$ and $y$ and not only $z$.  The local sub-algebra must  contain local
transformations that belong to the Cartan involution invariant subgroup of $E_{11}$ that depend in an
arbitrary way on  $z$ and $y$. However, we must also have local transformations that belong to the  Borel
sub-algebra of $E_{11}$ that depend on $z$ but not in an arbitrary way on $y$ so as to leave the group
element in the desired form. We note that for the case of no dimensional reduction we have a local
sub-algebra that has only the Cartan involution invariant subgroup of $E_{11}$ which depend in an arbitrary
way on  $z$ and $y$. While for the dimensional reduction on a torus we have $\Theta^N_\alpha=0$ and so there
are no $y$ coordinates. It would be good to understand in more detail the local subgroup and the precise way
in which it is local given that we have two sets of coordinates and so the meaning of local is more subtle
that the usual case.

We have made no attempt in this paper to discuss what happens to the dependence of the fields on the $z$
coordinates other than the very lowest one which is that of the usual description of space-time. As suggested
in \cite{37}, it could be  that these may lead to more propagating degrees of freedom.  This is perhaps the
most important unanswered question in the non-linear realisation of $E_{11}\otimes_s l$.

Another aspect of the above discussion that requires further thought is the commutator between the $V^N$'s of
equation (5.3.4). While the commutators of the generators of $E_{11}$ are unmodified regardless of what
theory one is discussing, the commutators of the  generators  of $k$ part of the generalised space-time
appear to change if one is discussing a gauged theory as opposed to a the massless theory and from one gauged
theory to another. To understand what is going on it is useful to consider gravity and its formulation as a
non-linear realisation. This is the non-linear realisation of $SL(D,\mathbb{R})$ closed with the conformal
group. Of course the resulting theory is Einstein's general relativity with a possible cosmological constant
and so has no preferred background. However, the intermediate step using first only $SL(D,\mathbb{R})$, or
alternatively the conformal group, is linked to Minkowski space, or equivalently the Poincare algebra, but
when the two are combined one has a background independent formulation. However, as the final result is
general relativity with a possible cosmological term it also possesses anti-de Sitter space as a solution.
Indeed, one can instead start with an anti-de Sitter algebra which has non-commuting space-time translations
to form Lorentz transformations rather than the commuting relations of the Poincare algebra and enlarging
this to include $SL(D,\mathbb{R})$ and closing with the conformal group.  One  finds that the vierbein
becomes redefined to incorporate that of anti-de Sitter space and Einstein's general relativity is again the
result. The isometries of anti-de Sitter space emerge from the formulation based on the Poincare group as the
space-time translations corrected by higher generators that enter when one considers the closure with the
conformal group. Similarly in the case being considered here although the algebra of the $k$ generators seems
to depend on the gauged supergravity being considered the result after one completes the non-linear
realisation will be equivalent in the same sense.

In reference \cite{29} a first adhoc attempt to account for the dynamics of gauged supergravities using a
non-linear realisation based on $E_{11}$ but also including the space-time translation operator $P_a$ was
given. The commutator of $P_a$ with generators of $E_{11}$ was taken to lead to another generator of $E_{11}$
and the Jacobi identities were used to find these commutators given the lowest one between $R^{aN}$ and $P_b$
which was given by $[R^{aN}, P_b]=\delta_b^a \Theta^N_\alpha R^\alpha$ where $\Theta^N_\alpha$ are constants.
As noted in that paper this was only correct when viewed form a suitable perspective. A similar adhoc
approach was taken when deriving the massive IIA supergravity theory as a non-linear realisation \cite{23}
and in this case one recovers the correct theory with all the required terms using this method.

However, when the generalised space-time was introduced in reference \cite{31} the commutator between
generators in $E_{11}$ and the $l$ multiplet was taken to be a member of the $l$ multiplet with a structure
constant that is determined  using the fact that the $l$ multiplet is a representation of $E_{11}$. This is
the case in the construction used in this paper {\it i.e.} equations (5.1.5)- (5.1.13). There is however a
relationship between the two approaches which is most easily seen by examining equation (5.2.15) which is a
commutator between an element of $F_{11}^{(1)\perp}$ and $P_a$ which results in an element of $k^{(0)}$. This
particular element is identified with the map $\Psi$ with the element $\Theta^N_\alpha R^\alpha$ of
$F_{11}^{(0)}$ which is the result in the alternative approach. This is indeed the general pattern and one
can recover the commutators of the adhoc approach from those of the correct approach of this paper in this
way. We note that   the relationship between the two approaches only applies to the commutator of the
generators of $F_{11}^{\perp}$ and not all those of $E_{11}$. Indeed, if one tries to use it  more generally
as was noticed in \cite{29} the Jacobi identities are not satisfied in the adhoc approach while they are
guaranteed in the approach of this paper. The Romans theory can also be constructed using the approach of
this paper and similar comments hold for this construction and the adhoc approach  of reference \cite{23}.

It was observed \cite{29} that one could find the relations satisfied by $\Theta^N_\alpha$ and $W_{MN}$ in
the adhoc approach by using the Jacobi identities on a suitable set of generators. This can be recovered from
the correct approach of this paper. The commutator of an element $S\in F_{11}^{\perp}$ with $U\in k^\perp$ is
an element  $V\in k$, {\it i.e.} generically $[S,U]=V$. As the map $\Psi$ is   a one to one map from
$F_{11}$ onto $k$,  $\Psi^{-1}$ is also a  one to one  onto map in the other direction. We note that
  $$
  [T, \Psi^{-1}([S,V])]=\Psi^{-1} ([T,[S,V] ) = -\Psi^{-1} ([ S,[ V, T]]  )-\Psi^{-1} ([V, [T,S]])
  \eqno(5.3.49)
  $$
using the fact that the map $\Psi$ is  invariant and so also is its inverse. Taking $T=\Theta^N_\alpha
R^\alpha$, $S=S^{aM}$ and $V=P_a$ we do indeed find the constraint of equation (5.2.10). Taking other choices
of generator one can find the other constraints. This is to be expected as one is using the invariance of
$\Psi$ which leads to the constraints in the method of this paper.

\section{An explicit example and the physical meaning of the map $\Psi$}
In order to make the constructions in this paper  more concrete we consider an explicit example that is well
known,  namely the gauged supergravity that arises from the IIB supergravity theories dimensionally reduced
on a five sphere with gauge group $SO(6)$. In doing so the physical meaning of the  the subspaces $F_{11}$,
$k$ and their complements will become readily apparent and seen to apply to any gauging.

We first recall the generators of $E_{11}$ labeled according to the preferred $SL(10,\mathbb{R})$ algebra
that leads to the IIB theory \cite{12,24};
$$
K^{\hat a}{}_{\hat b}, R^0, R^+, R^{\hat a_1\hat a_2\alpha}, R^{\hat a_1\dots \hat a_4}, R^{\hat a_1\dots
\hat a_6\alpha}, R^{\hat a_1\dots \hat a_7,\hat b}, R^{\hat a_1\dots \hat a_8 (\alpha\beta )}, R^{\hat
a_1\dots \hat a_{10}\alpha}, R^{\hat a_1\dots \hat a_{10}(\alpha\beta\gamma) },\ldots \eqno(6.1)$$ where $
\hat a, \hat b=1,\ldots ,10$ and  $R^-, R^0$ and  $R^+$ are the generators of the manifest $SL(2,\mathbb{R})$
of the IIB theory whose locally realised  $SO(2)$ subgroup is given by $R^+ - R^-$.  We denote the  indices
of the vector representation of $SL(2,\mathbb{R})$  by  $\alpha, \beta=1,2$.  The $l$ multiplet for the IIB
theory is given by \cite{39}
$$
P_{\hat a}, Z^{\hat a\alpha}, Z^{\hat a_1\dots \hat a_3}, Z^{\hat a_1\dots \hat a_5\alpha}, Z^{\hat a_1\dots
\hat a_6,\hat b},
  Z^{\hat a_1\dots \hat a_7},
Z^{\hat a_1\dots \hat a_7 (\alpha\beta)}, Z^{\hat a_1 \dots \hat a_9\alpha}, Z^{\hat a_1 \dots \hat
a_9(\alpha\beta\gamma) },\ldots \eqno(6.2)$$ We note that if we delete a space-time index from the generators
of $E_{11}$ of equation (6.1) we find those of the $l$ multiplet in equation (6.2) as expected.

It is instructive to first examine how the $E_{11}$ generators and members of the $l$ multiplet of the five
dimensional theory arise form these multiplets in  the IIB the dimensional theory given in equations (6.1)
and (6.2).  To find this we split the indices range of $\hat a$ etc  into $\hat a=a, a=1,\dots , 5$ and $\hat
a=i+5 ,a=6,\dots , 10$. The $i,j$ indices  transform under $SL(5,\mathbb{R})$ in an obvious way.
\par
The $E_6$ internal symmetry group in five dimensions  comes from the $E_{11}$ generators $K^{i}{}_{j}$,
$R^{-}$, $R^0$, $R^+$, $R^{ij\alpha}$, $R^{i_1\dots i_4}$ where $i,j=1,\ldots ,5$ as well as  the negative
root generators $R_{ij\alpha}$, $R_{i_1\dots i_4}$. The maximal compact, or equivalently Cartan involution
invariant, subgroup of $SL(5,\mathbb{R})$ is  $SO(5)$ which are just the Lorentz transformations in the upper
five dimensions. Under this $SO(5)$  the generators $K^{i}{}_{j}$ decompose to  $K^{(i j)}$ and $K^{[ij]}$,
which are the ${\bf 10}$ and ${\bf 5}$ representations, the latter being just the Lorentz generators. The
decomposition of the other fields is obvious.  The local $USp(8)$ symmetry consists of the 36 generators
$K^{[ij]}$, $R^-$, $R^{ij\alpha}- R_{ij\alpha}$ and $ R^{i_1\dots i_4}-R_{i_1\dots i_4}$. The remaining
generators of $E_6$ lead in the non-linear realisation to the 42 scalars of the theory.

The 1-form $E_{11}$ generators in the five dimensional theory are easily seen from equation (6.1) to be given
by
  $$
  K^a{}_i ({\bf 5,1}), R^{ai\alpha} ({\bf \overline{5},2}), R^{a i_1\ldots i_3} ({\bf 10,1}),
  R^{a i_1\ldots a_5 \alpha} ({\bf 1,2}),
  \eqno(6.3)
  $$
which make up the $\bf  \overline{27}$ of $E_6$ that is $R^{aN}$. The numbers in brackets denote the
$SL(5,\mathbb{R})\otimes SL(2,\mathbb{R})$ representations. The 2-form $E_{11}$ generators are easily seen to
be given by
$$
R^{a_1a_2 \alpha} ({\bf 1,2}), R^{a_1a_2ij} ({\bf \overline{10},1}),R^{a_1a_2\ldots a_5 i_1\ldots i_4\alpha}
({\bf 5,2}), R^{a_1a_2\ldots a_5 i_1\ldots i_5 j} (\bf \overline{5},1) \eqno(6.4)$$ which is the $\bf 27$ of
$E_6$ {\it i.e.} $R^{a_1a_2}_N$.

Examining equation (6.2) we find the Lorentz scalar members of the $l$ multiplet are given by
$$
P_i ({\bf 5,1}), Z^{i\alpha} ({\bf \overline{5}, 2}), Z^{ijk} ({\bf 10,1}), Z^{i_1\ldots i_5\alpha} ({\bf
1,2}) \eqno(6.5)$$ which make up the $\bf \overline{27}$ of $E_6$ i.e$ Z^N$, while the one forms are given by
$$
Z^{a\alpha} ({\bf 1, 2}), Z^{a ij} ({\bf \overline{10},1}),  Z^{a i_1\ldots i_4,\alpha}({\bf 5,2}) , Z^{a
i_1\ldots i_5,\alpha}({\bf  \overline{5},1}) \eqno(6.6)$$ which make up the $\bf 27$ of $E_6$, {\it i.e.}
$Z^a_N$.

The supergravity with gauge group $SO(6)$  has a cosmological constant resulting from a non-zero field
strength for the 4-form gauge field $A_{a_1\ldots a_4}$. This is the self-dual field strength of the IIB
theory and this is an  $SL(2,\mathbb{R})$ singlet the gauge group $SO(6)$ commutes with the manifest
$SL(2,\mathbb{R})$ symmetry of the IIB theory in ten dimensions. We will have to reorganise  all the above
fields into representations of $SO(6)\otimes SL(2,\mathbb{R})$. This is straightforward once one realises that
the this $SO(6)$ has an $SO(5)$ sub-algebra that is just the Lorentz transformations in the upper five
dimensions and so  the Cartan involution invariant sub-algebra of $SL(5,\mathbb{R})$. Since all  the above
generators transform under $SL(5)\otimes SL(2)$ we just perform the decomposition of the generators under the
first factor to $SO(5)$ and then reconstitute the resulting generators into those of $SO(6)\otimes
SL(2,\mathbb{R})$.

To find the gauged supergravity of interest we must take $F_{11}^{(0)}= SO(6)$  as the  gauge algebra is just
$F_{11}^{(0)}$. As such, in this case, $F_{11}^{(0)}$ only contains the $\bf 15$ of $SO(6)$ out of all the
$SO(6)$ representations in the adjoint ($\bf 78$) of $E_6$. We note the the  $\bf 78$ of $E_6$ decomposes into
the $\bf (1,3)\oplus (20,2) \oplus(35,1)$ of $SL(6,\mathbb{R})\otimes SL(2,\mathbb{R})$.  In fact the $\bf
(35,1)$ decomposes under $SO(6)\otimes Sl(2,\mathbb{R})$ to contain the $\bf (20,1)$ and the $\bf (15,1)$ and
it is the latter which is the adjoint of $SO(6)$. Examining the $E_{11}$ generators that lead to $E_6$ we
find that the $SO(6)$ algebra consists of the generators $F_{11}^{(0)}=\{ K^{[ij]} , R^{i_1\dots
i_4}-R_{i_1\dots i_4}\}$ which belong to the $\bf (10,1)$ and $\bf (5,1)$ of $SO(5)\otimes SL(2,\mathbb{R})$
respectively.

The map $\Psi$ maps  $F_{11}^{(0)}=SO(6)$ to a  $\bf (15,1)$ of $l^{(0)}$ which can only consist of the
$SL(2,\mathbb{R})$ invariant generators in equation (6.5) and so
$$
k^{(0)}= \{P_i,Z^{ijk}\} \eqno(6.7)$$ which are the $\bf (5,1)$ and $\bf (10,1)$ of $SO(5)\otimes
SL(2,\mathbb{R})$ and so indeed belong to the $\bf (15,1)$ of $SO(6)\otimes SL(2,\mathbb{R})$.  The
complement contains the generators $k^{(0)\perp}= \{ Z^{i\alpha}, Z^{i_1\ldots i_5\alpha}\}$ which belong to
the  $\bf (5,2)$ and $\bf (1,2)$ of $SO(5)\otimes SL(2,\mathbb{R})$ and so the $\bf (6,2)$ of $SO(6)\otimes
SL(2,\mathbb{R})$.

We recall that  $k^{(0)\perp}$ consists of the objects $W_{MN}Z^N$ and as this is the same projector that
defines $F^{(1)}_{11}$  we conclude that this latter space is also  the $\bf (6,2)$ of $SO(6)\otimes
SL(2,\mathbb{R})$ and so is given by
$$
F^{(1)}_{11}=\{ R^{ai\alpha} ,R^{a i_1\ldots a_5 \alpha} \} \quad .\eqno(6.8)$$ As a result the complementary
space belongs to the $\bf (15,1)$ of $SO(6)\otimes SL(2,\mathbb{R})$ and is given by
  $$
  F^{(1)\perp}_{11}=\{ K^{a}{}_i , R^{a i_1\ldots i_3}\} \quad .\eqno(6.9)
  $$

We note that $\Psi$ maps  $F^{(1)}_{11}$ to $k^{(1)}$ and so this and its complementary space are given by
$$
k^{(1)}= \{Z^{a\alpha} , Z^{a i_1\ldots i_4,\alpha}  \} \ \ {\rm and }\ \ k^{(1)\perp}= \{  Z^{a ij} ,  Z^{a
i_1\ldots i_5,\alpha}\} \eqno(6.10)$$ which belong to the $\bf (6,2)$ and  $\bf (15,1)$  of $SO(6)\otimes
SL(2,\mathbb{R})$ respectively.

By carrying out the commutators of generators of $F^{(1)}_{11}$ with themselves we find that
$$
F^{(2)}_{11}=\{ R^{a_1 a_2ij} ,R^{a_1 a_2 i_1\ldots a_5,j } \} \eqno(6.11)$$ which belongs to the $\bf
(15,1)$ of $SO(6)\otimes SL(2,\mathbb{R})$, while the commutators of $F^{(1)}_{11}$ with $F^{(1)\perp }_{11}$
imply that
$$
F^{(2)\perp}_{11}=\{ R^{a_1 a_2\alpha} ,R^{a_1 a_2 i_1\ldots a_4\alpha } \} \eqno(6.11)$$ which belongs to
the $\bf (6,2)$ of $SO(6)\otimes SL(2,\mathbb{R})$.

We now comment on the physical meaning of the above spaces. As we have mentioned $F^{(0)}_{11}$ is just the
gauge group and it included the $SO(5)$ Lorentz rotations in the upper five dimensions  as well as
transformations that originate from a four index generator  in the upper directions. The subspace $k$ of the
$l$ multiplet contains the generators that lead to  the coordinates which are active in the gauged theory. At
the lowest level these are in the adjoint representation of the gauged group representation and the
generators are given in equation (6.7). These consist of the space-time generators of the internal space
$P_i$ and the $Z^{ijk}$. The corresponding coordinates are $y^i$ and   the $y_{ijk}$.   The former are those
of space-time and can be thought of as belonging to the coset $SO(6)/SO(5)$, while the latter belong to
$SO(5)$. Thus we see that even in this case of gauged supergravity, which unlike most cases is obtainable
from a conventional super gravity by dimensional reduction, the techniques of this paper adds extra
coordinates which make more manifest the underlying gauge symmetry.

At the next level we find in $E_{11}$ the 1-form generators which are in one to one correspondence with the
vector fields of the theory. In particular, the generators in  $F^{(1)\perp}_{11}$ correspond to the vectors
that form the Yang-Mills theory with gauge group $SO(6)$ while those in $F^{(1)}_{11}$ correspond to vectors
in the $\bf (6,2)$ of $SO(6)\otimes SL(2,\mathbb{R})$. The latter can be eaten by the 2-forms whose
associated generators are in $F^{(2)\perp}_{11}$. The 2-forms associated with $F^{(2)}_{11}$ can then be
eaten by the 3-forms etc. The eating process is apparent from the transformations of equations
(5.3.41)-(5.3.43) where one finds that the projectors that define $F^{(n)}_{11}$ occur acting on the naked
gauge parameter of rank $n$. For example,  we find in $\delta A_{aN}$ the term $4g W_{NM} \Lambda^M_a$ and so
we may gauge away the 1-forms in the space projected by $W_{NM}$, that is those in $F^{(1)}_{11}$. Similarly
in $\delta A_{a_1a_2}^N$ there occurs the term $-3g \Theta^M_\alpha \Lambda_{a_1a_2}^\alpha$ implying that we
may gauge away the 2-forms associated with $F^{(2)}_{11}$ etc. This would leave just the fields associated
with $F^{(n)\perp}_{11}$. It is simple to understand why this is the case for any gauging.  The generators in
$k$ lead to  rigid transformations that can be identified with the space-time independent part of the gauge
transformation. As such the $k$ transformations can be identified with the gauge parameters that appear in
naked form, that is without space-time derivatives. When this occurs in the variation of a field we can gauge
it away. However, the  map $\Psi$ identifies $k$ with $F_{11}$ and so it is the fields associated with the
latter that can be gauged away.

The active coordinates can also be given a physical meaning. The 1-form fields which are physical are those
associated with the generators in $F^{(1)\perp}_{11}$ of equation (6.8). They couple to the point particle
and the D3 brane as seen from ten dimensions. The corresponding charges are just found by looking in the $l$
multiplet for  an object with one less space-time index and they are in this case found in $k^{(0)}$ of
equation  (6.7). The corresponding coordinates are just the scalar coordinates that are active as this is the
role of $k^{(0)}$ in the group element of the non-linear realisation. This is very natural as it means that
the generalised  space-time used in this paper  includes just the coordinates corresponding to the branes
which are active. We find the analogous relations between the fields associated with $F^{(2)\perp}_{11}$, the
branes to which they couple and  their corresponding coordinates in  $k^{(1)}$. It is then not surprising
that the generators in $k$  obey equation (5.3.4) as this is just the algebra expected for the brane charges.

One can map all the fields and coordinates of the IIB theory to the eleven dimensional theory just using the
fact that the underlying $E_{11}$ symmetry is unique \cite{12,25}. One finds that the 4-form field
$A_{a_1a_2a_3a_4}, a_1,a_2..=1,\ldots 9$ responsible for the $SO(6)$ gauge field gets mapped over to
$A_{a_1\ldots a_2 10 \ 11}$ which  is part of the six form field. This corrects the statement made by the
authors in reference \cite{29}. The mistake made was to assume that the $SO(6)$ gauge symmetry was related to
the gravity $SL(6,\mathbb{R})$ symmetry which occurs on the reduction from eleven dimensions to five
dimensions. This error is most readily apparent when one considers the way the preferred gravity sub-algebras
of the eleven and IIB theory occur in the $E_{11}$ Dynkin diagram and the fact that the $SO(6)$ symmetry
commutes with the manifest $SL(2,\mathbb{R})$ symmetry of the IIB theory.

Clearly, $D-1$ forms that arise from the compactification of $E_{11}$ fields that are beyond the traditional
fields of supergravity lead to massive theories that can not be found by usual geometric compactification
procedures on traditional supergravities. An example of this is the IIA theory of Romans, whose mass
parameter is dual to a 9-form that arises from the eleven-dimensional field $A_{a_1 \dots a_{10},(bc)}$.
However, as the five-dimensional case examined in this section shows, the gauged $SO(6)$ theory, when seen as
arising from eleven dimensions, involves the 6-form which is a traditional field of eleven dimensional
supergravity. However this does not lead to a geometric interpretation from eleven dimensions as a
non-vanishing 7-form field strength does not admit an decomposition in terms of invariant objects in five
dimensions. Thus the notion of geometric compactification is more restrictive.

\section{Conclusions}
In this paper we have derived the fields, transformations and dynamics of all the  five dimensional gauged
supergravities  from a formulation based on $E_{11}$ and separately by viewing it as a traditional
supergravity and using its local  supersymmetry algebra. The results are in precise agreement providing a
very precise check of the $E_{11}$ programme. The five dimensional case was selected for this test as it
shares with the lower dimensional cases a very rich group structure, but it also possesses all the main
duality features involving fields of higher rank of the supergravities in higher dimensions.

The $E_{11}$ formulation has a field content of form fields, that is fields with one set of totally
antisymmetrised indices, which is democratic that is for a physical degree of freedom of the theory described
by a $p$ form we also find its dual field that is a $5-p-2$ form.  Thus the scalars are dual to 3-forms, the
vectors to 2-forms and we also have  form fields of rank 4 and 5, which are not dual to any physical degree
of freedom of the system but lead to the gauged supergravities and space-filling branes respectively. In
section 2 we derived the $E_{11}$ transformation of these fields for the ungauged theory that is the massless
maximal supergravity theory and so arrived at the gauge transformation of these fields. The dynamics is given
by equating the field strength of a gauge field to that of its dual using the $\epsilon$ symbol with the
field strength for the 4-form gauge field being zero. In section 3 we showed that the supersymmetry algebra
closes precisely when one adds the form fields predicted by $E_{11}$ and that the gauge transformations this
requires are in precise agreement.

The rest of the paper concerned the gauged supergravity theories. In section 4 we deformed  the supersymmetry
algebra to find all the possible gauged supergravities in the framework of the democratic formulation. This
formulation is particularly suited to incorporating the dynamics of the gauged supergravities in that the
dynamics is of almost the same structure as the ungauged case except that the field strengths now contain
additional terms and the five form field strength is dual to the mass deformation parameters suitably
contracted with the scalars.

We then derive the field transformations and the dynamics of the gauged supergravity from the $E_{11}$
viewpoint. An essential role  is played by the generalised space-time associated with the first fundamental
representation $l$ of $E_{11}$. In particular we consider the non-linear realisation $E_{11}\otimes_s l$. An
essential step in the construction of the dynamics is the existence of a linear map $\Psi$ from $E_{11}$ onto
a subspace $k$ of the representation $l$ such that the image is the adjoint representation of a sub-algebra,
denoted $F_{11}$ of $E_{11}$. This map is invariant under $F_{11}$ and it preserves the Lorentz character of
the elements on which it acts. Such a map does not exist in eleven dimensions, however, there is such a map
in the IIA theory and this is responsible for the theory of Romans in ten dimensions. Such maps also exists
in any dimension below ten. The map provides a projection from $E_{11}$  into $F_{11}$ and so splits $E_{11}$
into $F_{11}$ and its complement $F_{11}^\perp$. It also follows that $F_{11}$ is isomorphic to $k$. The
generators of $k$ as well as $E_{11}$ are active in the non-linear realisation and as such one finds  a
space-time with coordinates arising from the presence of $k$ in addition to those of the familiar space-time.
This also implies that we have additional transformations resulting from the presence of $k$ which become
identified with the space-time independent components of the gauge transformations. The latter can be used to
gauge away some of the fields of $E_{11}$, which as a result of the identification of $k$ and $F_{11}$, are
just the fields associated with $F_{11}$. Thus the fields which can not be gauged away are those
corresponding to $F_{11}^\perp$. The additional coordinates do not appear in the final dynamical equation but
their presence is very natural in that they are associated with the branes that couple to the latter fields.
Some of these additional coordinates are just those of the usual space-time,  but in the upper dimensions.
These correspond to the presence of components of the graviton in the upper directions and so to point
particles.

The existence of an invariant map between $E_{11}$ and $l$ divides the generalised coordinates in two sets
one of which is closely associated with the gauging. It also specifies the group which is gauged and the
corresponding constraints on the embedding tensor. The resulting gauge transformations and so the
corresponding dynamics agree precisely with that found in section 4 using supersymmetry.  A very special case
of this technique is that of the Scherk-Schwarz reduction \cite{40}, however, the technique used in this
paper is much more general.

In \cite{bernard} it has been pointed out that the quadratic constraint of the embedding tensor can be
associated to (some of) the representations of the $D$ forms. This is clear in the five-dimensional example
carried out in this paper, given that the quadratic constraints of the embedding tensor project out the ${\bf
{27}} \oplus {\bf {1728}}$ of the product $\Theta \Theta$ \cite{20}, which are the complex conjugates of the
representations of the 5-form fields. The authors of \cite{41} observe that one can interpret the $D$-forms
as Lagrange multipliers whose field equations produce the quadratic constraint of the embedding tensor. In
the five dimensional case this observation can be checked explicitly determining the field strength of the
4-form at order $g$, which contains the 5-forms. This analysis has not been carried out in this paper. It
would be interesting to further investigate in this direction. A month after this paper was originally
submitted, it was explicitly shown in \cite{dWNS} in the case of maximal supergravity in three dimensions
that the field equations of the 3-forms precisely lead to the quadratic constraint of the embedding tensor.

As mentioned earlier there is considerable evidence for the $E_{11}$ part of the non-linear realisation and
for $l$ being the multiplet of brane charges however, there has so far been very little evidence for the $l$
part of the non-linear realisation that is the generalised space-time that $l$ leads to. However, in this
paper we have seen that it is essential for the construction of the gauged supergravities. In particular it
directly leads to the terms in the dynamics that contain no space-time derivatives such as the non-Abelian
terms in the Yang-Mills field strength and the gauge transformations that contain no space-time derivatives.
Indeed, the former can be traced back to derivatives in the Cartan forms with respect to the extra
coordinates while the latter arise from  transformations in the extra coordinates. While there is much that
remains to be understood about the role of the  $E_{11}$ generalised space-time, at least part of it  has
been confirmed indicating that the rest also has a required purpose.

As has already been noted \cite{37} the use of the $E_{11}$ generalised space-time \cite{31} has some
features in common with the more recent generalised geometry \cite{42,43} which also adds structure to that
of traditional space-time. The  $E_{11}$ approach automatically adds to the usual spacetime all the necessary
coordinates and in particular those required to ensure $U$ duality and all the higher symmetries in $E_{11}$.
Those at low level are just the coordinates corresponding to the charges of table 1 \cite{33,34}. Indeed the
necessity of adding the scalar charges in the first column was specifically commented on in reference
\cite{34}. The procedure spelt out in this paper also includes all the effects from higher level field
strengths, or fluxes, and coordinates which occur at the higher levels of $E_{11}$ and the $l$ multiplet,
indeed the map $\Psi$ involves generators associated with all the gauge fields and coordinates which are not
Lorentz scalars.

The generalised geometry programme \cite{42,43} has largely concentrated on the coordinates required for $T$
duality introduced in a systematic way first in \cite{44}. From the $E_{11}$ perspective these are those
found by decomposing to the $O(10,10)$ symmetry and keeping the lowest level coordinates which for the IIA
theory for example are $P_a$ and $Z^{a11}$ corresponding to $x^a$ and $y_a$ respectively \cite{34}. Indeed
one can formulate the string from the $E_{11}$ perspective using these coordinates, however to formulate the
eleven dimensional membrane and five brane one requires more of the coordinates contained in the $l$
representation \cite{34}.

The $E_{11}\otimes l$ non-linear realisation studied in this paper includes as a very special case the old
Scherk-Schwarz dimensional reduction technique \cite{40}. The latter exploited the existence of a rigid
internal symmetry by giving the transformations some limited  dependence on the  upper coordinates. However
in the  $E_{11}\otimes l $ approach  a vast symmetry {\it i.e.} $E_{11}$  can be used in conjunction with all
the coordinates in the $l$ multiplet. We note that this includes symmetries related to vector and higher rank
fields. Indeed, the Romans IIA theory can be found using such a symmetry.

The conformal group applied to $E_{11}\otimes l$ results in the usual coordinates of space-time having
general coordinate transformations. It would be good to understand what the conformal group implies for the
higher coordinates and indeed what is their corresponding geometry. Particularly in this context it would be
good to see how the $E_{11}$ and generalised geometry approaches compare and  what they can learn from each
other. That the generalised geometry required addition coordinates beyond those of the $x^a$ and $y_a$ of the
doubled torus of was readily apparent from the $E_{11}$ picture \cite{34,37}. However, it would be
interesting to see how the geometrical aspects of the generalised geometry programme appear when viewed from
an $E_{11}$ perspective.

One advantage of the $E_{11}$ approach is that it unifies many aspects of supergravity and so string theory.
The gauged supergravities are such examples, while some can be obtained by dimensional reduction of the ten
and eleven dimensional supergravity theories there are many others which have no higher dimensional origin.
However, each gauged supergravity is associated with a non-trivial $D-1$ form and it is part of the unifying
$E_{11}$ non-linear realisation \cite{29}. Previously the gauged supergravities which had no higher
dimensional supergravity origin could only be obtained by deforming the supersymmetry algebra and so were
outside the framework of M theory as usually envisaged. It is straightforward to apply the $E_{11} \otimes_s
l$ non-linear realisation described in this paper to all the other cases and obtain all the gauged maximal
supergravities in any dimension.

\vskip 2cm

\section*{Acknowledgments}
The research of P.W. was supported by a PPARC senior fellowship PPA/Y/S/2002/001/44. The work of both authors is
also supported by a PPARC rolling grant PP/C5071745/1 and the EU Marie Curie, research training network grant
HPRN-CT-2000-00122.

\vskip 2cm
\begin{appendix}
\section{5-forms in 5 dimensions from $E_{11}$}
In this appendix we want to extend the analysis of section 2 to include a 5-form generator. This operator is
associated to a field with five antisymmetric indices in five dimensions, which has no field strength and
therefore has no propagating degrees of freedom. Fields with $D$ antisymmetric indices in $D$ dimensions are in
general associated to spacetime-filling branes, that have a crucial role in the construction of orientifold
models.

The 5-form generator $R^{a_1 \ldots a_5 , \alpha}_M$ occurs in the commutator
  \begin{equation}
  [ R^{a_1 a_2 }_M , R^{a_3 a_4 a_5 , \alpha} ] = R^{a_1 \ldots a_5 , \alpha}_M \quad , \label{A.1}
  \end{equation}
and the Jacobi identity between the operators $R^{a,M}$, $R^{bc}_N$ and $R^{de}_P$ leads to the commutation
relation
  \begin{equation}
  [R^{a,M}, R^{bcde}_{NP}] = -2 D^\alpha_{[N}{}^M R^{abcde , \beta}_{P]} g_{\alpha\beta} \quad . \label{A.2}
  \end{equation}
A constraint on this 5-form operator comes from the Jacobi identity between the operators $R^{a ,M}$, $R^{b,N}$
and $R^{cde , \alpha}$, which is
  \begin{equation}
  d^{MNP} R^{abcde , \alpha}_P + 4 D^\beta_P{}^{(M} S^{\alpha N) [PQ]} g_{\beta \gamma} R^{abcde , \gamma}_Q =0
  \quad .\label{A.3}
  \end{equation}
The representation of the 5-form generator is contained in the ${\bf 78 \otimes 27} = {\bf 27 \oplus 351
\oplus 1728}$, as can be seen from its $E_6$ index structure, and it can be shown that eq. (\ref{A.3})
restricts this generator to be in the ${\bf 27 \oplus 1728}$ of $E_6$, in exact agreement with \cite{29,30}.

In order to determine the gauge transformations of the field associated to this generator, we have to extend
the form of the group element of eq. (\ref{2.37}), and we therefore write
  \begin{eqnarray}
  & &   g_A = {\rm exp} (A^{M}_{a_1 \ldots a_5, \alpha} R^{a_1 \ldots a_5 ,\alpha }_M ) \
  {\rm exp} (A^{MN}_{a_1 \ldots a_4}
  R^{a_1 \ldots a_4 }_{MN} ) \nonumber \\
  & &  {\rm exp} (g_{\alpha\beta} A^{\alpha}_{a_1 \ldots a_3} R^{a_1 \ldots a_3, \beta } ) \ {\rm exp}
  (A^{M}_{a_1 a_2} R^{a_1 a_2 }_M ) \ {\rm exp} (A_{a,M} R^{a,M } ) \quad .
  \label{A.4}
  \end{eqnarray}
Acting with
  \begin{equation}
  g_0^{(5)} = {\rm exp} (a^{M}_{a_1 \ldots a_5, \alpha} R^{a_1 \ldots a_5 ,\alpha }_M ) \label{A.5}
  \end{equation}
leads to a transformation of the 5-form field
  \begin{equation}
  \delta A^{M}_{a_1 \ldots a_5, \alpha} = a^{M}_{a_1 \ldots a_5, \alpha} \quad , \label{A.6}
  \end{equation}
while acting with the group element of eq. (\ref{2.42}) leads to
  \begin{equation}
  \delta A^{M}_{a_1 \ldots a_5, \alpha} = a^{M}_{[a_1 a_2} A_{a_3 a_4 a_5 ], \alpha} \label{A.7}
  \end{equation}
and acting with the one of eq. (\ref{2.44}) leads to
  \begin{eqnarray}
  \delta A^{M}_{a_1 \ldots a_5, \alpha} & =& -2 a_{[a_1 , N}
  A_{a_2 \dots a_5 ]}^{PM} D^\beta_P{}^N g_{\alpha
  \beta} + {1 \over 2} A_{[a_1 a_2 }^{M} A_{a_3 a_4 }^{N} a_{a_5 ] , P} D^\beta_N{}^P g_{\alpha
  \beta} \nonumber \\
  & + & {2 \over 5!} A_{[a_1 , N} A_{a_2 , P} A_{a_3 , Q} A_{a_4 , R} a_{a_5 ] , S}
  d^{RST} D^\gamma_T{}^Q S^{\delta P [UM]} D^\beta_U{}^N g_{\gamma \delta } g_{\alpha \beta} \nonumber \\
  & - & {1 \over 6} A_{[a_1 a_2 }^{M} A_{a_3 , N} A_{a_4 , P} a_{a_5 ], Q} d^{PQR}
  D^\beta_R{}^N g_{\alpha \beta} \quad . \label{A.8}
  \end{eqnarray}

From eq. (\ref{A.4}) one can compute the part of the Maurer-Cartan form which is proportional to $R^{a_1
\ldots a_5 , \alpha}_M$. The result is
  \begin{eqnarray}
  G_{\mu a_1 \dots a_5 , \alpha}^M & =& \partial_\mu A^{M}_{a_1 \ldots a_5, \alpha} + 2 A_{[a_1 , N}
  \partial_\mu A^{PM}_{a_2 \dots a_5 ]} D^\beta_P{}^N g_{\alpha \beta} - A^{M}_{[a_1 a_2 } \partial_\mu
  A^{\beta}_{a_3 a_4 a_5 ]} g_{\alpha \beta} \nonumber \\
  & +& A_{[a_1 , N} A_{a_2 , P} \partial_\mu A^{\gamma }_{a_3 a_4 a_5 ]} S^{\delta P [ QM]}
  D^\beta_Q{}^N g_{\gamma \delta} g_{\alpha \beta} \nonumber \\
  &-& {1 \over 3} A_{[a_1 , N} A_{a_2 , P}
  A_{a_3 , Q} \partial_\mu A^{R}_{a_4 a_5 ]} D^\gamma_R{}^Q S^{\delta P [SM]} D^\beta_S{}^N g_{\gamma
  \delta} g_{\alpha \beta} \nonumber \\
  & + & {2 \over 5!} A_{[a_1 , N} A_{a_2 , P}
  A_{a_3 , Q} A_{a_4 , R} \partial_\mu A_{a_5 ] , S} d^{RST} D^\gamma_Q{}^T S^{\delta P[UM]}
  D^\beta_U{}^N g_{\gamma \delta } g_{\alpha \beta} \quad . \label{A.9}
  \end{eqnarray}

As was already discussed in section 2, consistency requires that the fields transform properly under the
closure of $E_{11}$ with the conformal group \cite{9}. This corresponds to promoting the global
transformations to local ones, which leads to eq. (\ref{2.56}) and
  \begin{equation}
  a_{a_1 \dots a_5 , \alpha}^M = 5 \partial_{[a_1 } \Lambda_{a_2 \dots a_5 ] , \alpha}^M \quad . \label{A.10}
  \end{equation}
The resulting gauge transformations are the ones of the 5-forms on maximal five-dimensional supergravity,
that is the gauge transformations that one would obtain imposing the closure of the supersymmetry algebra on
the 5-forms in five dimensions. The corresponding field-strength would result from eq. (\ref{A.9}) with all
the indices antisymmetrised, but this object vanishes identically because it has six indices. Thus the 5-form
fields have no field strength, and they do not correspond to any propagating degree of freedom.

\section{Generalised coordinates in a toy model}
In this paper we have seen how generalised coordinates have played a crucial role in formulating the dynamics
of the gauged supergravities. Such coordinates have not been used in this way before and in this appendix we
will illustrate some of the steps for a simple model so that the reader can gain some familiarity with the
techniques without all the complications of the five dimensional gauged supergravity theory. We will see that
a very simplified case of the toy model is just the Scherk-Schwarz dimensional reduction procedure \cite{40}.

We consider an algebra that has the generators $P_a$ and $V^\alpha$ and $R^{a\alpha}$ and $R^\alpha$. They
obey the relations
  \begin{equation}
  [V^\alpha,V^\beta]=-gf^{\alpha\beta}{}_\gamma V^\gamma,\ \ [R^\alpha, V^\beta ]= f^{\alpha\beta}{}_\gamma
  V^\gamma,\ \  [R^{a \alpha}, V^\beta ]=0,\ \ [R^{a \alpha},P_b]=\delta_b^a V^\alpha \label{B.1}
  \end{equation}
and
  \begin{equation}
  [R^\alpha,R^\beta]=f^{\alpha\beta}{}_\gamma R^\gamma,\ \ [R^\alpha,R^{b \beta}]=f^{\alpha\beta}{}_\gamma R^{b
  \gamma} \quad . \label{B.2}
  \end{equation}
We note that if we define $Y^\alpha= V^\alpha +gR^\alpha$ then
  \begin{equation}
  [Y^\alpha,Y^\beta]=g f^{\alpha\beta}{}_\gamma Y^\gamma,\ \ [R^\alpha,Y^\beta]=f^{\alpha\beta}{}_\gamma
  Y^\gamma,\ \ [V^\alpha,Y^\beta]=0 \quad . \label{B.3}
  \end{equation}
The generators $P_a$ and $V^\alpha$ are to be associated with a generalised space-time while $R^\alpha$
generate the group $G$ and the generators  $R^{a\alpha}$ belong to the adjoint representation of $G$.

The group element has the form
  \begin{equation}
  g=e^{x^aP_a} e^{y_\alpha Y^\alpha} e^{A_{a\alpha}(x)R^{a\alpha}}g_\varphi(x) \label{B.4}
  \end{equation}
where $g_\varphi(x)=e^{\varphi _\alpha R^\alpha}$ is a group element of $G$ and $x^a$ and $y_\alpha$ are the
coordinates of the generalised space-time. The fields $A_{a\alpha}$ and $\varphi$ depend only on the
coordinates $x^a$ and not on the $y_\alpha$'s. In doing so we assume that the local subgroup of the
non-linear realisation possess $y$ dependent transformations that can be used to bring the group element to
the above form from the most general form. We will also assume that the part of the local subgroup that
depends only on the coordinates $x^a$  is the group that contains the  identity  element.  The above model
emerges from that of the five dimensional gauged supergravity theory if we truncate to the above fields and
coordinates, take those that remain to transform in the adjoint representation and set
$\Theta_\alpha^N=\delta_\alpha^N, W_{MN}=0$.

To calculate the Cartan forms we need that
  \begin{equation}
  e^{-y_\alpha Y^\alpha}de^{y_\alpha Y^\alpha}= dy_\alpha e^\alpha{}_\beta Y^\beta \label{B.5}
  \end{equation}
where $ e^\alpha{}_\beta $ are the vierbeins, or Cartan forms, for the group $G$. The Cartan forms are then
easily found to be given by
  \begin{eqnarray}
  g^{-1}dg &=& dx^\mu E_\mu{}^a P_a +dy_\alpha E^{\alpha}{}_\beta V^\beta +dx^\mu E_{\mu,\beta} V^\beta+dy_\alpha
  E^{\alpha ,a}P_a \nonumber \\
  &+&dx^\mu G_{\mu, \alpha}R^\alpha +dx^\mu G_{\mu, a\alpha}R^{a\alpha} +dy_\alpha
  G^\alpha_{,\beta}R^\alpha+dy_\beta G^\alpha_{,a\beta}R^{a\beta} \label{B.6}
  \end{eqnarray}
where
  \begin{equation}
  E_\mu{}^a=\delta_\mu^a \ ,\ \ E^{\alpha}{}_\beta=e^{\alpha}{}_\gamma(e^{-\varphi\cdot f})^\gamma{}_\alpha \ ,\ \
  E_{\mu,\alpha}= -A_{\mu\beta}(e^{-\varphi\cdot f})^\beta{}_\alpha \ ,\ \ E^{\alpha ,a}=0 \label{B.7}
  \end{equation}
and
  \begin{eqnarray}
  & & G_{\mu, \alpha} R^\alpha =g_\varphi^{-1}\partial_\mu g_\varphi \ ,\ \ G_{\mu, a\alpha}= \partial_\mu
  A_{a\beta}(e^{-\varphi\cdot f})^\beta{}_\alpha \ ,\nonumber \\
  & &  G^\alpha_{,\beta}=g e^\alpha{}_ \gamma  (e^{-\varphi\cdot
  f})^\gamma{}_\beta \  , \ \ G^\alpha_{,a\beta}
  R^{a\beta}=g e^\alpha{}_ \gamma A_{a\delta}f^{\gamma\delta}{}_\epsilon (e^{-\varphi\cdot
  f})^\epsilon{}_\beta \quad . \label{B.8}
  \end{eqnarray}
In carrying out this calculation we have used the fact that
  \begin{equation}
  e^{-\varphi_\beta R^\beta} R^\alpha e^{\varphi_\beta R^\beta} = (e^{-\varphi\cdot f})^\alpha{}_\beta
  R^\beta \label{B.9}
  \end{equation}
where $(\varphi\cdot f)^\alpha{}_\beta= \varphi_\gamma f^{\gamma \alpha}{}_\beta$ which contains the only
dependence on $y_\alpha$.

In the method of non-linear realisations one usually uses the inverse vierbein to make the first index on the
$G$'s ``flat'' that is $\hat G_{a,\bullet}= (E^{-1})_a{}^\mu  G_{\mu,\bullet} +(E^{-1})_a{}_\gamma
G^\gamma{}_{,\bullet}$ where $\bullet$ stands for the indices on the $R$'s. The $\hat G$'s are inert under
the rigid transformations $g\to g_0 g$ up to possible compensating local transformations which maintain the
form of the group element.   One finds that
  \begin{equation}
  \hat G_{a, \alpha} R^\alpha = g_\varphi^{-1}(\delta _a^\mu \partial_\mu +gA_{a\alpha }R^\alpha )g_\varphi
  \label{B.10}
  \end{equation}
while
  \begin{equation}
  \hat G_{a, b\alpha}= (\delta _a^\mu \partial_\mu A_{b\beta}+ g  A_{a\delta}A_{b\gamma}
  f^{\delta\gamma}{}_\epsilon )(e^{-\varphi\cdot f})^\epsilon{}_\alpha \quad . \label{B.11}
  \end{equation}
We note that the ``flat'' Cartan forms do not contain the factor $e^\alpha{}_ \gamma $ and as a result are
independent of $y$. Usually the dynamics is constructed from the  $\hat G$'s in which case the dynamics is
independent of $y_\alpha$, however,  in this toy model this is not quite the case but the conclusion is the
same.

The transformations of the fields are a little more lengthy  to calculate. Using the commutation relations of
eqs. (B.1), (B.2) and (B.2) and taking the rigid group element of the form
  \begin{equation}
  g_0= e^{b_\alpha V^\alpha} e^{a_\alpha R^\alpha}e^{a_{a\alpha} R^{a\alpha}} \label{B.12}
  \end{equation}
we must  evaluate $g_0g$ to lowest order in the parameters of $g_0$. We find that $x^a$ is unchanged,
$y_\alpha$  becomes a complicated function of $y_\alpha$ and the parameters of $g_0$, while the fields
transform as
  \begin{equation}
  A^\prime_{a\alpha}=A_{a\alpha} +a_{a\alpha}+c_\gamma A_{a\delta } f^{\gamma\delta}{}_\alpha \ , \ \
  g_{\varphi^\prime}=e^{c_\gamma R^\gamma} g_\varphi \label{B.13}
  \end{equation}
where
  \begin{equation}
  c_\gamma= -g(b_\gamma+x^c a_{c\gamma}) + a_\gamma \quad . \label{B.14}
  \end{equation}
In carrying out this calculation we have used $y$ dependent compensating transformations to maintain the form
of the group element as in eq. (B.4). The $x$ dependent part of $c_\alpha$ arises from passing
$e^{a_{a\alpha}R^{a\alpha}}$ past $e^{x^aP_a}$ to create a $V^\alpha$ transformation and then processing
this. We note that $- g a_{a\alpha}= \partial_a c_\alpha$.

By explicitly calculating the variation of the fields using eqs. (B.13) and (B.14) one finds that  the
covariant objects are given by
  \begin{equation}
  \hat G_{a, \alpha} R^\alpha = g_\varphi^{-1}(\delta _a^\mu \partial_\mu +gA_{a\alpha }R^\alpha )g_\varphi
  \label{B.15}
  \end{equation}
and
  \begin{equation}
  F_{a b\alpha}= 2(\delta _{[a}^\mu \partial_\mu A_{b]\beta}+ {g\over 2}  A_{a\delta}A_{b\gamma}
  f^{\delta\gamma}{}_\epsilon )(e^{-\varphi\cdot f})^\epsilon{}_\alpha \label{B.16}
  \end{equation}

The invariant action is then given by
  \begin{equation}
  \int d^Dx ({1\over 2}\hat G_{a, \alpha}\hat G_{a, \beta}+{1\over 4}F_{a b\alpha}F^{a b}_\beta
  )g^{\alpha\beta} \quad , \label{B.17}
  \end{equation} which is the Yang-Mills action coupled to scalars which are in a
non-linear realisation of $G$.

We note that $F_{a b\alpha}$ is not quite $2\hat G_{[a ,b]\alpha}$ since there is a factor of 2 out on the
$AA$ term. This discrepancy arises from the fact that the $\hat G$'s still transform under compensating local
transformations that are $y$ dependent. Taking this into account one arrives at the above covariant
expressions.  This point is explained in detail in section 5.

We will now explain that if one takes a particularly simple case one finds the dimensional reduction of
Scherk and Schwarz. We consider a theory that has undergone a dimensional reduction with the result that it
contains some scalars $\varphi$ in a non-linear realisation, gravity which we neglect and a Kaluza-Klein
vector $A_a{}^\star$ which we keep. Let $x^a$ be the coordinates of the remaining space-time after the
dimensional reduction and $y_\star$ one of the other coordinates that lies in the same direction as the
vector field. In this case we can identify $V=P_\star$ $R^a=-K^a{}_\star$, $y=y_\star$ and $A_a{}^\star=A_a$
where the $K$'s belong to the $SL(D,\mathbb{R})$ algebra associated with gravity in the higher dimension.
All these generators are singlets under the group $G$ to which the scalars belong and one has
$[-K^a{}_\star,P_b]=\delta_a^b P_\star$ as required.  The group element of equation (B.4)  takes the form
  \begin{equation}
  g=e^{x^aP_a} e^{y Y} e^{A_{a}(x)R^{a}}g_\varphi(x) \label{B.18}
  \end{equation}
where now $Y=V+gT$,  $T=m_\alpha R^\alpha$ is just a specific element of $G$ and $m_\alpha$ are constants.
Clearly, the dynamics is $y$ independent as the Cartan forms $g^{-1}dg$ do not contain this coordinate. The
reason being in this case that $Y$ form an Abelian algebra. In general in the Scherk-Schwarz dimensional
reduction,  and indeed in this case,  one finds a mass term for the scalars. The reason it is absent in the
above toy model is that $\Theta^\beta_\alpha=\delta^\beta_\alpha$ and so is rather trivial. It is
straightforward to generalise the toy model to the case of a non-trivial $\Theta$ as is the case for the
gauged supergravities of sections five.

\end{appendix}

\end{document}